\documentclass[aps,prx,onecolumn,groupedaddress,nofootinbib,floatfix,12pt]{revtex4-2}
\usepackage{amsmath,amssymb,amsthm,bm}
\usepackage{dsfont}
\usepackage{graphicx}
\graphicspath{{./}}
\usepackage{hyperref}
\hypersetup{colorlinks=true,linkcolor=blue,citecolor=blue,urlcolor=blue}
\usepackage{xcolor}
\usepackage{placeins}
\usepackage{booktabs}
\newcommand{\Jz}{\hat J_z}
\newcommand{\Jx}{\hat J_x}

\newcommand{\JSD}{\mathrm{JSD}}
\newcommand{\TV}{\mathrm{TV}}
\newcommand{\Tr}{\mathrm{Tr}}
\newcommand{\Sinst}{S_{\rm inst}}
\newcommand{\Tobs}{T_{\rm obs}}
\newcommand{\BN}{\mathcal{B}_N}

\newcommand{\mstar}{m_*}
\newcommand{\Pdb}{\hat P_{\rm db}}
\newcommand{\Hdb}{\hat H_{\rm db}}
\newcommand{\sigmax}{\sigma_x^{(\rm par)}}
\newcommand{\sigmay}{\sigma_y^{(\rm par)}}
\newcommand{\sigmaz}{\sigma_z^{(\rm par)}}
\newtheorem{proposition}{Proposition}

\newtheorem{remark}{Remark}
\makeatletter
\def\Dated@name{}%
\let\@date\@empty%
\makeatother
\newcommand{\plateauExcess}{+0.0147}
\newcommand{\plateauErr}{0.0009}
\newcommand{\plateauChiSq}{1.08}
\newcommand{\plateauTraj}{384{,}000}
\newcommand{\plateauRep}{+0.015}
\newcommand{\tobsDiff}{-0.0006}
\newcommand{\tobsDiffErr}{0.0011}
\newcommand{\tobsChiSq}{0.87}
\newcommand{\tobsNxi}{0.007}
\newcommand{\tobsNxiHi}{0.29}
\newcommand{\tobsPhaseLo}{0.05}
\newcommand{\tobsPhaseHi}{1.9}
\newcommand{\tobsNcount}{13}
\newcommand{\closureExcess}{-0.0001}
\newcommand{\closureErr}{0.0020}
\newcommand{\closureSigma}{0.1}
\newcommand{\closureTraj}{80{,}000}

\newcommand{\Nclose}{1000}
\newcommand{\xiClose}{0.005}
\newcommand{\xiPlateauEdge}{0.02}
\newcommand{\xiKneeHalf}{0.009}
\newcommand{\totalTraj}{1.48\times10^{6}}
\newcommand{\totalCells}{77}
\newcommand{\secFMQdeepA}{-0.0009}
\newcommand{\secFMQdeepB}{-0.0006}
\newcommand{\secSigDeep}{0.2}
\newcommand{\secNlistdeep}{350,\,370,\,400}
\newcommand{\secXideep}{12.3,\,9.2,\,6.1}
\newcommand{\secNlistcross}{440,\,470,\,500}
\newcommand{\secXicross}{3.5,\,2.3,\,1.5}
\newcommand{\secNfivehundredExc}{+0.013}
\newcommand{\secNfivehundredSig}{3.3}
\newcommand{\secTrajPer}{16{,}000}

\begin{document}
\setlength{\emergencystretch}{3em}
\raggedbottom
\title{Reading Weakly, Acting Strongly:\\[6pt]
\large A Static Parity Horizon and its Dynamical Bypass\\ in the Monitored Lipkin--Meshkov--Glick Model}
\author{Stavros Mouslopoulos}
\affiliation{Faculty of Science and Engineering, University of Nottingham Ningbo China, Ningbo 315100, China}


\begin{abstract}

We study the broken-symmetry phase of the Lipkin--Meshkov--Glick (LMG) model, where the two lowest states form a near-degenerate parity doublet. Each member of the doublet is built from two macroscopically separated magnetisation wells, and the exponentially small splitting between them is generated by tunnelling. The first point of the paper is that the same instanton action $\Sinst$ which controls this splitting also controls the amount of parity information that can be extracted from a static $\hat J_z$ readout.

This is not the obvious conclusion one would draw from the measured observable alone. The operator $\hat J_z$ measures magnetisation, not parity. Thus a $\hat J_z$ histogram distinguishes the two wells very efficiently, but is almost blind to the relative sign between the two well amplitudes, namely to the parity label of the doublet. Using WKB barrier arguments together with exact diagonalisation, we find that the parity information is carried by the exponentially small tunnelling-barrier region, including the exact parity-forbidden central node. As a result, the spectral gap, the total variation distance, and the nonlinear distinguishability measures --- the Jensen--Shannon divergence and the Chernoff information --- share a single instanton exponent, although a naive small-deviation expansion in the lobes would suggest twice this exponent. Exact diagonalisation on grids extending to $N=4500$, with reliability-masked fitting windows, supports a common leading exponent for the gap and for the three histogram distinguishability measures; in the largest reliable fitting windows the fitted exponents lie within a few percent of the WKB instanton value.

The same coupling has a very different meaning on the backaction side. Although the static $\hat J_z$ record reads the parity label only exponentially weakly, the operator $\hat J_z$ acts strongly within the parity doublet: its off-diagonal matrix element grows linearly with $N$. Hence the bath can disturb the parity label much more strongly than it can read it from a frozen histogram. We call this separation the static parity horizon. It is a benchmark for the idealised static $\hat J_z$ channel with $\hat H=0$, not a universal bound on time-resolved monitoring.

We finally restore the full monitored LMG dynamics and ask whether a continuous record can bypass this static limitation. Continuous-monitoring simulations, using 1.48 million full-LMG trajectories with matched QND controls across 77 independent settings, show that a time-resolved homodyne record can extract parity information which is hidden from the single-shot static histogram. The bypass appears over a finite window of system sizes, organised most clearly by the ratio between coherent doublet rotation and measurement-induced dephasing. The excess over the static benchmark is statistically resolved in this window and closes again under strong measurement.

\end{abstract}

\maketitle
\section{Introduction}
\label{sec:intro}
\paragraph*{Physical scenario.} Let us start from an ideal finite-$N$ collective-spin double well: the broken phase of a $\mathbb{Z}_2$-symmetric Hamiltonian, with the ferromagnet below $T_c$ as the familiar picture. At mean-field level the system has two broken-symmetry branches of opposite magnetisation, $\mu=\pm\mstar$, while at finite $N$ these branches communicate only by tunnelling. The lowest spectrum is therefore a parity doublet, $|E_0\rangle$ (even) and $|E_1\rangle$ (odd), with an exponentially small gap $\Delta E\sim e^{-N\Sinst}$ and with $\Jz$-distributions spread over \emph{both} wells in symmetric or antisymmetric combination. A $\Jz$ histogram easily tells the two wells apart. What it cannot resolve efficiently is the relative \emph{sign} between the two well amplitudes, namely the exact conserved $\mathbb{Z}_2$ parity label of the doublet. Two facts set up everything that follows: the parity-conditioned histograms differ only in the exponentially thin barrier overlap, while $\Jz$ has an extensive off-diagonal element inside the same doublet. We take this mismatch as the starting point of the paper.

One caveat belongs here at the outset. The parity eigenstates $|E_0\rangle,|E_1\rangle$ are not what a large symmetry-broken body normally settles into by itself: in a realistic environment, or after an explicitly symmetry-breaking preparation, the body sits in a well-localised magnetisation branch ($|P\rangle$ or $|R\rangle$), for which a $\Jz$ histogram is trivially informative. The parity-eigenstate input studied here is therefore a prepared input, requiring engineered preparation or non-demolition post-selection (Sec.~\ref{sec:discussion}); the ``ferromagnet'' is used as the physical double-well analogy, not as a claim about an ordinary laboratory magnet. The distinction we study is between reading a single well and reading the conserved parity label that combines the two wells; it is formalised in Sec.~\ref{sec:two-roles}. A companion study~\cite{SweepPaper} treats a separate coherence-detectability window of the same model.

We now ask what this same setup implies for the \emph{nonlinear} information measures of the two histograms, and what follows operationally. On the analytical side (Sec.~\ref{sec:readout}), a Wentzel--Kramers--Brillouin (WKB) path-additivity argument carries the tunnelling (instanton) exponent $\Sinst$ to the doublet splitting and, immediately, to the total variation distance, which is linear in the histogram difference. The harder point, supported below by the WKB barrier mechanism and ED, is that the \emph{nonlinear} measures inherit the \emph{same} exponent rather than the doubled exponent $e^{-2N\Sinst}$ which a small-deviation expansion would predict in the lobes. These measures --- the Jensen--Shannon divergence (JSD) and the Chernoff information --- quantify distinguishability by weighting the two distributions against each other, not merely by subtracting them. Two effects produce the single exponent. In the barrier, the mean probability $\bar p$ is itself $\sim e^{-N\Sinst}$, so the quadratic surrogate $(p^{(0)}-p^{(1)})^2/\bar p$ recovers one factor of $e^{-N\Sinst}$. For the JSD, moreover, the exact parity-forbidden node $p^{(1)}_0=0$ breaks the small-deviation premise outright and supplies a closed-form term linear in $e^{-N\Sinst}$ that anchors the exponent from below. For the Chernoff information the same node and barrier region provide an instanton-scale deficit at the observed interior optimiser $s^*$, with the measure-dependent prefactor handled numerically below. On the operational side (Secs.~\ref{sec:readout}--\ref{sec:backaction}), we take the static $\Jz$-readout channel ($\hat H=0$), the simplest commuting-record benchmark for a $\Jz$-monitored LMG bath. There the asymptotic information recorded by the bath about parity --- the accessible (Holevo) information --- equals \emph{exactly} the Jensen--Shannon divergence of the parity-resolved $\Jz$-distributions (Proposition~\ref{prop:horizon}), while the same operator acts transversely inside the doublet with $|J_{01}|\to N\mstar/2$. The two sides are strongly asymmetric: the static parity \emph{distinguishability} is instanton-suppressed, $\sim e^{-N\Sinst}$, whereas the backaction is an extensively large \emph{rate}, $\Gamma_z^{(\rm par)}=2\gamma_\phi|J_{01}|^2\to\gamma_\phi N^2\mstar^2/2$. This is the sense in which the bath reads weakly and acts strongly. It is not a failure of pointer-state theory; it is a statement about a label, parity, that is not the pointer observable of a $\Jz$-coupled bath.

\paragraph*{The single organising mechanism.} The paper is organised by one simple ingredient: $\Jz$ is the operator through which the spin system is monitored, and the same coupling has two inequivalent roles. The first role is the static readout role. In the static channel ($\hat H=0$), the bath record is diagonal in the $\Jz$-eigenbasis and amounts to a phase-blind $\Jz$ histogram. This histogram separates the two magnetisation wells easily, but it carries only exponentially small information about the parity label, because $|E_0\rangle$ and $|E_1\rangle$ differ in their probabilities only through the forbidden barrier region and the exact central node. This is the content of Proposition~\ref{prop:horizon}, together with the instanton analysis, and it defines the static parity horizon. The second role appears when the LMG Hamiltonian is restored and the full time-resolved record is retained. The record is then no longer a frozen histogram; it contains temporal correlations. The same $\Jz$ coupling which reads parity weakly acts strongly inside the doublet, where its transverse matrix element grows as $|J_{01}|\to N\mstar/2$. This is why the time-resolved record can contain dynamical information which the static histogram has discarded.

The dynamical bypass lives entirely in this second, time-resolved setting. By \emph{dynamical bypass} we mean a finite-time excess in parity discrimination of the specified homodyne record of the full monitored LMG evolution over the static $\Jz$ benchmark. The phrase does not mean that the static Holevo/JSD identity of Proposition~\ref{prop:horizon} is violated, and it does not mean that arbitrary bath measurements have been optimised. The numerical claim is deliberately narrower: the static benchmark is defined by the readout geometry, while the bypass is generated by the backaction geometry of Sec.~\ref{sec:backaction} and by the full monitored dynamics studied below.

In decoherence theory, environmental monitoring is the mechanism by which classical-looking observables emerge from quantum dynamics~\cite{Zurek2003,JoosZeh}. A bath operator $\hat L$ has, in this language, two roles: it \emph{reads}, because it imprints information about the system into the environment, and it \emph{acts}, because the same coupling produces backaction on the system. Usually these two roles are discussed in the same pointer basis. Here we use the LMG doublet to exhibit a more selective question: what happens when the monitored coordinate is the order parameter, but the label we want to infer is a conserved symmetry label of the doublet?

For a $\Jz$-coupled bath the einselected pointer basis is the magnetisation/well basis. In that basis the bath both reads and disturbs the system, and there is no mismatch in the usual einselection sense. The asymmetry studied here concerns a different label, namely the parity of the finite-$N$ LMG doublet~\cite{LMG65,BotetJullien83}. Parity is a relational sign between the two wells. It is not the local well coordinate itself. The bath therefore reads the coordinate in which the two lobes are obvious, but it is almost blind to the sign which combines them into even and odd states.
\[
\{\hat{\mathcal{P}},\Jz\}=0.
\]
This has two consequences. First, the diagonal matrix elements in the parity basis vanish,
\[
\langle E_i|\Jz|E_i\rangle=0,
\]
so the first moment of $\Jz$ does not distinguish the parity label; the sharper static-channel statement below is that the accessible information in the $\Jz$ histogram reduces to the Jensen--Shannon divergence of the two parity-conditioned distributions, whose leading decay sits at the instanton rate $\Sinst(g)$ in the WKB/ED analysis. Second, the same selection rule \emph{permits} a transverse matrix element inside the doublet,
\[
J_{01}=\langle E_0|\Jz|E_1\rangle.
\]
Its magnitude is not fixed by the selection rule alone; it follows from the semiclassical separation of the two wells,
\[
|J_{01}|\;\to\;\frac{N\mstar}{2}
\]
up to subleading corrections characterised in Sec.~\ref{sec:backaction}.

Thus the readout suppression comes from exponentially small lobe overlap, while the backaction strength comes from macroscopic lobe separation. In the readout role the bath record carries exponentially little information about the relative sign; in the backaction role the same operator connects the two parity states with an extensive matrix element. What happens to this benchmark once the Hamiltonian is restored during monitoring --- whether a continuous record recovers what the single-shot histogram cannot --- is the subject of Sec.~\ref{sec:dynamical}.

This separation is the core of the parity horizon. We define the readout role of $\Jz$ through the conditional distributions $p_m^{(i)}=|\langle m|E_i\rangle|^2$ and the backaction role through the matrix element $|J_{01}|^2$. Proposition~\ref{prop:horizon} shows that the static $\Jz$-readout channel produces commuting record states, so the Holevo quantity for the parity-record ensemble reduces exactly to the Jensen--Shannon divergence of the parity-conditioned $\Jz$-distributions. The backaction matrix element $|J_{01}|^2$ has no such suppression: in the broken-symmetry regime it approaches the semiclassical scale $(N\mstar/2)^2$, growing as $N^2$. Exact diagonalisation supports these scalings on grids extending to $N=4500$ at $g=0.95$, with a parallel multi-$g$ sweep across several couplings (fit windows and reliability masks are documented in Apps.~\ref{app:reproducibility} and~\ref{app:ed_supp}). Through the two faces of the parity-odd selection rule, then, the single operator $\Jz$ generates an exponentially weak parity-readout channel and a macroscopically strong parity-backaction channel --- a feature invisible to readout-only diagnostics such as the static accessible (Holevo) information or the optimal single-shot (Helstrom) discrimination error~\cite{Helstrom1976}.

\paragraph*{Relation to prior work.}
The conceptual ingredients of the parity horizon appear across several literatures. Spontaneous symmetry breaking with environmental selection has been discussed by van Wezel and collaborators~\cite{vanWezel2005,vanWezel2007} for thin-spectrum decoherence in macroscopic SSB systems; decoherence/einselection and envariance-based accounts of the Born rule by Zurek and collaborators~\cite{Zurek2003,Zurek2005}; and the basis-mismatch geometry of macroscopic superpositions in~\cite{mouslopoulos2026basis}. These are conceptual background and do not formulate the static $\Jz$ parity-readout benchmark or the readout/backaction split for the LMG doublet. One point is worth flagging, since the setting is easily mistaken for textbook cat-state decoherence: the standard spin-coherent-state estimate of the collective dephasing rate uses localised-state matrix elements and gives the well-coherence scale $G_{\rm loc}=(N\mstar)^2/2$, whereas the exact eigenstate-coherence rate is set by $G_{01}=\tfrac12(\langle E_0|\Jz^2|E_0\rangle+\langle E_1|\Jz^2|E_1\rangle)$. The two differ because parity forbids the cross-term the localised estimate implicitly includes; the companion analysis~\cite{mouslopoulos2026basis} makes this quantitative. The exponentially small \emph{readout} information studied here is a separate, sharper consequence of the same selection rule. The closed-system LMG spectrum is well characterised: the thermodynamic-limit spectrum and density of states have been obtained by Majorana-polynomial and spin-coherent-state methods~\cite{RibeiroVidalMosseriPRL07,RibeiroVidalMosseriPRE08}, while finite-size corrections to the broken-symmetry ground-state sector have been studied systematically via the Holstein--Primakoff representation~\cite{HolsteinPrimakoff40} and continuous unitary transformations~\cite{DusuelVidalPRL04,DusuelVidalPRB05}; the latter provide the finite-size framework underlying our $|J_{01}|\to N\mstar/2$ derivation. Our use of the model is complementary: we take this controlled spectral structure as the starting point for an open-system question --- how a parity-odd collective bath operator can read the parity label only weakly in a static $\Jz$ histogram while acting strongly inside the parity doublet.

We extend this line in three ways. First, we promote the matrix element $|J_{01}|^2$ from a quantity computed en route to the WKB instanton to the central object on the backaction side. Second, we show that the readout--backaction separation can be read off directly from exact diagonalisation, with no stochastic-master-equation simulation. Third, and this is the main new scaling result, we give a WKB barrier argument, supported by ED, that the instanton exponent carries from the linear observables (doublet splitting, total variation), where it appears immediately, to the \emph{nonlinear} information functionals (JSD, Chernoff), where it does not. A small-deviation expansion in the semiclassical lobes would predict the doubled exponent $e^{-2N\Sinst}$; the single exponent survives through the barrier-weighting and parity-odd-node mechanism of Sec.~\ref{sec:readout}. The WKB path-additivity identity itself is standard in the spin-WKB/instanton literature~\cite{Garg1998,DusuelVidalPRL04,DusuelVidalPRB05} (see~\cite{ColemanInstantons,Landau3} for background); what is new is recognising that it controls the nonlinear functionals and showing, by ED on grids extending to $N=4500$ with observable-dependent reliability masks, that the four fitted exponents approach a common leading instanton exponent within the accessible window. Proposition~\ref{prop:horizon} is the operational static-channel anchor: the standard accessible-information (Holevo) reduction for commuting records, applied to the parity-conditioned $\Jz$-distributions, with the physical content being the choice of \emph{parity-eigenstate} hypotheses for a phase-blind record. Beyond these static results, Sec.~\ref{sec:dynamical} adds a direct numerical study of the continuously monitored dynamics: the specified classifier shows a finite-time excess over the static benchmark across a finite window of system sizes and returns to it under strong measurement. The distinct \emph{coherence-detectability} window of the same model under continuous driving --- where macroscopic quantum coherence stays testable by a Leggett--Garg inequality --- is treated in the companion analysis~\cite{SweepPaper} and is not used here.

\paragraph*{Outline.} The organisation is as follows. Section~\ref{sec:model} sets up the open LMG model. Section~\ref{sec:two-roles} formalises the readout and backaction roles of $\Jz$. Section~\ref{sec:readout} treats the readout geometry, with Proposition~\ref{prop:horizon} as the static-channel anchor and a WKB instanton calculation as the asymptotic mechanism. Section~\ref{sec:backaction} treats the backaction geometry, deriving the doublet-qubit form $\Pdb\Jz\Pdb=J_{01}\sigmax$ and the leading semiclassical $(N\mstar/2)^2$ scaling of $|J_{01}|^2$. Section~\ref{sec:dynamical} is the dynamical core: it introduces the continuous-monitoring simulation and the Girsanov-optimal record classifier, and reports that continuous monitoring opens a finite-size window in which it exceeds the static benchmark. Section~\ref{sec:separation} reports the joint exact-diagonalisation evidence on grids extending to $N=4500$, with observable-dependent reliability masks (App.~\ref{app:reproducibility}). Sections~\ref{sec:operational}--\ref{sec:scope_regimes} draw the operational consequences and place the static and dynamical results in a common regime map. Section~\ref{sec:discussion} discusses spinor BEC and trapped-ion platforms; Section~\ref{sec:conclusion} concludes.

\paragraph*{Scope of this paper.} The paper has two pillars. The first is an exact static-channel benchmark with its semiclassical and exact-diagonalisation (ED) analysis (Secs.~\ref{sec:readout}--\ref{sec:separation}). The second is a direct numerical study of the continuously monitored dynamics (Sec.~\ref{sec:dynamical}). In order to keep the claim structure clear, we separate what is exact, what is WKB/ED-supported, and what remains open. Throughout we use ``extensive'' (or, loosely, ``macroscopic'') for quantities scaling with positive powers of $N$, ``mesoscopic'' for the finite but large-$N$ window in which doublet and monitoring effects coexist, and ``thermodynamic'' for the strict $N\to\infty$ limit.
\begin{itemize}
\item \emph{Established here:} (a) the static-channel identity $\chi_\infty^{\rm static}=\JSD(p^{(0)},p^{(1)})$ as an exact equality at all $N$ (Prop.~\ref{prop:horizon}, App.~\ref{app:proof}); (b) the asymptotic scaling $\chi_\infty^{\rm static}\sim e^{-N\Sinst}$, supported by the analytical node-bin mechanism, WKB barrier/path-additivity arguments, and ED evidence that the spectral-gap exponent and the three histogram-distinguishability exponents ($S_{\rm TV}$, $S_{\rm JSD}$, $S_{\rm Chernoff}$) are consistent, within finite-$N$ corrections, with a common leading exponent at $g=0.95$ and across $g\in[0.70,0.98]$; (c) the backaction scaling $|J_{01}|^2\to (N\mstar/2)^2$ from a Holstein--Primakoff (HP) derivation and ED verification; (d) the structural readout/backaction split as the central static message --- exponentially weak readout, extensively strong backaction, one operator; (e) a numerical demonstration (Sec.~\ref{sec:dynamical}) that the specified continuous-monitoring classifier exhibits a finite-time excess over the static single-shot benchmark across a finite window of system sizes and returns to that benchmark under strong measurement.
\item \emph{Open:} (i) whether the dynamical closure follows a universal power law in the control ratio --- we report the closure as a robust numerical fact and leave the \emph{exponent} open (App.~\ref{app:closurelaw}); (ii) the full Holevo limit for arbitrary measurements on the entire bath, a different mathematical object from the static commuting-record benchmark; (iii) a specific experimental protocol --- spinor Bose--Einstein condensates (BECs) and trapped-ion arrays are discussed as platforms with the right ingredients (Sec.~\ref{sec:discussion}), not as turnkey tests, and the natural symmetry-broken preparation is a well state, a distinct state-discrimination problem (Secs.~\ref{subsec:qnd_scope},~\ref{sec:discussion}). One structural caveat: the doublet-projected analysis is a \emph{relative} statement --- the relative leakage $\ell_{\rm leak}=L_{\rm leak}/G_{01}$ vanishes algebraically (effective exponent $\approx 1.0$--$1.2$ over the accessible window), while the \emph{absolute} off-doublet variance $L_{\rm leak}=G_{01}\ell_{\rm leak}$ grows ($\approx N^{0.8}$--$N^{1.0}$) at fixed $\gamma_\phi$, so the doublet is spectrally isolated but not absolutely dynamically closed in the thermodynamic limit (Sec.~\ref{subsec:leakage}).
\end{itemize}

\paragraph*{Idealisations and their physical fragility.} Two features of the model are exact \emph{within the model} but only approximate in a physical realisation, and we state both explicitly because they bound the regime in which the horizon is sharp. First, parity. In the ideal LMG Hamiltonian parity is an exact symmetry by construction, $[\hat{\mathcal{P}},\hat H_{\rm LMG}]=0$; a real device may contain weak parity-breaking perturbations (a residual longitudinal field, disorder, imperfect collective coupling). A parity-odd perturbation mixes the doublet once its doublet matrix element becomes comparable to the splitting, $|\langle E_0|\delta H|E_1\rangle|\sim\Delta E$. For a longitudinal perturbation $\delta H=h\Jz$, the relevant element is $h|J_{01}|$, so the criterion is $h\,|J_{01}|\sim\Delta E$, i.e.\ a tolerable field $h^\ast\sim\Delta E/|J_{01}|\sim 2\Delta E/(N\mstar)$. Since $|J_{01}|\sim N$ grows while $\Delta E\sim e^{-N\Sinst}$ collapses, $h^\ast\sim e^{-N\Sinst}/N$ shrinks exponentially in $N$, with an additional algebraic $1/N$ factor from the growing matrix element. The parity label is therefore sharp only while the physical parity-breaking scale stays below this rapidly falling bound; at large $N$, even a tiny fixed parity-odd term destroys it. Second, the symmetric sector. The restriction to $j=N/2$ is protected by the collective, permutation-symmetric form of $\hat H_{\rm LMG}$ (which commutes with $\hat{\mathbf{J}}^2$), not by the angular-momentum algebra alone. Non-collective terms --- single-site fields, disorder, inhomogeneous couplings, non-collective noise --- do not commute with $\hat{\mathbf{J}}^2$ and leak population into lower-$j$ sectors. Neither idealisation is a flaw of the analysis: the LMG symmetry is exact within the model, and real systems may break it weakly. We spell this out so that the horizon is read as a sharp property of the idealised collective model, with an explicit validity window in any implementation.


\section{Open LMG model and doublet structure}
\label{sec:model}
We consider the LMG Hamiltonian $\hat H_{\rm LMG}\equiv \hat H$ in the symmetric Dicke sector $j=N/2$,
\begin{equation}
\hat H = -\frac{2J}{N}\Jz^2 - 2\Gamma\Jx,\qquad g\equiv\Gamma/J,
\label{eq:H_LMG}
\end{equation}
on $N$ pseudospins-$1/2$. We restrict to even $N$ throughout the paper, so $j=N/2$ is integer. The model has a $\mathbb{Z}_2$ parity symmetry generated by the spin-flip operator
\begin{equation}
\hat{\mathcal{P}}=\prod_{r=1}^{N}\hat\sigma_x^{(r)},\qquad \hat{\mathcal{P}}|j,m\rangle=|j,-m\rangle,
\label{eq:parity_def}
\end{equation}
acting in the symmetric Dicke basis as pure reflection $m\mapsto -m$ with no Wigner phase factor. By construction $\hat{\mathcal{P}}\Jz\hat{\mathcal{P}}^\dagger=-\Jz$ and $\hat{\mathcal{P}}\Jx\hat{\mathcal{P}}^\dagger=+\Jx$, so that $[\hat{\mathcal{P}},\hat H_{\rm LMG}]=0$. The operator $\hat{\mathcal{P}}$ is Hermitian and unitary, with $\hat{\mathcal{P}}^2=\mathds{1}$ and eigenvalues $\pm 1$. (Equivalently $\hat{\mathcal{P}}=\prod_r\sigma_x^{(r)}=\exp[i\pi(\Jx-N/2)]$, so $\hat{\mathcal{P}}|j,m\rangle=|j,-m\rangle$ in the Dicke basis; it differs from the collective rotation $e^{i\pi\Jx}$ only by the global sign $(-1)^{j}=i^N$, not by an $m$-dependent factor. We use this Hermitian reflection form~\eqref{eq:parity_def} throughout, matching the numerical projection in the code, App.~\ref{app:reproducibility}.) The quantum critical point is at $g_c=1$. For $g<1$ the mean-field ground state breaks parity along the $z$-axis, with order parameter $\mstar=\sqrt{1-g^2}$, so that $\langle\Jz\rangle/j=\pm\mstar$ at the two saddle points. We use the dimensionless coordinate $\mu=2m/N\in[-1,1]$, so the broken-symmetry wells lie at $\mu=\pm\mstar$. Throughout we work in the broken phase, fixing $g=0.95$, $\mstar=0.312$ as the reference unless stated otherwise. With the real parity-eigenvector convention used in App.~\ref{app:LMG}, the off-diagonal matrix element $J_{01}\in\mathbb{R}$; the corresponding well-combination identities $|E_{0,1}\rangle=(|P\rangle\pm|R\rangle)/\sqrt{2}$ are stated in Sec.~\ref{subsec:WKB}, Eq.~\eqref{eq:PR_def}.

The system is coupled to a Markovian bath through the collective jump operator $\hat L=\sqrt{\gamma_\phi}\Jz$, giving the Lindblad equation
\begin{equation}
\dot\rho = -\frac{i}{\hbar}[\hat H,\rho] + \gamma_\phi\bigl(\Jz\rho\Jz - \tfrac12\{\Jz^2,\rho\}\bigr).
\label{eq:Lindblad_eq}
\end{equation}
Unless otherwise stated, $\gamma_\phi$ is an external dephasing-rate parameter; since $\Jz\sim N$, different thermodynamic scalings of $\gamma_\phi$ give different absolute decoherence and leakage rates, discussed in Sec.~\ref{subsec:leakage}. This collective-dephasing model has been studied~\cite{vanWezel2005,Landsman2017,LandsmanReview} as a minimal realisation of broken-symmetry monitoring along the order-parameter axis.

In the broken phase, the lowest two energy eigenstates $|E_0\rangle$ (even parity) and $|E_1\rangle$ (odd parity) form a near-degenerate \emph{parity doublet} with exponentially small splitting $\Delta E=E_1-E_0\sim e^{-N\Sinst(g)}$~\cite{DusuelVidalPRL04}. We resolve the doublet states explicitly by diagonalising the parity operator inside the two-dimensional low-energy subspace returned by the eigensolver (see App.~\ref{app:LMG}); this matters numerically at large $N$, where $\Delta E$ falls below machine precision and the raw eigensolver may return an arbitrary rotation within the near-degenerate doublet.

\paragraph*{Even-$N$ choice and odd-$N$ remark.} We use even $N$ so that the parity-odd node at the central Dicke bin $m=0$ is an exact single-bin diagnostic of parity ($p_0^{(1)}=0$ rigorously). For odd $N$ the innermost bins are $m=\pm 1/2$; parity acts on Dicke states as $\hat{\mathcal{P}}|m\rangle=|-m\rangle$, relating the innermost amplitudes by a sign in the odd-parity eigenstate while leaving the probabilities symmetric, $p_{+1/2}^{(i)}=p_{-1/2}^{(i)}$. The same central-barrier parity-information mechanism therefore persists at odd $N$, but the exact even-$N$ single-bin node is replaced by an innermost two-bin barrier contribution. Since the sharp $m=0$ node is absent at odd $N$, the rigorous JSD lower bound derived in Sec.~\ref{subsec:WKB} (which uses $p_0^{(1)}=0$) requires even $N$. The asymptotic scaling claims of Sec.~\ref{sec:separation} are insensitive to the parity of $N$; we present even-$N$ results for clarity.

\section{The two roles of $\Jz$}
\label{sec:two-roles}
\paragraph*{Why probe through $\Jz$ and not directly through $\hat H$?} There are two reasons. (i) \emph{Operational and conceptual.} In the model studied here the environmental coupling operator is $\hat L=\sqrt{\gamma_\phi}\Jz$, not $\hat H$; this is the natural bath-coupling structure for monitoring a broken-symmetry order parameter, and we do not claim that $\hat H$ is inaccessible in principle to every possible bath. (ii) \emph{Spectral resolution.} $\Jz$ is the collective magnetisation, measurable atom by atom in spinor BECs and trapped-ion arrays, whereas a direct projective energy measurement would face the exponentially small splitting $\Delta E\sim e^{-N\Sinst}$ and so require a resolution time $\hbar/\Delta E\sim e^{+N\Sinst}$. This spectral-resolution problem is a different operational setting from the static $\Jz$-readout argument of this paper. We keep the two separate: the smallness of $\Delta E$ is a spectral fact independent of the chosen probe, while the parity horizon of Proposition~\ref{prop:horizon} concerns specifically the asymptotic classical information in the static $\Jz$ record.

\paragraph*{Conserved label versus monitored coordinate.} Which quantity is conserved and which quantity is measured are different questions, and their mismatch is the whole content of the horizon. Parity is the good quantum number of the closed LMG Hamiltonian, $[\hat{\mathcal{P}},\hat H_{\rm LMG}]=0$, and the two doublet states are distinguished by their parity eigenvalue. In the symmetric Dicke sector parity is the global spin flip $\hat{\mathcal{P}}=\prod_i\sigma_x^{(i)}$ --- equivalently, up to a global phase, a $\pi$ rotation about the $x$-axis --- acting on magnetisation eigenstates as $|m\rangle\mapsto|-m\rangle$, so that $\Jz\mapsto-\Jz$ while $\Jx$ is left invariant. The monitored variable $\Jz$ is \emph{not} conserved by the physical dynamics: for $\Gamma\neq 0$, $[\Jz,\hat H_{\rm LMG}]\neq 0$. A bath record diagonal in the $\Jz$ basis therefore means only that the environment has classicalised the measured pointer coordinate; it is not a measurement of a conserved quantum number. The parity horizon is precisely this mismatch: the bath reads the order-parameter coordinate $\Jz$, while the label to be inferred is the conserved parity eigenvalue. (In the static benchmark $\hat H=0$ is set by construction, so there the $\Jz$ record commutes trivially; this is a channel idealisation, not a claim that $\Jz$ is conserved in the physical LMG dynamics.)

The bath operator $\Jz$ plays two distinct roles in the open dynamics~\eqref{eq:Lindblad_eq}.

\emph{(a) Readout role.} For any fixed system state $\rho_S$, the classical $\Jz$-record produced by the static channel (the idealised channel with internal Hamiltonian set to $\hat H=0$, defined in Sec.~\ref{subsec:qnd_scope}) is the distribution $p_m=\mathrm{Tr}_S[\rho_S\,\hat\Pi_m^{S}]$, where $\hat\Pi_m^{S}=|m\rangle_S\langle m|_S$ acts on the system Hilbert space and $|m\rangle_S$ is a Dicke $\Jz$-eigenstate. The bath record state for outcome $m$ is a separate pointer state $|m\rangle_E$, related to but not identified with $|m\rangle_S$. The parity-conditioned distributions
\begin{equation}
p_m^{(i)} = |\langle m|E_i\rangle|^2,\qquad i\in\{0,1\}
\label{eq:p_m}
\end{equation}
carry, within this static commuting-record channel, all the information the bath record contains about the parity label. The point used below, and shown in Sec.~\ref{sec:readout}, is that the asymptotic classical Holevo quantity for parity over this static channel equals the Jensen--Shannon divergence $\JSD(p^{(0)},p^{(1)})$ of these distributions.

\emph{(b) Backaction role.} The same operator $\Jz$ generates the conditional system disturbance through the homodyne stochastic master equation~\cite{Wiseman2009,Korotkov2001,Diosi1988,Belavkin1988}. Within the parity doublet it takes a clean two-level form. Let $\Pdb=|E_0\rangle\langle E_0|+|E_1\rangle\langle E_1|$ be the doublet projector. The parity selection rule (Sec.~\ref{sec:backaction}) gives $\langle E_i|\Jz|E_i\rangle=0$ for $i\in\{0,1\}$, so the only non-vanishing matrix element inside the doublet is
\begin{equation}
J_{01}=\langle E_0|\Jz|E_1\rangle,\qquad \BN\equiv|J_{01}|^2,
\label{eq:J01}
\end{equation}
and, with the real parity-eigenvector convention used throughout (see App.~\ref{app:LMG} and the discussion below Eq.~\eqref{eq:Hdb}), the doublet-projected operator is
\begin{equation}
\Pdb\Jz\Pdb = J_{01}\sigmax,
\label{eq:Pdb_Jz_Pdb}
\end{equation}
with the Pauli operators $\sigmax=|E_0\rangle\langle E_1|+|E_1\rangle\langle E_0|$ and $\sigmaz=|E_0\rangle\langle E_0|-|E_1\rangle\langle E_1|$ acting in the parity basis $\{|E_0\rangle,|E_1\rangle\}$.

The doublet-projected Hamiltonian, in the same basis,
\begin{equation}
\Hdb = E_0 \,|E_0\rangle\langle E_0| + E_1\,|E_1\rangle\langle E_1| = \frac{E_0+E_1}{2}\,\hat I_{\rm db} - \frac{\Delta E}{2}\,\sigmaz,
\label{eq:Hdb}
\end{equation}
gives precession in the parity basis at angular frequency $\omega_{01}=\Delta E/\hbar$. The coupling within the doublet is therefore the standard transverse-noise qubit. The Hamiltonian $\Hdb$ has only a $\sigmaz$ non-trivial component (the identity term in Eq.~\eqref{eq:Hdb} is an irrelevant overall energy shift), while $\Pdb\Jz\Pdb = J_{01}|E_0\rangle\langle E_1| + J_{01}^*|E_1\rangle\langle E_0| = \mathrm{Re}(J_{01})\sigmax - \mathrm{Im}(J_{01})\sigmay$. With the real parity-eigenvector convention (App.~\ref{app:LMG}), $J_{01}\in\mathbb{R}$, so this reduces to $\Pdb\Jz\Pdb=J_{01}\sigmax$.

\emph{(c) Physical summary.} The single bath operator $\hat L=\sqrt{\gamma_\phi}\Jz$ acts through two distinct geometric roles. Role~1 (readout): the first moment of a $\Jz$ record vanishes for both parity eigenstates, $\langle E_i|\Jz|E_i\rangle=0$, by the parity selection rule, so the first-moment record is blind to the parity population imbalance; the stronger static-channel result below is that the full commuting $\Jz$ record carries only exponentially small information about the parity label. Role~2 (backaction): $\Pdb\Jz\Pdb=J_{01}\sigmax$ is a logical $X$ on the parity basis, producing transverse parity backaction at rate scale $\Gamma_z^{(\rm par)}=2\gamma_\phi |J_{01}|^2$ in the projected doublet.

The parity horizon extends Role~1 from the first-moment record to the full static bath record. Proposition~\ref{prop:horizon} shows that the classical Holevo information accessible in the static-channel bath record is
\[
\chi_\infty^{\rm static}=\JSD(p^{(0)},p^{(1)}).
\]
The WKB and ED analysis then supports the leading scaling
\[
\chi_\infty^{\rm static}\sim e^{-N\Sinst},
\]
with the same exponent appearing in the histogram readout measures considered here (TV, JSD, Chernoff; Sec.~\ref{subsec:WKB}). These two roles are the two layers of the analysis: Role~1 defines the static horizon (Sec.~\ref{sec:readout}), while Role~2 supplies the extensive backaction that drives the dynamical bypass (Sec.~\ref{sec:dynamical}).

\section{Readout geometry}
\label{sec:readout}
\subsection{Scope of the static restriction: three conditions to distinguish}
\label{subsec:qnd_scope}
Proposition~\ref{prop:horizon} below concerns a static commuting-record benchmark. Before stating it we separate three operating conditions which are often lumped together as quantum non-demolition (QND) measurement, although they are logically and physically distinct:
\begin{enumerate}
\item[(i)] \emph{Strict QND ($[\hat H, \hat A]=0$).} The standard textbook definition: a measurement of an observable $\hat A$ is quantum non-demolition if the system Hamiltonian commutes with $\hat A$, so that successive $\hat A$-measurements yield consistent results~\cite{Braginsky1980,Braginsky1996}. For the LMG with $\hat A=\Jz$, strict QND readout is impossible because $\hat H_{\rm LMG}$ contains the noncommuting transverse term $-2\Gamma\Jx$, so $[\hat H_{\rm LMG},\Jz]\neq 0$.
\item[(ii)] \emph{Static $\Jz$-readout channel ($\hat H=0$).} A cleaner but more restrictive idealisation: we set the system Hamiltonian to zero entirely. The bath then couples through $\Jz$ to a system that does not evolve, and successive bath records over $[0,\Tobs]$ become simultaneously diagonal in the $\Jz$-eigenbasis. The reduced bath states $\rho_E^{(0)},\rho_E^{(1)}$ commute, and the Holevo quantity collapses to a classical functional of the parity-resolved $\Jz$-distributions $p_m^{(i)}=|\langle m|E_i\rangle|^2$. \emph{This is the channel actually used in the proof of Proposition~\ref{prop:horizon}}; we refer to it throughout as the static $\Jz$-readout channel.
\item[(iii)] \emph{Secular regime ($\xi=\omega_{01}/\Gamma_{01}\gg 1$, where coherent rotation is fast compared with measurement-induced dephasing).} A condition on the \emph{full} LMG dynamics rather than on the channel definition: the doublet Bohr oscillation is fast compared to bath dephasing. In this regime, standard adiabatic (Bloch--Redfield secular) averaging is expected to suppress dissipator cross-terms between populations and coherences, so the static channel should provide the natural benchmark for the slowly varying part of the bath record. Section~\ref{sec:dynamical} tests this expectation directly by simulating the monitored dynamics.
\end{enumerate}
Condition (i) is impossible for the physical LMG. Condition (ii) is the operative restriction in Proposition~\ref{prop:horizon}: it replaces the physical Hamiltonian by $\hat H=0$, a mathematical idealisation in which the records commute by construction and the JSD identity is exact. Condition (iii) is not a channel definition. It is the regime in which the static result is expected to be a useful benchmark for the full monitored LMG dynamics, which we study numerically in Sec.~\ref{sec:dynamical}.

\paragraph*{The channel under analysis.} Within condition (ii), the channel is the ideal measurement map
\begin{equation}
\rho_S \;\longmapsto\; \mathcal{M}_{\rm static}(\rho_S) = \sum_m \mathrm{Tr}_S[\hat\Pi_m^{S}\rho_S]\,|m\rangle_E\langle m|_E,
\qquad \hat\Pi_m^{S} = |m\rangle_S\langle m|_S,
\label{eq:static_channel}
\end{equation}
applied to the two possible system preparations $|E_0\rangle_S$ and $|E_1\rangle_S$, with parity-resolved $\Jz$-Born distributions
\[
p_m^{(i)} = \mathrm{Tr}_S\bigl[\hat\Pi_m^{S}\,|E_i\rangle_S\langle E_i|_S\bigr] = \bigl|{}_S\langle m|E_i\rangle_S\bigr|^2.
\]
In this restricted channel, the bath record is diagonal in a common environment pointer basis $\{|m\rangle_E\}$ and the Holevo quantity reduces exactly to the Jensen--Shannon divergence of the two distributions $p_m^{(i)}$.

It is important not to confuse this construction with the full continuous Lindblad evolution of a single monitored system under
\[
\dot\rho = -\frac{i}{\hbar}[\hat H_{\rm LMG},\rho]+\gamma_\phi\mathcal{D}[\Jz]\rho.
\]
Although $|E_0\rangle$ and $|E_1\rangle$ are stationary under the unitary part $\hat H_{\rm LMG}$, they are not stationary under the full Lindblad evolution with the $\Jz$ dissipator~\cite{Braginsky1980,Braginsky1996}. Inside the doublet,
\[
\Pdb\Jz\Pdb=J_{01}\sigmax,
\]
so the dissipator contains a transverse logical-noise component that drives parity-population mixing. The static distributions $p_m^{(i)}=|\langle m|E_i\rangle|^2$ are therefore Born-rule input distributions for the static commuting-record channel. They are not the time-dependent distributions of a single continuously monitored trajectory after backaction has acted.

The quantum-Zeno regime (strong, frequent measurement) also needs care. Strong $\Jz$ monitoring projects onto $\Jz$-eigenstates, i.e.\ individual Dicke bins. It can therefore localise the system within the magnetisation wells and suppress coherent motion relative to $\Jz$, but it does not protect the parity eigenstates $|E_0\rangle,|E_1\rangle$. Thus the static-channel JSD result below is a commuting-record benchmark and an independent-copy $\Jz$-tomography benchmark; it is not, by itself, the discrimination achievable under continuous monitoring with $\hat H_{\rm LMG}\neq 0$. Section~\ref{sec:dynamical} computes that monitored discrimination directly and finds that it exceeds the static benchmark over a finite window of system sizes, before returning to it deep in the Zeno regime.

The full dynamical Holevo problem, including arbitrary measurements on the time-correlated bath output, measurement backaction, and leakage outside the doublet, is a distinct open-system problem and is not solved here. What we compute directly in Sec.~\ref{sec:dynamical} is the discrimination performance of the specified continuously monitored homodyne record, analysed with the classifier described in App.~\ref{app:sme}. That classifier shows a finite-time excess over the static single-shot benchmark over a finite $\xi$ window and returns to that benchmark in the Zeno tail. The relationship between the static result and the full LMG monitored dynamics across the three sub-regimes (secular/window/Zeno) is developed in Sec.~\ref{sec:dynamical} and summarised in Sec.~\ref{sec:scope_regimes}.

\paragraph*{Preparation vs channel: two independent restrictions.} The static-channel condition $\hat H=0$ of (ii) above is one restriction; the choice of input ensemble is a logically separate one. Proposition~\ref{prop:horizon} takes the symmetric parity-eigenstate mixture $\rho=\tfrac{1}{2}(|E_0\rangle\langle E_0|+|E_1\rangle\langle E_1|)$ as input. This second restriction plays a role distinct from the static-channel condition, and we name it explicitly. For parity-eigenstate inputs, the single-time $\Jz$-Born distributions $p_m^{(i)}=|\langle m|E_i\rangle|^2$ are unchanged by free evolution under $\hat H_{\rm LMG}$, since energy eigenstates acquire only global phases. This justifies using the same histograms in the static benchmark whether $\hat H=0$ or $\hat H=\hat H_{\rm LMG}$; it does \emph{not} imply equivalence between the static channel and the full continuously monitored dynamics with $\hat H_{\rm LMG}$ active. For a \emph{superposition} input, by contrast, the Born distribution does evolve. A natural broken-symmetry preparation in the present model is a well state $|P\rangle=(|E_0\rangle+|E_1\rangle)/\sqrt{2}$ or $|R\rangle=(|E_0\rangle-|E_1\rangle)/\sqrt{2}$, which is a superposition of energy eigenstates rather than an energy eigenstate itself. Under $\hat H_{\rm LMG}\neq 0$, the well-state Born distribution acquires a Bohr-oscillation term. In general,
\begin{equation}
p_m^{|P\rangle}(t) = \tfrac{1}{2}p_m^{(0)}+\tfrac{1}{2}p_m^{(1)} + \mathrm{Re}\!\left[e^{-i\omega_{01}t}\langle E_0|m\rangle\langle m|E_1\rangle\right];
\label{eq:wellstate_Born}
\end{equation}
with the real parity-eigenvector convention used here, this reduces to
\[
p_m^{|P\rangle}(t) = \tfrac{1}{2}p_m^{(0)}+\tfrac{1}{2}p_m^{(1)} + \mathrm{Re}\!\left[\langle E_0|m\rangle\langle m|E_1\rangle\right]\cos(\omega_{01}t),
\]
the oscillating component carrying the relative phase between the two parity sectors. The horizon as stated in Proposition~\ref{prop:horizon} therefore corresponds to one cell of a $2\times 2$ structure (preparation $\times$ channel): parity-eigenstate input on the static channel. The diagonally opposite cell --- well-state input under the full $\hat H_{\rm LMG}$ dynamics --- is a different operational problem (it discriminates wells rather than parity, and is operationally trivial under $\Jz$-readout), and is not what the proposition addresses. The relevance of the present analysis to a typical SSB experiment, which naturally produces well states rather than parity-eigenstate mixtures, therefore requires either engineered preparation or post-selection (Sec.~\ref{sec:discussion}).

\subsection{Static $\Jz$-readout: parity-horizon identification}
\label{subsec:prop_horizon}
\paragraph*{Operational setting (binary quantum hypothesis testing).} Throughout this section we use the standard binary state discrimination framework~\cite{Helstrom1976,NielsenChuang,OhyaPetz}. The physical scenario is that the system has been prepared in one of two known parity eigenstates $|E_0\rangle$ or $|E_1\rangle$ with equal prior probability. We adopt the symmetric prior as the unbiased binary-discrimination setting; for this prior the static-channel Holevo quantity equals the equal-weight Jensen--Shannon divergence exactly (Proposition~\ref{prop:horizon} below). A symmetric Gibbs preparation restricted to the doublet would approximately realise this prior (up to corrections of order $\Delta E/k_BT$) in the temperature window $\Delta E\ll k_BT\ll\hbar\omega_{\rm HP}$; the Lindblad model used here is pure dephasing rather than thermalisation, so this thermal interpretation is context only, not a derivation. The density matrix
\begin{equation}
\rho \;=\; \tfrac{1}{2}\bigl(|E_0\rangle\langle E_0|+|E_1\rangle\langle E_1|\bigr)
\label{eq:prior}
\end{equation}
encodes this \emph{classical} prior over \emph{quantum} preparations; the system itself is fully quantum at all times. Equation~\eqref{eq:prior} is not a quantum superposition of indefinite parity, but the experimenter's Bayesian ignorance about which of two definite quantum preparations was carried out. The Holevo quantity $\chi_\infty$ then asks how much classical information about the preparation label the bath can transmit. The asymptotic exponential rate $\Sinst$ is prior-robust: for any fixed prior bounded away from $0$ and $1$, the leading exponential rate is the same.

We isolate the readout role of $\Jz$ by restricting attention to the static $\Jz$-readout channel: the system Hamiltonian is set to zero ($\hat H=0$, condition (ii) of Sec.~\ref{subsec:qnd_scope}), and the bath record asymptotically resolves the $\Jz$ pointer value. In this restricted setting the asymptotic accessible information about the parity label is the classical Holevo quantity for the parity ensemble.

\begin{proposition}[Parity-horizon identification]
\label{prop:horizon}
Consider the open LMG~\eqref{eq:Lindblad_eq} in the static $\Jz$-readout channel ($\hat H=0$, Sec.~\ref{subsec:qnd_scope}), in the symmetric Dicke sector $j=N/2$. For the prior ensemble $\rho=\tfrac12(|E_0\rangle\langle E_0|+|E_1\rangle\langle E_1|)$, the asymptotic classical Holevo quantity at the bath equals the Jensen--Shannon divergence of the parity-conditioned $\Jz$-distributions:
\begin{equation}
\chi_\infty^{\rm static} = \JSD\bigl(p^{(0)},p^{(1)}\bigr) = H(\bar p) - \tfrac12 H(p^{(0)}) - \tfrac12 H(p^{(1)}),
\label{eq:prop_horizon}
\end{equation}
where $\bar p=\tfrac12(p^{(0)}+p^{(1)})$, $H(\cdot)$ denotes Shannon entropy, and $p_m^{(i)}=|\langle m|E_i\rangle|^2$.
\end{proposition}

\paragraph*{Claim-status calibration.} Proposition~\ref{prop:horizon} establishes the \emph{equality} $\chi_\infty^{\rm static}=\JSD(p^{(0)},p^{(1)})$ exactly, at all $N$, under the stated static-channel condition. The \emph{asymptotic scaling} $\chi_\infty^{\rm static}\sim e^{-N\Sinst}$ is a separate semiclassical statement. The parity-odd node supplies an analytical lower-bound mechanism at the instanton scale: at even $N$, $p_0^{(1)}=0$, so the exact $m=0$ contribution to JSD is $\tfrac12 p_0^{(0)}\ln 2$, and WKB path additivity gives $p_0^{(0)}\sim e^{-N\Sinst}$. This prevents the JSD from decaying faster than $e^{-N\Sinst}$. The absence of a slower contribution from other bins is supported by the WKB barrier analysis and by ED evidence for equality of the four fitted exponents $\{S_{\rm gap},S_{\rm TV},S_{\rm JSD},S_{\rm Chernoff}\}$ --- the spectral gap exponent and the three histogram distinguishability exponents --- at $g=0.95$ and across multiple $g$ (Sec.~\ref{sec:separation}). We do not provide a fully rigorous matched-asymptotic derivation of the JSD prefactor across the barrier; the exponent claim should therefore be read as a controlled WKB/ED-supported asymptotic result rather than as a theorem.

\paragraph*{Operational taxonomy.} The parity-discrimination statement of Proposition~\ref{prop:horizon} has three operationally distinct readings:
\begin{itemize}
\item[(A)] \emph{Mean-signal $\Jz$ readout} is blind to parity: $\langle E_i|\Jz|E_i\rangle=0$ for both $i$ by parity. This is true but only a first-moment statement; it does not yet address the full histogram or the JSD structure.
\item[(B)] \emph{Many-shot $\Jz$-tomography} on independently prepared parity eigenstates: discriminating $|E_0\rangle$ from $|E_1\rangle$ at fixed error probability $\epsilon$ requires
\[
K\sim\frac{|\ln\epsilon|}{C}\sim e^{N\Sinst}
\]
independent preparations and $\Jz$ measurements, because the single-shot Chernoff information scales as $C\sim e^{-N\Sinst}$. This is the operational reading most directly relevant to a laboratory experimenter.
\item[(C)] \emph{Static-channel bath observer} reading commuting records:
\[
\chi_\infty^{\rm static} = \JSD(p^{(0)},p^{(1)}) \sim e^{-N\Sinst}.
\]
\end{itemize}
Statements (B) and (C) are tightly linked --- repeated uses of the static channel on independently prepared copies are statistically equivalent to repeated ideal $\Jz$-tomography in the asymptotic commuting-record limit (a single static record from one copy yields one resolved outcome, not a tomographic ensemble). The exact Holevo/JSD identity establishes (C); the Chernoff scaling in (B) has the WKB/ED-supported status described above. Statements about the full quantum Holevo capacity with arbitrary bath access (no static restriction, full $\hat H_{\rm LMG}$ dynamics) are not made here.

\paragraph*{Visibility, not quantumness.} The Holevo-to-JSD reduction of Proposition~\ref{prop:horizon} is standard, and therefore it should not be mistaken for the result. The LMG model is quantum from the outset: the ground doublet, the instanton splitting, the noncommutation $[\hat H_{\rm LMG},\Jz]\neq 0$, and the measurement backaction are all defined within a quantum many-body Hamiltonian. The question here is narrower and operational: given this quantum substrate, how much information about the \emph{parity} label is visible in a \emph{specified} bath record? A record can be classical even when the underlying state is quantum. In the static $\Jz$ channel the bath record is a classical magnetisation histogram, and Proposition~\ref{prop:horizon} identifies the information in that record with the Jensen--Shannon divergence of the two parity-conditioned histograms. The nontrivial physics is not the reduction itself, but the fact that these two histograms are \emph{exponentially close} for the LMG parity doublet. The parity horizon is therefore a visibility statement: the quantum label exists in the model, but the static $\Jz$ record is almost blind to it.

As a physical picture, the parity horizon is neither a collapse mechanism nor a decoherence mechanism. The finite-$N$ doublet remains a coherent quantum object: the even and odd states are exact parity eigenstates ($\hat{\mathcal P}|E_0\rangle=+|E_0\rangle$, $\hat{\mathcal P}|E_1\rangle=-|E_1\rangle$), not incoherent mixtures of the two broken-symmetry branches $|P\rangle,|R\rangle=(|E_0\rangle\pm|E_1\rangle)/\sqrt2$. Nevertheless, under the natural collective readout $\Jz$ their outcome distributions are exponentially close both to each other and to that of a classical equal-weight branch mixture,
\begin{equation}
p_m^{(0)}\;\approx\;p_m^{(1)}\;\approx\;\tfrac12 p_m^{P}+\tfrac12 p_m^{R},
\label{eq:looks_classical}
\end{equation}
with the residual differences confined to the instanton-suppressed barrier/interference region and scaling as $e^{-N\Sinst}$ (the parity-conditioned histograms differ by a Jensen--Shannon divergence $\approx 0.026$ at the benchmark $N=370$, with the total-variation distance falling from $\approx 14\%$ at $N=370$ to below $1\%$ by $N\approx 10^3$). A closed, coherent quantum system can therefore look classical under a restricted macroscopic observable without becoming classical. This is a statement about \emph{visibility under $\Jz$}, not a standalone witness of non-classicality: because the match in Eq.~\eqref{eq:looks_classical} is at the level of $\Jz$ statistics, a classical equal-weight mixture over the two branches reproduces these single-time statistics, and distinguishing the coherent parity doublet from such a mixture requires access to an observable off-diagonal in the $\Jz$ basis (for example the parity operator $\hat{\mathcal P}$ itself, or $\Jx$) or to a temporal-correlation protocol of the kind studied in the companion analysis~\cite{SweepPaper}. The present paper makes no non-classicality claim of this stronger type; it characterises the readout/backaction structure and the information bypass.

The standard reduction by itself does not know whether the measured basis is aligned with the label being inferred. Table~\ref{tab:basis_mismatch} contrasts a basis-matched parity readout with the $\Jz$ readout studied here. In both cases the static commuting-record Holevo quantity is a JSD; the difference is the \emph{value} of that JSD --- order unity ($\ln 2$) for a direct parity readout, instanton-suppressed for the $\Jz$ record.

\begin{table}[h]
\centering
\caption{The Holevo-to-JSD reduction is basis-agnostic; the physics is not. Both columns describe a static commuting-record readout of the same parity doublet, and in both the asymptotic Holevo quantity equals a Jensen--Shannon divergence. They differ only in whether the measured basis is aligned with the conserved label, and hence in the \emph{size} of that JSD. For the basis-matched column, ``direct parity readout'' means the two-outcome projective measurement of the parity operator $\hat{\mathcal{P}}$ itself, whose parity-conditioned outcome distributions are $(1,0)$ and $(0,1)$, giving $\JSD=\ln 2$ exactly.}
\label{tab:basis_mismatch}
\small
\setlength{\tabcolsep}{5pt}
\begin{tabular}{@{}l l l@{}}
\toprule
Static readout & Basis matched: & LMG horizon: \\
setting & direct parity readout & $\Jz$ readout \\
\midrule
Hidden label & parity $|E_0\rangle/|E_1\rangle$ & parity $|E_0\rangle/|E_1\rangle$ \\
Measured record & parity eigenvalue & magnetisation $m$ \\
Record basis & parity basis & $\Jz$/Dicke basis \\
Conditional dists. & disjoint & almost identical in wells \\
Holevo reduction & $\chi_\infty=\JSD$ & $\chi_\infty=\JSD$ \\
Size of JSD & $O(1)$, $=\ln 2$ & $\sim e^{-N\Sinst}$ \\
Physics & reads the label directly & reads the pointer, not the label \\
\bottomrule
\end{tabular}
\end{table}

\noindent Proposition~\ref{prop:horizon} supplies the common information-theoretic container; the parity/$\Jz$ mismatch supplies the physics.

The proposition rests on two ingredients: a \emph{physical identification} that the static $\Jz$ channel produces classical-mixture record states diagonal in a common pointer basis $\{|m\rangle_E\}$, and the \emph{textbook reduction} that the Holevo quantity for such an ensemble collapses to the classical mutual information $H(\bar p)-\tfrac12 H(p^{(0)})-\tfrac12 H(p^{(1)})$. The reduction is standard~\cite{NielsenChuang,OhyaPetz}; we do not claim it as new. For the static commuting-record channel the Holevo bound is saturated by the natural pointer measurement, so the Holevo quantity, the accessible information, and the classical Shannon mutual information all coincide here; we use these phrases interchangeably within this channel only. What makes the statement physically informative is the \emph{hypothesis choice}: the two states being discriminated are the parity eigenstates $|E_0\rangle,|E_1\rangle$, while the measurement alphabet is the full $\Jz$ Dicke basis. In words, the proposition says that a phase-blind $\Jz$ record can recover only the JSD of two histograms identical except in their exponentially thin barrier overlap --- a parity question asked of a record that discards relative sign. We retain the full Dicke alphabet (rather than coarse-graining) precisely so that the resulting smallness is not a resolution artefact. The proof, including the identification, the entropy reduction, and the asymptotic saturation by the pointer measurement, is given in App.~\ref{app:proof}.

\paragraph*{Weak readout is not parity protection.} The smallness of $\chi_\infty^{\rm static}\sim e^{-N\Sinst}$ should not be read as protection of the parity qubit. The same operator $\Jz$ that is weak when used to infer the parity label from a static $\Jz$ histogram acts transversely inside the doublet, and therefore produces measurement backaction on the parity population at rate $\Gamma_z^{({\rm par})}=2\gamma_\phi |J_{01}|^2\to\gamma_\phi N^2\mstar^2/2$. Weak readout and strong backaction coexist.

\paragraph*{Evolution of $p_m^{(i)}$ with $N$ in three distinct regions.} The $\Jz$-distributions of the two parity eigenstates have three structurally different regions, evolving with $N$ as follows (Fig.~\ref{fig:jz_distributions} illustrates this at $N=370$). The two semiclassical lobes condense into mirror-image Gaussians of width $\sigma_\mu\sim 1/\sqrt{N}$ in the dimensionless magnetisation $\mu=2m/N$, centred at $\mu=\pm\mstar=\pm\sqrt{1-g^2}$. In the lobes, $p_m^{(0)}$ and $p_m^{(1)}$ become identical to all orders in $1/N$; the lobes carry essentially all the probability but no parity information. The barrier interior is exponentially suppressed with rate set by the WKB action. The exact parity-odd node at $m=0$ remains structurally distinct at all $N$: $p_{m=0}^{(1)}=0$ rigorously for even $N$ by parity, while $p_{m=0}^{(0)}\sim e^{-N\Sinst}$ by the WKB/path-additivity mechanism discussed in Sec.~\ref{subsec:WKB}. The parity information is concentrated in the central forbidden/barrier region, anchored by the parity-odd node at $m=0$ --- a region with vanishingly small probability mass --- and this is what controls the bath's ability to distinguish the two sectors. As $N$ grows, the lobes converge, but the central-node bump-vs-dip structure persists in form while shrinking exponentially in magnitude.

\subsection{Asymptotic WKB structure: a single instanton exponent}
\label{subsec:WKB}
In the broken-symmetry regime, define the well states
\begin{equation}
|P\rangle\equiv\frac{|E_0\rangle+|E_1\rangle}{\sqrt 2},\qquad |R\rangle\equiv\frac{|E_0\rangle-|E_1\rangle}{\sqrt 2},
\label{eq:PR_def}
\end{equation}
with $\langle P|R\rangle=0$ by construction. Inverting,
\begin{equation}
|E_{0,1}\rangle = \frac{1}{\sqrt{2}}\left(|P\rangle \pm |R\rangle\right),
\label{eq:E_LR}
\end{equation}
exactly. In the broken-phase large-$N$ regime, the wavefunctions $\psi_P(m)=\langle m|P\rangle$ and $\psi_R(m)=\langle m|R\rangle$ are localised near $m=+N\mstar/2$ and $m=-N\mstar/2$ respectively (the broken-symmetry lobes), with a small exponential tail of the opposite-lobe partner in the barrier. The semiclassical HP+Bogoliubov saddle-point states (Sec.~\ref{subsec:semiclassical} below) provide a separate analytical approximation to $|P\rangle,|R\rangle$; we denote them $|\tilde P\rangle,|\tilde R\rangle$ to distinguish them from the exact combinations~\eqref{eq:PR_def}. The overlap of the HP saddle-point states is exponentially small in $N$, $\langle\tilde P|\tilde R\rangle=O(e^{-c(g)N})$ for some $c(g)>0$, so orthogonalisation affects prefactors but not the leading exponent (Sec.~\ref{sec:backaction}). The conditional distributions decompose as
\begin{equation}
p_m^{(0,1)} = \tfrac12\bigl(\psi_P(m)^2 + \psi_R(m)^2\bigr) \pm \psi_P(m)\psi_R(m),
\label{eq:pm_PR}
\end{equation}
where we take $\psi_{P,R}$ real: the LMG Hamiltonian is real symmetric in the Dicke basis, so the parity eigenvectors can be chosen real, and with the real-eigenvector and global-sign convention of App.~\ref{app:LMG}, the relevant barrier tails of $\psi_P$ and $\psi_R$ are chosen with the same sign. The symmetric leading part of \eqref{eq:pm_PR} cancels in the parity-conditioned difference, leaving the cross-term:
\begin{equation}
p_m^{(0)} - p_m^{(1)} = 2\,\psi_P(m)\,\psi_R(m).
\label{eq:diff_PR}
\end{equation}
This identity is \emph{exact} in the symmetrised well basis defined by~\eqref{eq:PR_def}, with $\langle P|R\rangle=0$ by construction. For the non-orthogonal HP saddle-point states $|\tilde P\rangle,|\tilde R\rangle$ introduced in Sec.~\ref{sec:backaction} below, analogous formulae acquire normalisation corrections of order $\langle\tilde P|\tilde R\rangle=O(e^{-c(g)N})$ that affect prefactors but not the leading exponent.

\paragraph*{Path additivity in the LMG barrier.}
In the WKB limit, the well states obey
\begin{equation}
\psi_P(m)\sim A_P(\mu_m)\,e^{-N s_P(\mu_m)},\qquad \psi_R(m)\sim A_R(\mu_m)\,e^{-N s_R(\mu_m)},
\label{eq:WKB_amp}
\end{equation}
where $\mu_m\equiv 2m/N$ is the continuum magnetisation coordinate evaluated at the discrete Dicke point $m$, $s_P(\mu)$ is the per-spin action accumulated from $\mu=+\mstar$ (the right-magnetised well where $|P\rangle$ is localised) to the point $\mu$, and $s_R(\mu)$ the per-spin action accumulated from $\mu=-\mstar$ (the left-magnetised well where $|R\rangle$ is localised) to the point $\mu$ (App.~\ref{app:WKB}). Here $s_{P,R}(\mu)\equiv-\log|\psi_{P,R}(\mu)|/N$ is the per-spin barrier exponent; in this normalisation it equals \emph{half} the imaginary-momentum integral $\int\tilde\phi\,d\mu'$, the factor $1/2$ arising from $j=N/2$ in the spin-WKB amplitude $\psi\sim e^{-jS_{\rm WKB}}$. With this convention, path additivity of the instanton trajectory (App.~\ref{app:WKB}) gives
\begin{equation}
\boxed{s_P(\mu)+s_R(\mu) = \Sinst(g) \quad\text{for all } \mu\in[-\mstar,\mstar],}
\label{eq:path_additivity}
\end{equation}
away from WKB turning-point boundary layers, where
\begin{equation}
\Sinst(g) = \int_0^{\mstar}\!\!d\mu\;\mathrm{arccosh}\!\left[\frac{1+g^2-\mu^2}{2g\sqrt{1-\mu^2}}\right]
\label{eq:Sinst_SCS}
\end{equation}
is the LMG spin-coherent-state instanton action; this is the explicit LMG form obtained from spin-coherent-state path integration~\cite{DusuelVidalPRL04,DusuelVidalPRB05,Braun2007spin,Garg1998}, with $\Sinst(0.95)=0.010787$, rounded to $0.01079$ in prose and figures. We adopt the convention in which $\Delta E\sim e^{-N\Sinst}$, so that $\Sinst$ is the WKB action integrated from the barrier centre $\mu=0$ to the turning point $\mu=\mstar$ --- half of the one-way under-barrier traversal from $+\mstar$ to $-\mstar$ (the one-way instanton action $W_1=2\Sinst$); the closed Euclidean bounce loop, with both halves traversed, has action $4\Sinst$ in standard instanton-calculus terminology, and the doublet splitting is $\Delta E\sim e^{-jW_1}=e^{-N\Sinst}$. Physically, the instanton is the under-barrier tunnelling trajectory connecting the two semiclassical magnetisation wells at $\mu=\pm\mstar$; its action produces the doublet splitting $\Delta E\sim e^{-N\Sinst}$. In the localised well basis this tunnelling appears as an off-diagonal coupling between the wells, whereas in the parity/energy basis the same physics is already diagonalised into the gap between $|E_0\rangle$ and $|E_1\rangle$: the parity eigenstates are the emergent normal modes of the instanton-coupled double well, and $\Delta E$ is the fingerprint of that coupling, not an off-diagonal matrix element of $\hat H$ in the energy basis. The WKB path-additivity structure \eqref{eq:path_additivity} is the key analytical mechanism by which the spectral instanton exponent $\Sinst$ is expected to appear in the spectral gap exponent and the three histogram distinguishability exponents. It implies that the product of the two well wavefunctions has the same leading exponent throughout the barrier interior, away from WKB boundary layers:
\begin{equation}
\psi_P(m)\,\psi_R(m) \sim A_P(\mu_m)A_R(\mu_m)\,e^{-N\Sinst},
\label{eq:product_uniform}
\end{equation}
up to smooth prefactors $A_{P,R}(\mu)$ and boundary-layer corrections. The same instanton action controls the spectral splitting~\cite{Garg1998,DusuelVidalPRL04,DusuelVidalPRB05}, while the product structure in~\eqref{eq:diff_PR} is the mechanism by which this exponent enters the histogram-based distinguishability measures TV, JSD and Chernoff. This is the content of the four-exponent equality tested below. We test the path-additivity identity numerically in App.~\ref{app:pathadd_numerics} (and Fig.~\ref{fig:pathadd}): the deviation $|s_P(\mu_m)+s_R(\mu_m)-\Sinst|/\Sinst$ evaluated at the Dicke points and averaged over the central $80\%$ of the barrier interior decays approximately as $\sim N^{-1.0\pm 0.1}$ over the accessible range, statistically compatible with an $O(1/N)$ correction (the product $N\times(\text{rel.~dev.})$ varies only weakly over this range, at the level expected from the limited $N$-window and finite numerical precision). We therefore read the data as evidence for convergence toward path additivity with subleading corrections close to $1/N$~\cite{Braun2007spin}, not as a precise exponent measurement. The test is restricted to a narrow $g$-window near the benchmark where barrier wavefunctions remain above the noise floor.

\paragraph*{Consequence for the fitted exponents.}
Combining \eqref{eq:diff_PR} with \eqref{eq:product_uniform}, the per-bin difference $|p_m^{(0)}-p_m^{(1)}|$ has the same leading exponential factor $e^{-N\Sinst}$ at every $m$ in the barrier interior. The total variation
\begin{equation}
\TV(p^{(0)},p^{(1)}) = \tfrac12\sum_m\bigl|p_m^{(0)}-p_m^{(1)}\bigr| \;\sim\; e^{-N\Sinst},
\end{equation}
where the number of contributing barrier bins contributes only a polynomial prefactor (set by the WKB amplitude $A_PA_R$ and the barrier width). For the Jensen--Shannon divergence, the situation is more subtle. At the parity-odd node $m=0$ exactly, $p_0^{(1)}=0$ by parity, and $p_0^{(0)}=2\psi_P(0)\psi_R(0)\sim 2A_P(0)A_R(0)e^{-N\Sinst}$ (since $\psi_P(0)=\psi_R(0)$ by reflection symmetry). The per-bin JSD contribution at $m=0$ is
\begin{equation}
\bigl[\JSD\bigr]_{m=0} = \tfrac12\,p_0^{(0)}\ln(p_0^{(0)}/\bar p_0)
= p_0^{(0)}\cdot\tfrac12\ln 2
\sim A_P(0)A_R(0)\,e^{-N\Sinst}\ln 2,
\label{eq:JSD_m0}
\end{equation}
which is at the same leading exponential as TV. The quadratic Taylor expansion $\JSD\approx\sum_m(p_m^{(0)}-p_m^{(1)})^2/(8\bar p_m)$ fails at $m=0$ because $p_0^{(1)}/p_0^{(0)}=0$, so the small-deviation assumption underlying the expansion is violated; the parity-odd node contributes at order $e^{-N\Sinst}$, not at order $e^{-2N\Sinst}$. For the Chernoff information $C=-\log\min_{s\in[0,1]}\sum_m (p_m^{(0)})^s(p_m^{(1)})^{1-s}$, the optimiser is observed numerically to remain interior, with $s^*\simeq 0.60$ at the benchmark parameters for the label ordering used here (swapping $p^{(0)}\leftrightarrow p^{(1)}$ sends $s^*\to 1-s^*$, leaving $C$ invariant). This interior value is distinct from the symmetric Bhattacharyya point $s=1/2$: numerically, $(C-\mathcal{B})/C$ stays around $3$--$4\%$ across $N\in[100,2500]$ at $g=0.95$. The same central barrier/node region supplies an instanton-scale deficit mechanism, with the matching leading exponent supported by the ED fits below. At the parity-odd node $m=0$, $p_0^{(1)}=0$ exactly (for even $N$); for $s\in(0,1)$ the integrand $(p_0^{(0)})^s\cdot 0^{1-s}=0$, so the node bin contributes nothing to the sum. The deficit from unity is estimated by expanding $(S_m+D_m)^s(S_m-D_m)^{1-s}$ around $S_m=(p_m^{(0)}+p_m^{(1)})/2$ and $D_m=(p_m^{(0)}-p_m^{(1)})/2$. To first order, $(S_m+D_m)^s(S_m-D_m)^{1-s}\approx S_m+(2s{-}1)D_m$. Because the node bin has been removed, the sums over $m\neq 0$ are $\sum_{m\neq 0}S_m=1-p_0^{(0)}/2$ and $\sum_{m\neq 0}D_m=-p_0^{(0)}/2$. Therefore $\sum_{m\neq 0}[S_m+(2s{-}1)D_m]=1-s\,p_0^{(0)}$, and for an interior optimum $s^*\in(0,1)$, as observed numerically here, the node produces a leading Chernoff deficit $s^*p_0^{(0)}\sim e^{-N\Sinst}$. This expansion is not uniformly valid across the central barrier, where $D_m/S_m$ can be $O(1)$ in bins immediately adjacent to the node. The node bin therefore provides an instanton-scale node mechanism for the $e^{-N\Sinst}$ exponent for the observed interior optimiser $s^*$; unlike the JSD node term it is not used here as a rigorous asymptotic lower bound. The numerically observed shift of the optimum away from the Bhattacharyya value $s=1/2$ reflects the asymmetry of the central-barrier statistics. A full uniform asymptotic derivation of the Chernoff prefactor is not attempted in this paper; the shared exponent across the four fitted quantities is supported numerically by the independent fits of Sec.~\ref{sec:separation}. The WKB mechanism and ED evidence below therefore support the asymptotic scaling claim
\begin{equation}
S_{\rm gap},\;S_{\rm TV},\;S_{\JSD},\;S_{\rm Chernoff}\;\xrightarrow{\,N\to\infty\,}\;\Sinst(g),
\label{eq:single_exponent}
\end{equation}
with finite-$N$ deviations set by the polynomial prefactor structure and the discrete-versus-continuum lattice details, not by any near-critical or quadratic-functional effect. In~\eqref{eq:single_exponent} the equalities $S_{\rm gap}=S_{\rm TV}$ follow directly --- the gap exponent is the Dusuel--Vidal instanton result~\cite{DusuelVidalPRL04,DusuelVidalPRB05} and TV is linear in the path-additive cross term --- while the relations involving the \emph{nonlinear} functionals $S_{\JSD}$ and $S_{\rm Chernoff}$ are the substantive claim, supported by the barrier-weighting/node mechanism above and the ED evidence below. Equation~\eqref{eq:single_exponent} is the main WKB/ED-supported readout-side claim of this paper. We test it directly in Sec.~\ref{sec:separation}, and we additionally show in Sec.~\ref{subsec:multig} that the finite-$N$ deviations collapse well when plotted against $N\cdot\Sinst(g)$ across $g\in[0.7,0.98]$, disfavouring an interpretation as a purely near-critical effect.

\subsection{Structural origin: why $\JSD(p^{(0)},p^{(1)})$ is exponentially small}
\label{subsec:structural_origin}
\begin{figure}[htbp]
\centering
\includegraphics[width=0.92\textwidth]{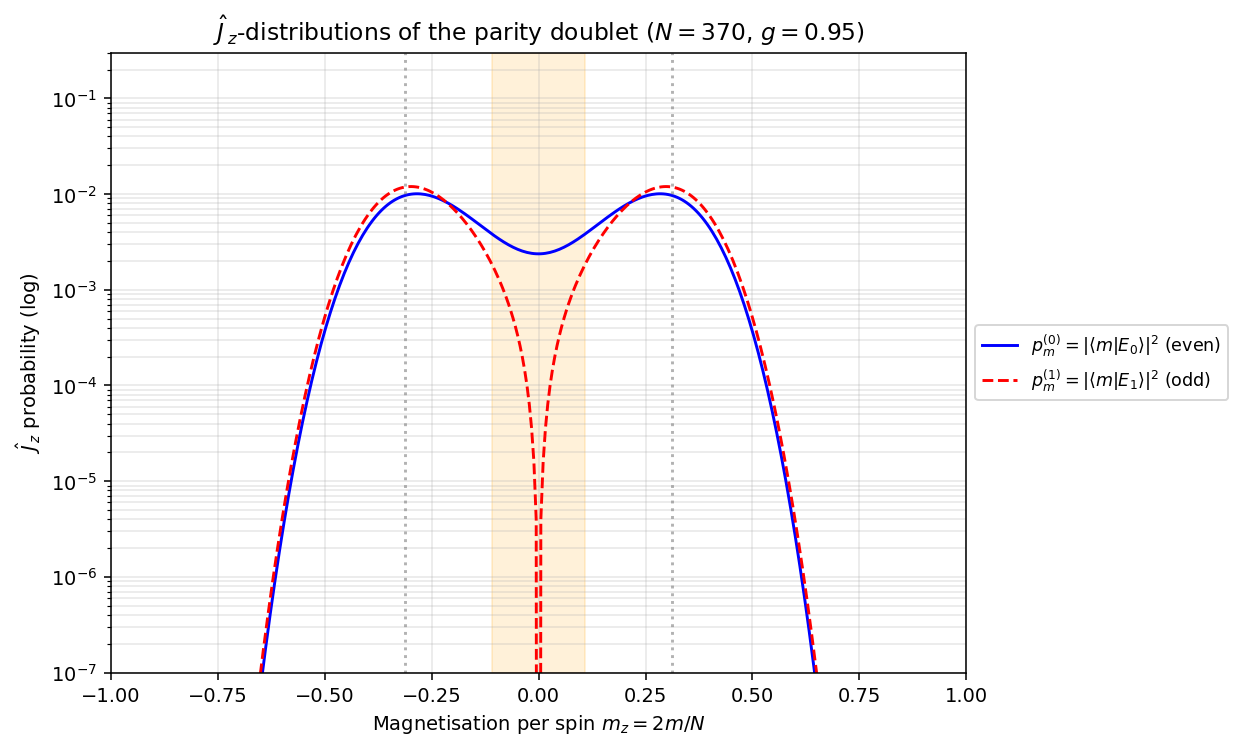}
\caption{\textbf{Why $\chi_\infty^{\rm static}$ is exponentially small: the $\Jz$-distributions of the parity doublet.} The $\Jz$-eigenvalue probability distributions $p_m^{(0)}=|\langle m|E_0\rangle|^2$ (parity-even, blue solid) and $p_m^{(1)}=|\langle m|E_1\rangle|^2$ (parity-odd, red dashed) for the LMG ground doublet at the benchmark $N=370$, $g=0.95$, plotted against magnetisation per spin $m_z=2m/N$ on a log scale. The broken-symmetry vacua at $m_z=\pm\mstar=\pm 0.312$ are marked (gray dotted lines). The two distributions are \emph{visually indistinguishable} in the broken-symmetry lobes; they differ measurably only in the central barrier region $|m|\leq 20$ (orange shading, $\approx 77\%$ of the total JSD but only a small fraction of the probability mass). The exact parity-odd node $p_m^{(1)}|_{m=0}=0$ (annotated) while $p_m^{(0)}|_{m=0}=2.37\times 10^{-3}$ is small but nonzero. The histogram distinguishability of $|E_0\rangle$ and $|E_1\rangle$ is concentrated in the barrier region; their exact orthogonality is carried by the relative sign structure that the classical $\Jz$ histogram largely discards.
\label{fig:jz_distributions}}
\end{figure}

The exponential suppression has a geometric origin, clearest from the two distributions at the benchmark $N=370$, $g=0.95$. Figure~\ref{fig:jz_distributions} plots $p_m^{(0)}$ and $p_m^{(1)}$ against magnetisation per spin $m_z=2m/N$ on a logarithmic scale. The two curves are \emph{visually indistinguishable} in the broken-symmetry lobes near $m_z=\pm\mstar$, where the bulk of the wavefunction sits; they differ measurably only in the central barrier region $|m|\lesssim 20$, which holds only a small fraction of the total probability mass.

The reason is Eq.~\eqref{eq:diff_PR}: $p_m^{(0)}-p_m^{(1)}=2\,\psi_P(m)\psi_R(m)$, the lobe-overlap cross term. Near the lobes, $\psi_P(m)$ is large but $\psi_R(m)\approx 0$, and vice versa; their product is nonzero only in the central barrier region where both lobes have exponentially small amplitude. The two parity eigenstates are perfectly orthogonal (they are energy eigenstates of opposite parity), but that orthogonality lives in the relative \emph{sign} between the lobes --- which the $\Jz$-basis classical record cannot see, because it discards phase. The static $\Jz$-record is sign-blind; the orthogonality lives in the sign.

\begin{figure}[htbp]
\centering
\includegraphics[width=0.80\textwidth]{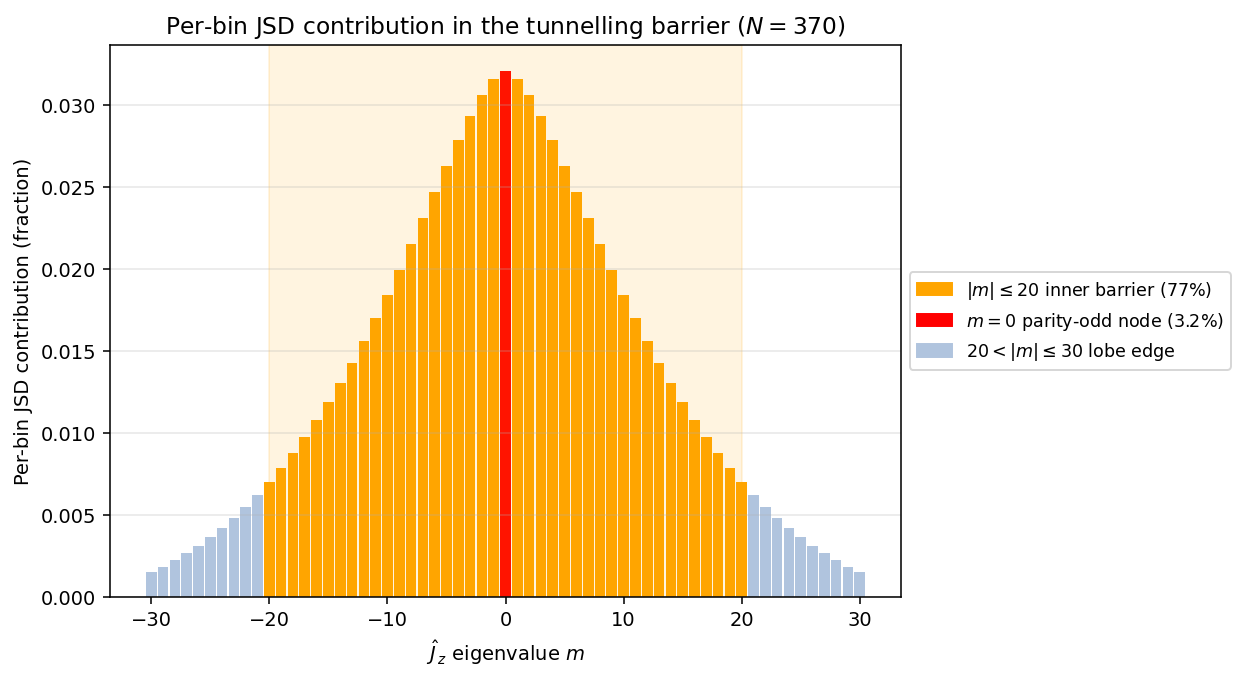}
\caption{\textbf{Per-bin JSD contribution in the tunnelling barrier ($N=370$).} Per-bin JSD contribution as a fraction of the total JSD, plotted versus $\Jz$ eigenvalue $m$ and restricted to $|m|\leq 30$. The contribution is tightly concentrated in the barrier region $|m|\lesssim 20$ (orange shading, dashed orange boundaries), forming a triangular peak centred on $m=0$. The $m=0$ bin (red bar) carries the parity-odd node contribution ($\approx 3.2\%$ of total JSD at $N=370$); the light-blue bins ($|m|>20$, lobe region) contribute essentially nothing despite dominating the probability mass. The probability imbalance $|p_m^{(0)}-p_m^{(1)}|$ carries the leading WKB factor $e^{-N\Sinst}$ throughout the barrier interior by path additivity~\eqref{eq:product_uniform}; the per-bin JSD contribution is a nonlinear function and behaves differently in equal-amplitude vs.\ unequal-amplitude regions. The leading total JSD contribution is concentrated in the central barrier/equal well-tail amplitude region around the parity-odd node, giving the same total exponent $\chi_\infty^{\rm static}\sim e^{-N\Sinst}$.
\label{fig:jz_perbin}}
\end{figure}

Figure~\ref{fig:jz_perbin} makes this quantitative at the benchmark $N=370$. The per-bin JSD contribution is concentrated in a triangular peak centred on $m=0$, with appreciable support over roughly $|m|\lesssim 20$: $\approx 77\%$ of the total JSD comes from $|m|\leq 20$. The exact parity-odd node $p_m^{(1)}|_{m=0}=0$ (annotated) is the sharpest single diagnostic; the $m=0$ bin alone contributes $\approx 3.2\%$ of the total JSD at $N=370$ (verifiable from $p_0^{(0)}=2.37\times 10^{-3}$ and total JSD $=0.0257$: $\tfrac12 p_0^{(0)}\ln 2/\text{JSD}\approx 0.032$). As $N$ grows, the inner barrier region (defined as $|m_z|<0.3\,\mstar$) absorbs an increasingly large fraction of the total JSD: in the ED data of Fig.~\ref{fig:jsd_anatomy} below, this fraction rises from $\approx 61\%$ at $N=200$ toward $\approx 100\%\;(99.99\%)$ at $N=2500$, while the lobe contribution decreases toward zero. The single $m=0$ bin, by contrast, contributes only a roughly constant $\approx 3$--$5\%$ of the total throughout; its fractional weight is a finite-bin diagnostic rather than the fundamental object. The robust physical feature is the concentration of the total JSD in the central barrier band.

\noindent The lobe decomposition explains both roles, but through \emph{different} matrix elements. The readout side is controlled by the exponentially small cross term $\psi_P(m)\psi_R(m)$ concentrated in the barrier. The backaction side is controlled by $J_{01}$, which is essentially the difference between the mean magnetisations of the two well states and hence scales as $N\mstar/2$. Small lobe overlap makes the readout weak; large lobe separation makes $J_{01}$ macroscopic. These are dual consequences of the same broken-symmetry geometry.

\subsection{Summary: the readout side}
The readout side of the parity horizon is tied to a single instanton scale. The spectral doublet splitting $\Delta E$ and the three histogram distinguishability measures --- total variation $\TV$, JSD, and Chernoff information --- are \emph{all} controlled by the same WKB instanton action $\Sinst(g)$. Proposition~\ref{prop:horizon} ties $\JSD$ exactly to the asymptotic classical accessible (Holevo) information in the static channel; the WKB and ED analysis then support the instanton scaling of that quantity.

Within the static $\Jz$-readout channel, the horizon is a suppression statement for the asymptotic classical information carried by the bath's commuting $\Jz$ record. Once the system Hamiltonian is restored during monitoring, the bath output is no longer a static commuting histogram but a time-correlated measurement record generated by dynamics with $[\hat H_{\rm LMG},\Jz]\neq 0$. The discrimination achievable from that record can exceed the static benchmark. We compute this directly in Sec.~\ref{sec:dynamical} before drawing the operational consequences in Sec.~\ref{sec:operational}.


\section{Backaction geometry}
\label{sec:backaction}
The backaction side is governed by a single matrix element. By the parity selection rule, $\langle E_0|\Jz|E_0\rangle=\langle E_1|\Jz|E_1\rangle=0$. The off-diagonal element
\begin{equation}
J_{01} = \langle E_0|\Jz|E_1\rangle
\label{eq:J01_def}
\end{equation}
encodes the strength with which $\Jz$ couples the two parity eigenstates. In this section we derive its semiclassical limit $|J_{01}|\to N\mstar/2$ from the Holstein--Primakoff (HP) construction of the well states, and we characterise the leading relative $1/N$ correction through an effective additive shift $\delta(g,N)$.

\subsection{Semiclassical scaling: $|J_{01}|^2\to (N\mstar/2)^2$}
\label{subsec:semiclassical}
\begin{remark}[Notation for the well states]
\label{rem:notation}
Throughout this paper, $|P\rangle$ and $|R\rangle$ are the \emph{exact orthogonal} doublet combinations $|P\rangle\equiv(|E_0\rangle+|E_1\rangle)/\sqrt{2}$, $|R\rangle\equiv(|E_0\rangle-|E_1\rangle)/\sqrt{2}$. They coincide with the well-localised states discussed in the related-basis preprint~\cite{mouslopoulos2026basis}. The label ``P'' is a well index (the right-magnetised well, $\langle\Jz\rangle/j\to+\mstar$), not a parity label. The HP saddle-point states $|\tilde P\rangle,|\tilde R\rangle$ constructed in this subsection from a Bogoliubov vacuum are not literally identical to $|P\rangle,|R\rangle$ at finite $N$; they approximate them with non-orthogonality $\langle\tilde P|\tilde R\rangle=O(e^{-c(g)N})$ for some $c(g)>0$ and matrix-element offset $\delta(g,N)$ defined in~\eqref{eq:J01_HP} below.
\end{remark}
Starting from the broken-symmetry mean-field minimum at polar angle $\theta_0$ on the Bloch sphere of $\mathbf{J}$, with $\cos\theta_0=\mstar$, $\sin\theta_0=g$ (the $z$-magnetisation at the minimum is $\langle\Jz\rangle/j=\cos\theta_0=\mstar$), we rotate the spin frame so that the new $z$-axis lies along the classical magnetisation. In the rotated frame, the HP transformation
\begin{equation}
J_z'=j-a^\dagger a,\qquad J_+'=\sqrt{2j-a^\dagger a}\,a,
\label{eq:HP}
\end{equation}
introduces a bosonic mode $a$ with $[a,a^\dagger]=1$. To leading order in $1/N$, the Hamiltonian~\eqref{eq:H_LMG} becomes a quadratic form in $(a,a^\dagger)$ which is diagonalised by a Bogoliubov transformation; the resulting vacuum is the displaced-well saddle-point state centred at the $+\mstar$ minimum. We denote this saddle-point state $|\tilde P\rangle$; its parity partner is $|\tilde R\rangle = \hat{\mathcal{P}} |\tilde P\rangle$.

Rotating back to the original spin frame, the saddle-point states have expectation
\begin{equation}
\langle\tilde P|\Jz|\tilde P\rangle = +\frac{N\mstar}{2} - \delta(g,N),\qquad \langle\tilde R|\Jz|\tilde R\rangle = -\frac{N\mstar}{2} + \delta(g,N),
\label{eq:L_R_Jz}
\end{equation}
where $\delta(g,N)$ is a subleading semiclassical correction. The non-orthogonality of the two saddle-point states is exponentially small in $N$, $\langle\tilde P|\tilde R\rangle=O(e^{-c(g)N})$. Here $c(g)>0$ is the spin-coherent-state Gaussian overlap exponent, set by the Holstein--Primakoff saddle width and the well separation, and is distinct from (and numerically larger than) the WKB instanton action $\Sinst(g)$ --- at $g=0.95$, $c\approx 0.051$ versus $\Sinst=0.0108$. These orthogonalisation effects affect subleading finite-$N$ details, but not the leading semiclassical scaling $|J_{01}|\to N\mstar/2$. The exact well states $|P\rangle,|R\rangle$ defined as exact orthogonal combinations of $|E_0\rangle,|E_1\rangle$ in Remark~\ref{rem:notation} differ from $|\tilde P\rangle,|\tilde R\rangle$ by corrections of the same type. The value of $\langle P|\Jz|P\rangle$ that enters $J_{01}$ via~\eqref{eq:J01_HP} below absorbs both the HP correction and the orthogonalisation correction into a single effective $\delta(g,N)$. We characterise the $N$-dependence of this combined $\delta$ in Sec.~\ref{subsec:delta_fit}.

The function $\delta(g{=}0.95,N)$ is non-monotonic: it rises from $\delta\approx 2.6$ at $N=100$ (where the doublet wavefunction overlap is not yet small and the HP+orthogonalisation picture has not formed), peaks at $\delta\approx 8.3$ near $N\sim 350$, and then decreases monotonically. The asymptotic intercept is therefore extracted from a fit restricted to $N\ge 1000$, where the HP+orthogonalisation picture is well within its domain of validity: the two-parameter model $\delta(N)=\delta_\infty+c/N$ gives $\delta_\infty\approx 5.81$ ($c=+656$), and the three-parameter extension gives $\delta_\infty\approx 5.85$, with the Akaike information criterion (AIC), used as a descriptive model-selection diagnostic, favouring both parametric models over a constant ($\Delta\mathrm{AIC}>140$). The statistical fit errors are small (sub-percent), but the M2-vs-M3 model spread is $\approx 0.04$; we therefore quote $\delta_\infty(g{=}0.95)=5.8\pm 0.1$ as a conservative rounded value, with the M2/M3 model spread dominating over formal least-squares errors.

A first-principles HP+Bogoliubov derivation of $\delta_\infty(g)$ --- which would require computing the subleading $1/N$ orthogonalisation correction between the HP saddle-point states $|\tilde P\rangle, |\tilde R\rangle$ and the exact symmetrised well states $|P\rangle, |R\rangle$ --- is the natural analytical next step and is not carried out in this paper. We flag it as an open problem on the backaction side, with the ED-extracted $\delta_\infty(g{=}0.95)=5.8\pm 0.1$ serving as a numerical benchmark for any future analytical calculation~\cite{DusuelVidalPRB05}.

With $|E_{0,1}\rangle=(|P\rangle\pm|R\rangle)/\sqrt{2}$ and the real parity-eigenvector convention, the cross terms $\langle P|\Jz|R\rangle$ and $\langle R|\Jz|P\rangle$ cancel, giving
\begin{equation}
J_{01} = \langle E_0|\Jz|E_1\rangle = \tfrac12\left[\langle P|\Jz|P\rangle - \langle R|\Jz|R\rangle\right] = \frac{N\mstar}{2} - \delta(g,N).
\label{eq:J01_HP}
\end{equation}
Hence
\begin{equation}
|J_{01}|^2 = \left(\frac{N\mstar}{2} - \delta(g,N)\right)^{\!2} = \left(\frac{N\mstar}{2}\right)^{\!2}\!\Bigg[1 - \frac{4\delta(g,N)}{N\mstar} + O\!\left(\frac{1}{N^2}\right)\Bigg],\qquad \mstar=\sqrt{1-g^2}.
\label{eq:J01_semicl}
\end{equation}
The relative deviation from the leading-order $(N\mstar/2)^2$ scaling is therefore $\sim 1/(N\mstar)$ at fixed $g$, with prefactor $4\delta(g,N)$. For $g=0.95$, $\mstar=0.312$, so the asymptotic value is $\BN \to N^2\mstar^2/4 \simeq 0.0244\, N^2$.

\paragraph*{Kinematic ceiling.} Define the secular Bloch--Redfield dephasing factor for the $E_0/E_1$ coherence by
\begin{equation}
G_{01}\equiv \frac{1}{2}\left(\langle E_0|\Jz^2|E_0\rangle+\langle E_1|\Jz^2|E_1\rangle\right),
\label{eq:G01_def}
\end{equation}
which equals the average variance because $\langle E_i|\Jz|E_i\rangle=0$. By Cauchy--Schwarz and resolution of identity, $|J_{01}|^2\le G_{01}$; trivially $G_{01}\le j^2=N^2/4$. The strict kinematic ceiling for the backaction strength is therefore $|J_{01}|^2\le G_{01}\le j^2=N^2/4$, attained only in the $g\to 0$ limit, where the two broken-symmetry wells approach the north and south poles of the collective Bloch sphere. At finite $g$, the wells sit at $\mu=\pm\mstar=\pm\sqrt{1-g^2}$, so the semiclassical attractor is lower by the factor $\mstar^2=1-g^2$: $|J_{01}|^2\to\mstar^2 j^2=(N\mstar/2)^2$. The benchmark value $\mstar(0.95)=0.312$ gives $|J_{01}|^2\to(0.312\,j)^2\approx 0.097\,j^2$, well below the kinematic ceiling.

\paragraph*{The $\Gamma_{01}/\Gamma_{\rm mix}\to 1$ asymptote.} A direct consequence of the backaction geometry is a universal asymptote for the eigenstate-coherence rate relative to the variance of a classical 50/50 statistical mixture of the two broken-symmetry wells (the \emph{mixture} reference, denoted ``mix''). The secular Bloch--Redfield coherence rate is $\Gamma_{01}=\gamma_\phi G_{01}$. A classical observer prepared with a 50/50 statistical mixture of well-localised states at $\langle\Jz\rangle=\pm N\mstar/2$ has variance
\begin{equation}
G_{\rm mix}\equiv\langle\Jz^2\rangle_{\rm mix} = (N\mstar/2)^2 = (N\mstar)^2/4,
\end{equation}
giving the reference dephasing rate
\begin{equation}
\Gamma_{\rm mix} \equiv \gamma_\phi\,G_{\rm mix} = \gamma_\phi\,(N\mstar)^2/4.
\label{eq:GammaMF_clean}
\end{equation}
In the broken-symmetry regime as $N\to\infty$, $G_{01}\to J_{01}^2\to(N\mstar/2)^2$ so
\begin{equation}
\frac{\Gamma_{01}}{\Gamma_{\rm mix}} = \frac{G_{01}}{(N\mstar)^2/4} \;\to\; 1.
\label{eq:half_asymp_clean}
\end{equation}
That is, the doublet asymptotically dephases at the same rate as the classical 50/50 mixture would. The vanishing diagonal elements $\langle E_i|\Jz|E_i\rangle=0$ follow from parity. The recovery of the classical-mixture variance, however, also requires the semiclassical broken-well structure, which gives $|J_{01}|\to N\mstar/2$. Thus the parity selection rule fixes the transverse structure of $\Jz$ inside the doublet, while the extensive ($\propto N$) magnitude of the transverse matrix element is supplied by the well separation.\footnote{We use $G_{\rm mix}=(N\mstar)^2/4$ for the variance of a classical 50/50 mixture over the two well centres. It is important not to confuse this with the localised branch-coherence dephasing scale $G_{\rm loc}=(N\mstar)^2/2$ (the central object of the companion basis-dependence analysis~\cite{mouslopoulos2026basis}), which refers to dephasing of coherence between the two localised branches and is larger by a factor of two because the branch separation is $N\mstar$. The two quantities are different physical reference scales answering different questions; no ED quantity is changed by this naming convention. With the present convention, $\Gamma_{01}/\Gamma_{\rm mix}\to 1$ as $N\to\infty$.} Numerically at $N=370$: $\Gamma_{01}/\Gamma_{\rm mix}=G_{01}/[(N\mstar)^2/4]=2839/3337=0.851$, approaching $1$ monotonically from below as $N$ grows. The approach to unity is controlled structurally by the two finite-$N$ corrections to $G_{01}$ characterised elsewhere in this section: writing $G_{01}=|J_{01}|^2+L_{\rm leak}$ with $|J_{01}|=N\mstar/2-\delta(g,N)$ [Eq.~\eqref{eq:J01_HP}] and $L_{\rm leak}=\ell_{\rm leak}G_{01}$, one has
\begin{equation}
1-\frac{\Gamma_{01}}{\Gamma_{\rm mix}} \;=\; \frac{4\,\delta(g,N)}{N\mstar} \;-\; \frac{4\,\delta^2}{(N\mstar)^2} \;-\; \ell_{\rm leak} \;+\;\dots,
\label{eq:gamma_ratio_struct}
\end{equation}
so the finite-$N$ deviation is a competition between the positive $\delta$-shift term and the negative leakage correction. At the accessible system sizes these terms are comparable and partly cancel; reconstructing Eq.~\eqref{eq:gamma_ratio_struct} from the independently fitted $\delta(g,N)$ and $\ell_{\rm leak}(N)$ reproduces the ED value at the benchmark ($1-\Gamma_{01}/\Gamma_{\rm mix}\approx 0.13$--$0.15$ at $N=370$). Because the non-monotonic $\delta(N)$ peaks near the benchmark, a direct single-power-law fit to $1-\Gamma_{01}/\Gamma_{\rm mix}$ over this window is not very informative; we therefore quote the structural form~\eqref{eq:gamma_ratio_struct} rather than a standalone exponent.

\subsection{Spectrum structure: doublet isolation from the HP phonon ladder}
\label{subsec:spectrum}
\begin{figure}[!ht]
\centering
\includegraphics[width=0.96\columnwidth]{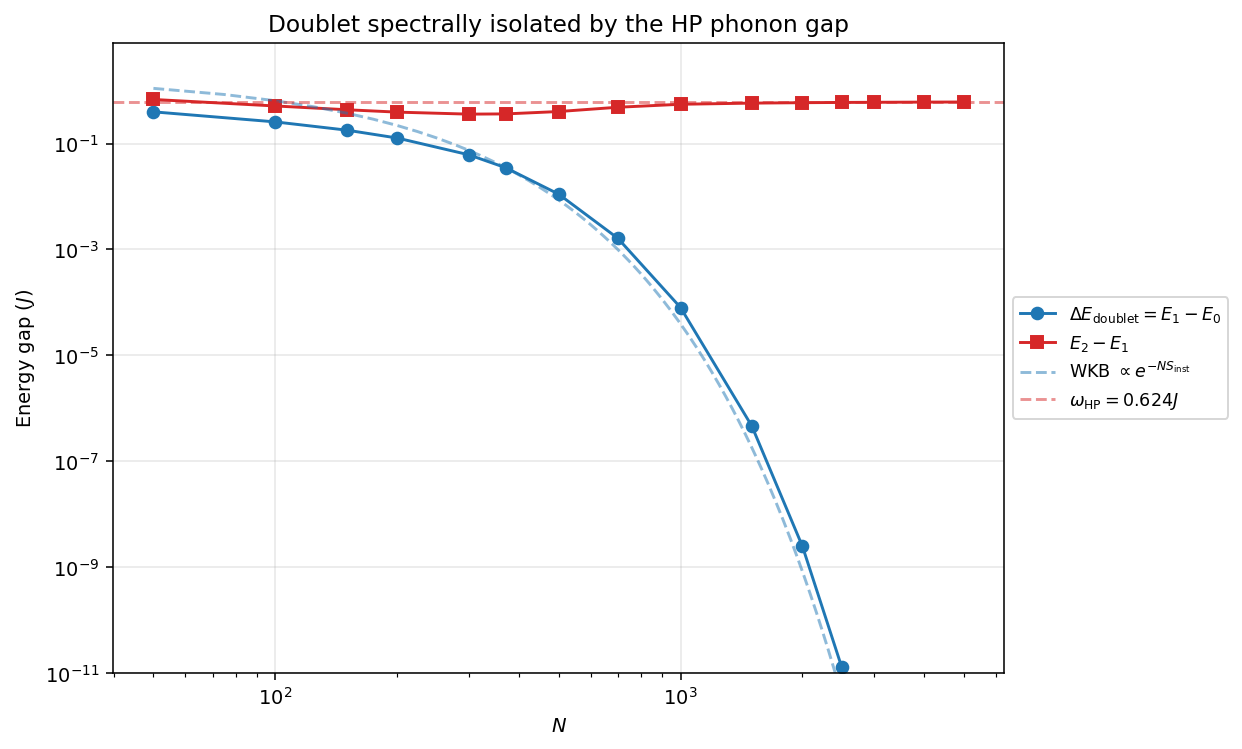}
\caption{\textbf{The two energy scales of the LMG broken phase ($g=0.95$).} The doublet splitting $\Delta E_{\rm doublet}=E_1-E_0$ (blue) decays exponentially with $N$ at the WKB instanton rate $\Sinst=0.0108$; the gap to the next eigenstate $E_2-E_1$ (red) approaches the Holstein--Primakoff gap $\hbar\omega_{\rm HP}=2J\sqrt{1-g^2}=0.624\,J$ with finite-$N$ corrections, asymptotically $N$-independent in contrast to the exponentially closing doublet. The two scales differ by $\sim 10$ orders of magnitude at $N=2500$. The ratio between the HP gap and the doublet splitting grows exponentially with $N$, confirming that the doublet is \emph{spectrally} isolated rather than embedded in a continuum (absolute dynamical isolation is a separate question, addressed in Sec.~\ref{subsec:leakage}). The asymmetric off-doublet $\Jz$-leakage measured in Sec.~\ref{subsec:leakage} is algebraic in $N$ rather than instanton-suppressed, consistent with coupling to HP phonon excitations through ordinary matrix-element corrections.
\label{fig:two_scales}}
\end{figure}

Above the doublet, the LMG spectrum is gapped by the small-oscillation frequency around the broken-symmetry minima,
\begin{equation}
\hbar\omega_{\rm HP}(g) \;=\; 2J\sqrt{1-g^2},
\label{eq:omegaHP}
\end{equation}
the Holstein--Primakoff phonon gap, which is independent of $N$ and equals $0.624\,J$ at our reference $g=0.95$ (equivalently, with $J/\hbar=3.72\times 10^4~\mathrm{rad\,s}^{-1}$, $\omega_{\rm HP}=2(J/\hbar)\sqrt{1-g^2}\simeq 2.32\times 10^4~\mathrm{rad\,s}^{-1}$, i.e.\ $\omega_{\rm HP}/(2\pi)\simeq 3.7~\mathrm{kHz}$)~\cite{DusuelVidalPRB05}. This gap is the scale that isolates the doublet from the rest of the spectrum (Fig.~\ref{fig:two_scales}). The doublet splitting $\Delta E\sim e^{-N\Sinst}$ is exponentially smaller than $\omega_{\rm HP}$ for any $N\gtrsim 1/\Sinst$, but the off-doublet leakage $\ell_{\rm leak}$ measured in Sec.~\ref{subsec:leakage} decays only algebraically in $N$ (effective exponent $\approx 1.0$--$1.2$ over the ED range), not at the instanton scale. The observed algebraic leakage is consistent with ordinary HP-phonon $\Jz$ matrix-element corrections rather than with tunnelling-instanton physics, although we treat the exponent as an empirical ED characterisation. The doublet is therefore \emph{spectrally} isolated from the phonon ladder, but not absolutely dynamically decoupled; the relative vs.\ absolute leakage distinction is made explicit in Sec.~\ref{subsec:leakage}.

\subsection{Variance budget: $G_{01}$ and the doublet weight}
\label{subsec:variance_budget}
The symmetric on-diagonal variance scale $G_{01}$ introduced in~\eqref{eq:G01_def} admits an equivalent expression in terms of the parity-resolved $\Jz$-distributions,
\begin{equation}
G_{01} = \tfrac12\left[(\Jz^2)_{00} + (\Jz^2)_{11}\right] = \tfrac12\sum_m m^2\left(p_m^{(0)} + p_m^{(1)}\right),
\label{eq:G01_pm_form}
\end{equation}
and, by completeness, decomposes as
\begin{equation}
G_{01} = |J_{01}|^2 + L_{\rm leak},\qquad L_{\rm leak} = \tfrac12\sum_{n\ge 2}\left[|\langle E_n|\Jz|E_0\rangle|^2+|\langle E_n|\Jz|E_1\rangle|^2\right],
\label{eq:variance_budget}
\end{equation}
where $L_{\rm leak}\ge 0$ measures the off-doublet variance of $\Jz$ --- i.e., the $\Jz$-matrix-element weight outside the parity doublet. We define the dimensionless \emph{doublet weight}
\begin{equation}
\eta_{\rm mismatch} \equiv \frac{|J_{01}|^2}{G_{01}} \in [0,1],
\label{eq:eta_def}
\end{equation}
which is the fraction of the $\Jz$-variance carried by the doublet itself, and the complementary leakage fraction $\ell_{\rm leak}=1-\eta_{\rm mismatch}=L_{\rm leak}/G_{01}$. In the broken-symmetry regime, the fraction of the $\Jz$ variance carried by matrix elements outside the doublet is suppressed, so $\eta_{\rm mismatch}\to 1$ and $\ell_{\rm leak}\to 0$. We quantify the convergence in Sec.~\ref{subsec:leakage}: over $N\in[200,6000]$ at $g=0.95$ the leakage fraction decays algebraically with an effective exponent $\approx 1.0$--$1.2$ (estimator- and window-dependent; see Sec.~\ref{subsec:leakage}). (The fit is asymptotic; at the mesoscopic benchmark $N=370$ it predicts a slightly smaller leakage than the exact ED value, $\ell_{\rm leak}\approx 0.130$ versus the ED value $\ell_{\rm leak}=0.137$ reported in Table~\ref{tab:benchmark}. The exact ED value at every $N$ in the table is the ground truth; the fit summarises the asymptotic trend.)

\subsection{Summary: the backaction side}
The backaction side of the parity horizon is therefore characterised by a single macroscopic matrix element $J_{01}$, or equivalently its weight $|J_{01}|^2$ growing as $N^2$, with semiclassical limit $(N\mstar/2)^2$ and no instanton suppression. The doublet weight $\eta_{\rm mismatch}=|J_{01}|^2/G_{01}$ approaches unity at large $N$ as a power law. The same operator $\Jz$ that acts transversely with a macroscopic matrix element as a backaction generator distinguishes parity exponentially weakly as a static readout.

A quantitative punchline at the mesoscopic benchmark ($N=370$, $g=0.95$, $\gamma_\phi=0.05\,\mathrm{s}^{-1}$): the intra-doublet parity-imbalance rate $\Gamma_z^{(\rm par)}=2\gamma_\phi J_{01}^2=245\,\mathrm{s}^{-1}$ exceeds the secular coherence decay rate $\Gamma_{01}=\gamma_\phi G_{01}=142\,\mathrm{s}^{-1}$ by a factor $\Gamma_z^{(\rm par)}/\Gamma_{01}=2\eta_{\rm mismatch}=1.73$. Within the projected doublet model, the parity-population imbalance relaxes at the rate scale $\Gamma_z^{(\rm par)}$, which is larger than the secular $E_0/E_1$ coherence-decay rate $\Gamma_{01}$ at this benchmark. This inversion of the intuitive ordering --- one might expect parity to be the protected quantity --- is a direct consequence of the doublet-$\sigma_x$ structure of $\Pdb\Jz\Pdb$. The leakage corrections break the simple projected-doublet symmetry. To quantify this parity-resolved second-moment asymmetry, we define the variance-imbalance parameter
\begin{equation}
\delta_{\rm var}\;\equiv\;\frac{|(\Jz^2)_{11}-(\Jz^2)_{00}|}{G_{01}},
\end{equation}
i.e.\ the imbalance of the two diagonal second moments normalised by their average $G_{01}=\tfrac{1}{2}[(\Jz^2)_{00}+(\Jz^2)_{11}]$. We find $\delta_{\rm var}\approx 0.186$ at the benchmark (the value $0.187$ in Table~\ref{tab:benchmark} is the ED ratio $0.1865$ rounded to three significant figures). Numerically, $(\Jz^2)_{00}=J_{01}^2+L^{(0)}_{\rm leak}$ with $L^{(0)}_{\rm leak}/J_{01}^2\approx 5\%$, and $(\Jz^2)_{11}=J_{01}^2+L^{(1)}_{\rm leak}$ with $L^{(1)}_{\rm leak}/J_{01}^2\approx 27\%$ (variance-budget ratios, not transition rates). The $|E_0\rangle$ sector is therefore well-described by the symmetric bit-flip model while $|E_1\rangle$ has a substantially larger leakage correction at this mesoscopic $N$.

$\delta_{\rm var}$ plays a qualitatively different role from $\eta_{\rm mismatch}$ and $\ell_{\rm leak}$ only at the mesoscopic benchmark, where it is an $O(0.2)$ asymmetry: the two parity eigenstates couple differently to the parity-resolved HP phonon ladder. Since $\Jz$ is parity-odd, $|E_0\rangle$ couples to odd-parity excited states while $|E_1\rangle$ couples to even-parity excited states; at $N=370$ the latter channel carries the larger leakage ($L^{(1)}_{\rm leak}/J_{01}^2\approx 27\%$ versus $L^{(0)}_{\rm leak}/J_{01}^2\approx 5\%$). This asymmetry is bounded by the total leakage, $\delta_{\rm var}=|L^{(1)}_{\rm leak}-L^{(0)}_{\rm leak}|/G_{01}\leq (L^{(0)}_{\rm leak}+L^{(1)}_{\rm leak})/G_{01}=2\ell_{\rm leak}$, and therefore \emph{vanishes} as $N\to\infty$ together with $\ell_{\rm leak}$: the ED values fall from $\delta_{\rm var}\approx 0.65$ at $N=100$ through $0.19$ at the benchmark $N=370$ to below $10^{-3}$ by $N=1000$. We quote $\delta_{\rm var}$ only as a benchmark-scale ($N=370$) parity-resolved second-moment asymmetry, not as a thermodynamic-limit quantity. In any case it does \emph{not} affect the asymptotic exponent of $\chi_\infty^{\rm static}$, which is controlled by the parity-odd node and is $\Sinst(g)$ regardless of $\delta_{\rm var}$.

\subsection{Unifying picture: a single WKB structure with two faces}
\label{subsec:unifying}
The exponential parity horizon and the extensive backaction arise from the same underlying WKB structure. The energy splitting $\Delta E\sim e^{-N\Sinst}$, the parity-distinguishing content of the $\Jz$-distributions $|p_m^{(0)}-p_m^{(1)}|\sim e^{-N\Sinst}$, and the spectral resolution time associated with direct energy discrimination are all governed by the same instanton action $\Sinst(g)$. In contrast, the on-doublet matrix element $\langle E_1|\Jz|E_0\rangle\to N\mstar/2$ --- the action of $\Jz$ as a parity-flipping operator within the doublet --- has no instanton suppression. The same operator is therefore exponentially weak as a parity \emph{reader} and macroscopically strong as a parity \emph{actor}. The kinematic asymmetry --- exponentially small tail overlap versus macroscopic lobe separation --- is a standard double-well property. What the parity horizon adds is the open-system mismatch built on top of it: the same operator that defines the static readout channel (Proposition~\ref{prop:horizon}) acts macroscopically inside the doublet as a backaction generator. This open-system mismatch is the central message of the present work.

\section{The dynamical bypass: a finite-time excess over the static benchmark}
\label{sec:dynamical}
The preceding sections establish a static fact and a kinematic mechanism. The static fact (Sec.~\ref{sec:readout}) is that the commuting $\Jz$-readout channel resolves the parity label only through an exponentially suppressed distinguishability, $\chi_\infty^{\rm static}\sim e^{-N\Sinst}$ (a dimensionless information measure, not a rate): this is the parity horizon. The kinematic mechanism (Sec.~\ref{sec:backaction}) is that the \emph{same} operator acts inside the doublet with an extensive transverse matrix element $|J_{01}|\to N\mstar/2$, generating strong backaction. These two facts raise a sharp question that the static analysis cannot answer: \emph{is the parity horizon a property of the system, or only of the static measurement?} A continuous monitoring record is a different, time-resolved object from a single-time histogram, and it can in principle carry additional information through temporal correlations generated by the monitored dynamics; the macroscopic backaction provides a dynamical channel through which the Hamiltonian could convert otherwise-hidden parity structure into such correlations. Whether it does so --- and over what range of $N$ --- is a question about the full monitored dynamics, not about the static channel.

We answer it here. Using stochastic-master-equation (SME) simulations of the continuously $\Jz$-monitored LMG doublet with a Girsanov-optimal record classifier (method below; full algorithm in App.~\ref{app:sme}), we find that time-resolved monitoring \emph{does} exceed the static single-shot Helstrom/TV benchmark, by a statistically resolved finite-$N$ margin, over a broad finite window of system sizes; that the window is organised by a single dimensionless control parameter $\xi=\omega_{01}/\Gamma_{01}$ rather than by the observation time; and that the bypass closes again --- the homodyne-record excess over the static benchmark returns to zero within uncertainty --- deep in the high-$N$, strong-measurement (Zeno) tail. We state what is demonstrated (the existence, control parameter, robustness, and closure of the bypass) separately from what is not (a universal closed-form closure law; see Sec.~\ref{subsec:closure} and App.~\ref{app:closurelaw}).

\subsection{Setup: continuous monitoring and the discrimination benchmark}
\label{subsec:sme_setup}
An observer continuously homodynes the collective dephasing channel $\hat L=\sqrt{\gamma_\phi}\,\Jz$. For the ideal unit-efficiency homodyne unraveling simulated here (no unobserved channels beyond the one being conditioned on), the full symmetric-sector conditional state obeys the diffusive SME
\begin{equation}
\mathrm{d}|\psi_c\rangle=\Big[-\tfrac{i}{\hbar}\hat H\,\mathrm{d}t
-\tfrac{\gamma_\phi}{2}\big(\Jz-\langle \Jz\rangle_c\big)^2\mathrm{d}t
+\sqrt{\gamma_\phi}\,\big(\Jz-\langle \Jz\rangle_c\big)\,\mathrm{d}W\Big]|\psi_c\rangle,
\label{eq:sme}
\end{equation}
with $\hat H$ the full LMG Hamiltonian and $\mathrm{d}W$ the Wiener increment of the homodyne current $\mathrm{d}y=2\sqrt{\gamma_\phi}\langle\Jz\rangle_c\,\mathrm{d}t+\mathrm{d}W$. The simulation propagates the full symmetric-sector dynamics under the exact $\Jz$ operator --- not a two-level truncation --- so that leakage out of the ground doublet is included by construction; only the two discrimination \emph{hypotheses} are the parity eigenstates $|E_0\rangle,|E_1\rangle$. The task is binary discrimination of these two hypotheses from the record over an observation window $\Tobs$. The optimal (Neyman--Pearson) decision is the sign of the log-likelihood ratio between the two hypotheses, which for a continuous diffusive record takes the Girsanov form, evaluated by propagating two hypothesis-conditioned quantum filters under the \emph{same} record; the success probability $P_{\rm full}$ is estimated by Monte Carlo over many records from each true preparation (App.~\ref{app:sme}).

The benchmark against which $P_{\rm full}$ is compared is the static parity horizon of Sec.~\ref{sec:readout}, expressed operationally. A single $\Jz$ measurement discriminates the two parity \emph{distributions} $p^{(0)}_m,p^{(1)}_m$ with optimal (Helstrom) success probability $\tfrac12(1+\mathrm{TV})$, where $\mathrm{TV}=\tfrac12\sum_m|p^{(0)}_m-p^{(1)}_m|\sim e^{-N\Sinst}$ is the total variation distance. Because $\hat H=0$ in the static channel and the parity-conditioned $\Jz$ statistics are time-independent there, this reference is independent of $\Tobs$. The quantity of interest throughout is the \emph{excess},
\begin{equation}
\Delta P(\,\cdot\,)\;\equiv\;P_{\rm full}-\tfrac12\big(1+\mathrm{TV}\big),
\label{eq:excess}
\end{equation}
the discrimination advantage that the time-resolved record buys over the static snapshot. A positive excess is a bypass of the static horizon \emph{in the single-shot Helstrom/total-variation metric}: it shows the monitored record discriminates parity better than the single-copy $\Jz$-histogram benchmark $(1+\TV)/2$. It is \emph{not} a claim against the static Holevo/JSD quantity of Proposition~\ref{prop:horizon}, which is the asymptotic accessible information of the commuting-record channel and a different object from the finite-time discrimination probability measured here.

\paragraph*{Validation and guardrail.} Every simulation reuses a single validated integrator whose energy-basis observables reproduce the ED benchmark of Sec.~\ref{sec:separation} to one part in $10^{10}$; as an absolute control we run the same pipeline with $\hat H=0$ (a quantum-non-demolition classifier), whose discrimination cannot exceed the static Helstrom reference. All error bars are conservative, defined as the larger of the binomial and batch-scatter standard error. The QND control, the timestep-convergence tests, the batch-level error analysis, and the full per-setting audit table are given in App.~\ref{app:audit}.

\subsection{The information window and its control parameter}
\label{subsec:window}
\begin{figure}[t]
\centering
\includegraphics[width=0.92\columnwidth]{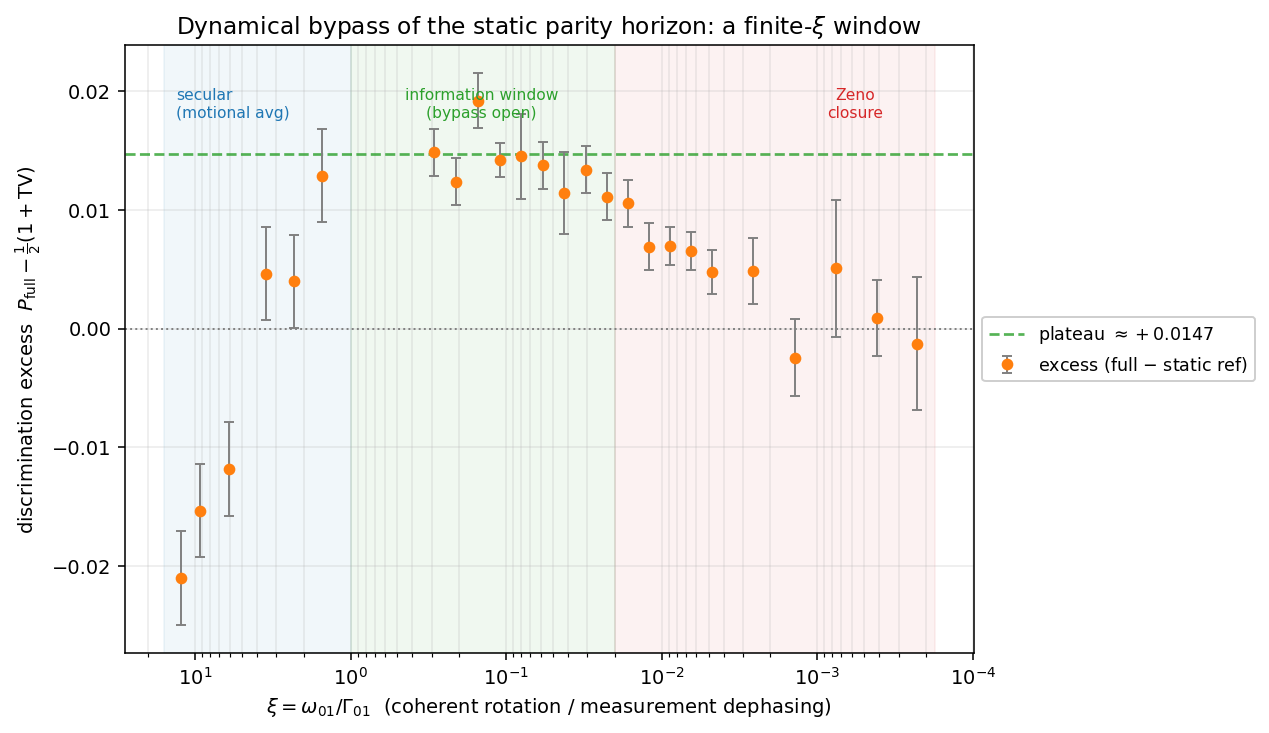}
\caption{\textbf{The dynamical bypass.} Discrimination excess $\Delta P=P_{\rm full}-\tfrac12(1+\mathrm{TV})$ of continuous monitoring over the static parity horizon, versus the control ratio $\xi=\omega_{01}/\Gamma_{01}$ (inverted log axis: large $\xi$ at left, small $\xi$ at right, so the secular regime is at left and the Zeno regime at right; the two baseline observation times pooled at converged timestep). The excess \emph{undershoots} the static reference at the most secular points (left), rises through zero and up to a broad plateau near $\plateauRep$ (Eq.~\ref{eq:plateau}) across the information window, and returns to the static reference in the Zeno tail (right). Error bars are conservative (App.~\ref{app:audit}). $g=0.95$, $\gamma_\phi=0.05~\mathrm{s}^{-1}$.}
\label{fig:dyn_window}
\end{figure}

Figure~\ref{fig:dyn_window} shows the excess \eqref{eq:excess} as a function of $\xi=\omega_{01}/\Gamma_{01}$, the ratio of the intra-doublet coherent rotation rate $\omega_{01}=\Delta E/\hbar$ to the measurement-induced dephasing rate $\Gamma_{01}=\gamma_\phi G_{01}$, with $G_{01}=\tfrac12(\langle E_0|\Jz^2|E_0\rangle+\langle E_1|\Jz^2|E_1\rangle)$. Three regimes are evident.

In the \emph{secular} regime ($\xi\gg 1$, small $N$), coherent rotation is fast compared with the measurement and the record sees only the parity-symmetric time-average of $\Jz$, so there is no positive bypass. At the most secular points the finite-time record in fact \emph{undershoots} the static single-shot benchmark (the asymptotic $\Tobs\!\to\!\infty$ Helstrom reference), and the excess rises through zero only as $\xi$ approaches the crossover near order unity. The monitored full record confers no advantage over the $\hat H=0$ (QND) control in this regime: both undershoot the asymptotic reference because finite-time monitoring cannot resolve the full parity-distinguishing structure.

In the \emph{information window} (intermediate $\xi$, mesoscopic $N$), the two rates are comparable --- a broad plateau over $\xiPlateauEdge\lesssim\xi\lesssim 10^{-1}$, with closure below a few $\times 10^{-3}$ --- and the measurement resolves parity-distinguishing structure before rotation averages it out and before measurement freezes the dynamics. The excess is robustly positive and approximately flat across the plateau, resolved down through the knee to $\xi\approx 0.0065$ before closure. Six representative system sizes ($N=625$--$750$) span the plateau and every one shows a positive excess; the pooled plateau statistic is
\begin{equation}
\Delta P_{\rm window}=\plateauExcess\pm\plateauErr,
\label{eq:plateau}
\end{equation}
a representative plateau level of $\plateauRep$ (per-setting significances, trajectory counts, and the error-bar construction are given in App.~\ref{app:audit}). Because the static departure $\tfrac12\mathrm{TV}$ falls across the window while the excess stays approximately flat, the \emph{relative} advantage grows with $N$: by $N\approx 750$ the excess is several times the static departure. In absolute terms the gain is about $1.5$ percentage points in success probability above the static benchmark --- small in size but statistically resolved; the significance is the existence of a finite-$N$ dynamical excess over the static commuting-record benchmark within the specified homodyne-record protocol and classifier, not a large absolute effect.

In the \emph{Zeno} regime ($\xi\ll 1$, large $N$), measurement-induced dephasing dominates coherent rotation, the dynamics are frozen into the measured ($\Jz$-pointer) structure, and the excess returns toward zero (Sec.~\ref{subsec:closure}). The window therefore has two walls, which close for two physically distinct reasons --- motional averaging on the secular side, measurement freezing on the Zeno side --- and it is this two-sided structure, not a one-sided anomaly, that justifies calling it a window.

\subsection{The control parameter is $\xi$, not the observation time}
\label{subsec:xicontrol}
Identifying $\xi$ as the organising variable requires ruling out the obvious alternatives. At fixed $N$ along the static family, $\xi$, the accrued coherent phase $\omega_{01}\Tobs$, and the record length in dephasing units $\Gamma_{01}\Tobs$ all vary together, so a scan in $N$ alone cannot distinguish ``the excess depends on $\xi$'' from ``the excess depends on how long, or how far, the doublet rotates during the record.'' These are physically different claims: the first is an intrinsic rate competition; the second would make the bypass an artifact of a finite observation window that longer monitoring would remove.

\begin{figure}[t]
\centering
\includegraphics[width=\columnwidth]{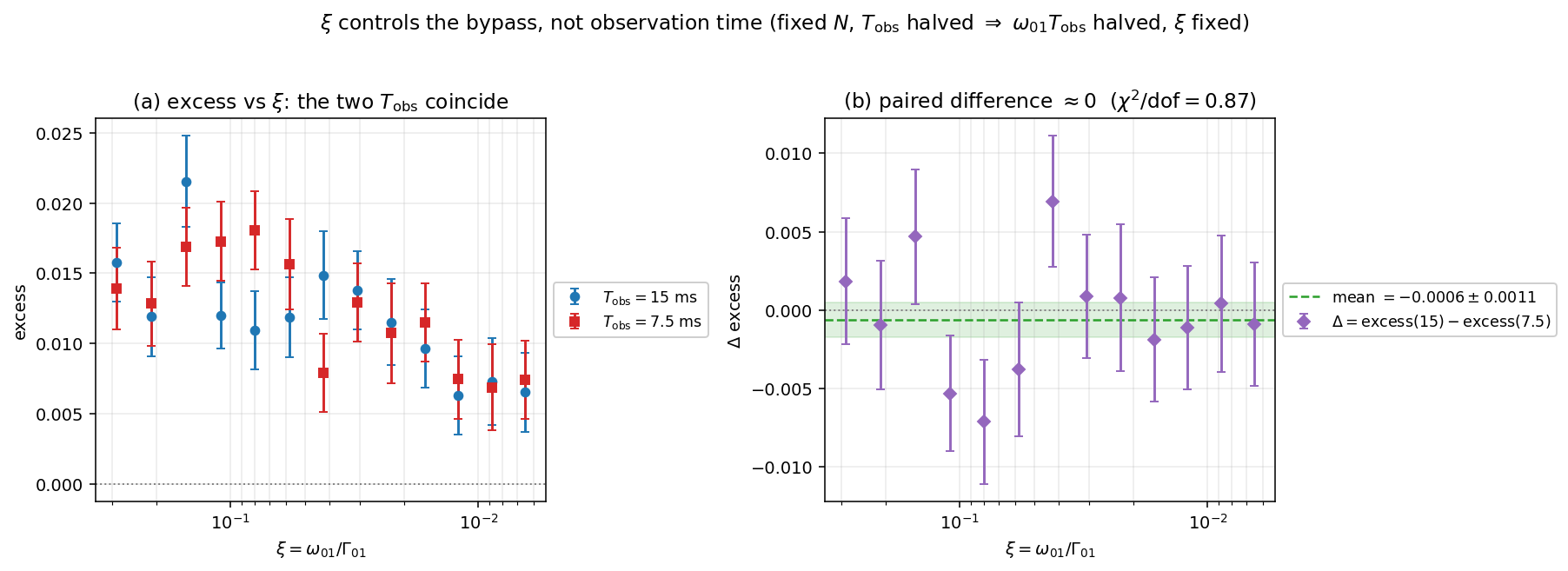}
\caption{\textbf{The rate ratio $\xi$, not the observation time, organises the bypass.} (a) Excess versus $\xi$ at $\Tobs=15$~ms (circles) and $7.5$~ms (squares): the two coincide at each $\xi$. (b) Paired difference $\Delta P(15)-\Delta P(7.5)$ at each $N$, consistent with zero (Eq.~\ref{eq:tobs}; band is the pooled mean $\pm$ s.e.). At fixed $N$, halving $\Tobs$ halves the accrued phase $\omega_{01}\Tobs$ and the record length $\Gamma_{01}\Tobs$ while leaving $\xi$ fixed; the excess is invariant, disfavouring both as the controlling variable.}
\label{fig:dyn_tobs}
\end{figure}

We break this degeneracy directly. Because $\xi$ is a ratio of rates it is independent of $\Tobs$, whereas $\omega_{01}\Tobs$ and $\Gamma_{01}\Tobs$ scale linearly with it. We therefore repeat the discrimination at each $N$ for two observation times, $\Tobs=15$ and $7.5$~ms, which halves both $\Tobs$-dependent quantities while holding $\xi$ fixed. Figure~\ref{fig:dyn_tobs} shows the result across the window and into the knee ($\xi$ from $\tobsNxiHi$ down to $\tobsNxi$, with $\omega_{01}\Tobs$ swept through unity over the range $[\tobsPhaseLo,\tobsPhaseHi]$). The two observation times give the \emph{same} excess at each $\xi$: the paired difference is
\begin{equation}
\begin{split}
\Delta P(15~\mathrm{ms})-\Delta P(7.5~\mathrm{ms})&=\tobsDiff\pm\tobsDiffErr,\\
\chi^2/\mathrm{dof}&=\tobsChiSq,
\end{split}
\label{eq:tobs}
\end{equation}

consistent with zero across the entire range, with no feature as $\omega_{01}\Tobs$ crosses unity. Across all $\tobsNcount$ system sizes tested at two observation times --- $N=625$, $650$, $675$, $700$, $725$, $750$, $775$, $800$, $825$, $850$, $875$, $900$, $925$ --- halving the observation time, and with it the accrued phase and the record length, leaves the excess statistically unchanged. This is strong evidence against the accrued-phase and record-length explanations and identifies $\xi$, the intrinsic rate ratio, as the variable that organises the bypass over the tested range. A physical corollary follows: the Zeno-side closure reflects an intrinsic competition between coherent rotation and measurement-induced dephasing, not a shortage of observation time.

\subsection{Zeno-side closure and the absence of a universal exponent}
\label{subsec:closure}
\begin{figure}[t]
\centering
\includegraphics[width=0.92\columnwidth]{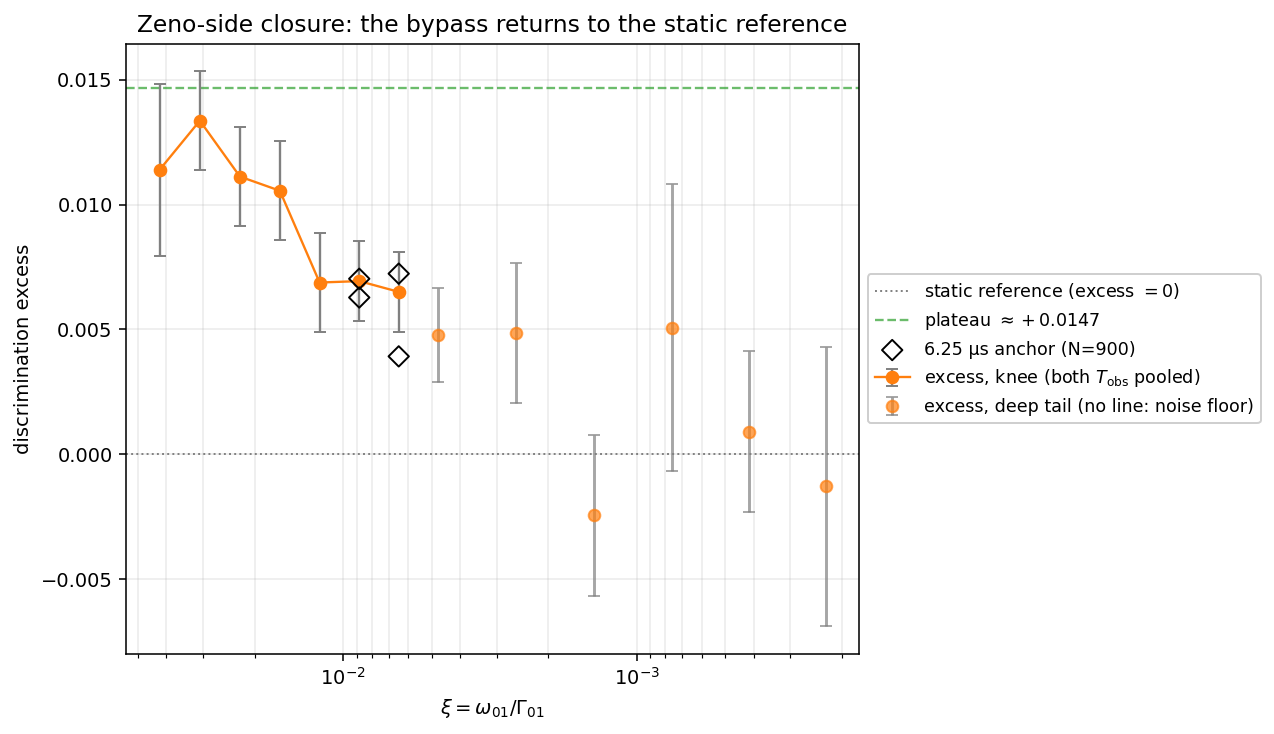}
\caption{\textbf{Zeno-side closure.} Detail of the descent: excess versus $\xi$ through the knee and into the deep tail (both $\Tobs$ pooled; line drawn only through the resolved knee, $\xi\gtrsim 0.006$; deep-tail points shown without a connecting line as they lie at the noise floor). Open diamonds: independent $6.25~\mu$s timestep anchors at $N=900$ and $N=925$, confirming timestep convergence. The excess returns to the static reference (Eq.~\ref{eq:closure}); the plateau level (dashed) and zero (dotted) are shown for reference.}
\label{fig:dyn_knee}
\end{figure}

Figure~\ref{fig:dyn_knee} resolves the Zeno wall. The plateau persists down to $\xi\simeq\xiPlateauEdge$; below this the excess descends through a knee, reaching approximately half the plateau value near $\xi\simeq\xiKneeHalf$, and is consistent with zero in the deep tail. Pooling the deepest converged batch-level settings ($N\geq 1050$, $\closureTraj$ full-LMG trajectories) by system size --- each $N$ one inverse-variance-weighted unit, as for the plateau --- gives
\begin{equation}
\Delta P_{\rm tail}=\closureExcess\pm\closureErr\qquad(\closureSigma\sigma),
\label{eq:closure}
\end{equation}
i.e.\ the homodyne-record excess over the static benchmark returns to zero within uncertainty: time-resolved monitoring confers no resolvable advantage once the dynamics are Zeno-frozen. The excess is strongly suppressed through the knee by $N\simeq\Nclose$ ($\xi\lesssim\xiClose$) and is statistically unresolved from zero only in the deeper Zeno tail, $N\gtrsim 1050$. The descent is gradual across the knee rather than an abrupt threshold, and the converged-timestep data are essential here: at coarser timestep the deep-tail excess is systematically overstated, and several apparently-positive deep points regressed to zero under increased statistics and finer timestep (App.~\ref{app:audit}).

The closure inherits the instanton exponent through $\omega_{01}$. Since $\omega_{01}\propto e^{-N\Sinst}$ collapses exponentially while $\Gamma_{01}\propto N^2\mstar^2$ grows only polynomially, the ratio $\xi(N)$ falls exponentially, and any fixed closure threshold $\xi\simeq\xi_c$ is reached at $N_{\rm close}\sim\Sinst^{-1}\ln[\,\cdot\,]$. The same instanton action $\Sinst(g)$ that sets the exponential suppression of the static readout (Sec.~\ref{sec:readout}) therefore enters the location where the dynamical bypass shuts off --- one instanton action governing both faces (Sec.~\ref{subsec:unifying}), now extended from the static horizon to the boundary of the dynamical window. This is the exponent entering through $\omega_{01}$, not a verified closure \emph{scaling law}: the dependence of the closure threshold on $g$ would require a multi-$g$ dynamical study, which we have not performed (all monitored runs are at $g=0.95$).

\paragraph*{What we do not claim.} It is tempting to seek a closed-form closure law $\Delta P\sim\xi^{p}$. We do not claim one. A heuristic argument based on the stationary power spectrum of the record suggests $p=2$, but that argument is not valid for the present classifier: by parity the ensemble-averaged $\langle\Jz\rangle$ vanishes in both hypotheses, the symmetric (classical) power spectrum is identical for the two hypotheses, and the parity-distinguishing information resides in the \emph{causal, conditional} filtering of the record --- a time-directed object not captured by a stationary spectral difference. The Girsanov-optimal record classifier exploits exactly this conditional structure. Empirically, the knee provides a resolved descent over which the excess falls from the plateau to the noise floor, but a finite-range descent does not establish an asymptotic exponent, and the deep tail is at the resolution floor where no slope can be fit. We therefore report the closure as a robust qualitative fact --- the bypass exists, is $\xi$-controlled, and closes --- and treat the closure \emph{exponent} as an open quantitative question (App.~\ref{app:closurelaw}), to be settled, if at all, by a dedicated high-statistics study of the knee at two observation times. No universal power law is claimed. All of these statements are made for the specified homodyne record and Girsanov-optimal record classifier, not for the full bath Holevo information. This is the conservative reading the data support, and it is independent of any particular closure law.

\subsection{Summary: the dynamical side}
\label{subsec:dyn_summary}
The static parity horizon is not a bound on time-resolved monitoring. Continuous $\Jz$ monitoring of the LMG doublet opens a broad, finite-$N$ dynamical bypass of the horizon, with a discrimination excess of $\plateauRep$ over the static benchmark across the information window (per-setting significances summarised in App.~\ref{app:audit}); the bypass is organised by the single rate ratio $\xi=\omega_{01}/\Gamma_{01}$ and is statistically consistent with $\Tobs$-insensitivity over the tested $7.5$--$15$~ms paired comparison (Eq.~\ref{eq:tobs}) and the wider $3$--$50$~ms sweep at $N=700$ (App.~\ref{app:audit}), so its closure reflects an intrinsic measurement-versus-tunnelling competition; and it closes in the high-$N$ Zeno tail at a location into which the same instanton action $\Sinst$ enters through the exponential collapse of $\omega_{01}$. The bypass is what makes the static benchmark consequential: the horizon is the exponentially weak baseline that a continuous record, in the right regime, exceeds by a statistically resolved margin in the tested window.

\section{Numerical evidence}
\label{sec:separation}
We now turn from analytical structure to exact-diagonalisation (ED) evidence for the static readout and backaction quantities. The aim of this section is to anchor the spectral gap, the histogram distinguishability measures, the backaction matrix element, and the benchmark variance budget in explicit ED calculations with controlled fitting and convergence. The ED grids extend to $N=4500$, with observable-dependent reliability masks; the full reproduction package is documented in App.~\ref{app:reproducibility}. The continuously monitored dynamics are treated separately in Sec.~\ref{sec:dynamical} and App.~\ref{app:sme}. A separate convergence study characterises the drift of the Chernoff optimum $s^*(N)$ across $N\in[100,2500]$ and verifies its stability at $s^*\simeq 0.60$ with drift $<0.02$ over the full range; the $(C-\mathcal{B})/C$ ratio cited in Sec.~\ref{subsec:WKB} as around $3$--$4\%$ derives from this convergence study.

\paragraph*{Benchmark parameter setup.} We adopt a single mesoscopic benchmark: $N=370$, $g=0.95$, $\gamma_\phi=0.05~\mathrm{s}^{-1}$, and $J/\hbar=3.72\times 10^4~\mathrm{rad\,s}^{-1}$. Here $g=0.95$ is in the broken phase but close enough to criticality to keep the instanton numerics accessible, with $\mstar(0.95)=0.312$. The rate $J/\hbar$ is a cold-atom-scale collective interaction rate used for numerical calibration, equivalently $J/h\simeq 5.9~\mathrm{kHz}$. The window-dependent operational scales (observation window $T_{\rm obs}$, $N_\pi$, $N_1$) and the intrinsic crossover scale $N_\xi$ are defined and used in Sec.~\ref{sec:operational}--\ref{sec:scope_regimes} below; they play no role in the static ED slopes of this section. The static readout/backaction separation is not tied to the numerical units chosen here, and the ED exponent analysis is checked across several $g$ values below. The monitored-dynamics demonstration, however, is carried out at the benchmark coupling $g=0.95$; extending the dynamical window and closure analysis across $g$ is a separate numerical problem.

We focus exclusively on $g=0.95$ unless otherwise stated, taking this as the near-critical broken-phase benchmark ($g=0.95$ is 5\% below $g_c=1$). The complementary $g$-sweep extends the picture across $g\in\{0.70,0.80,0.85,0.90,0.95,0.98\}$.

\subsection{Asymptotic convergence at $g=0.95$: a single instanton exponent}
\label{subsec:T11}
\begin{figure*}[t]
\centering
\includegraphics[width=0.96\textwidth]{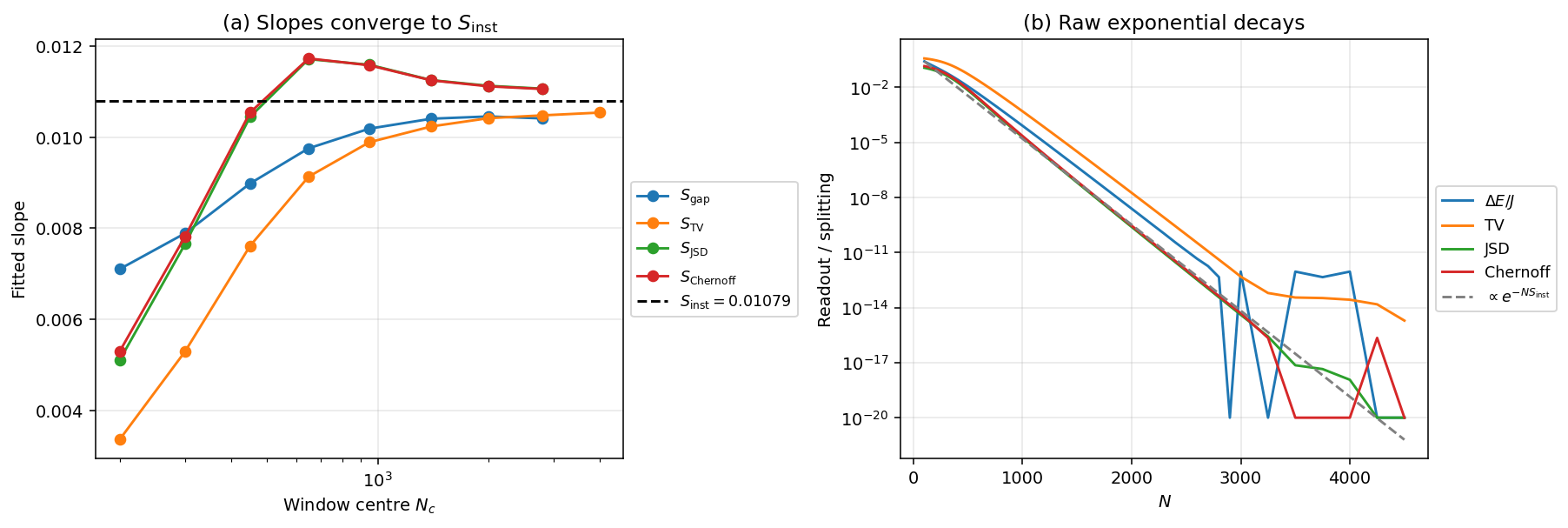}
\caption{\textbf{Asymptotic convergence to a single instanton exponent.} \emph{(a)} Rolling-window fits of $S_X = -d\ln Q_X/dN$ for $Q_X\in\{\Delta E, \TV, \JSD, \text{Chernoff}\}$ over log-spaced fitting windows $[N_c/1.5,\,1.5N_c]$ centred at $N_c\in[200,3000]$. All four slopes converge to the analytical instanton action $\Sinst(0.95)=0.01079$ (black dashed); $S_{\rm gap}$ and $S_{\rm TV}$ approach from below, $S_{\JSD}$ and $S_{\rm Chernoff}$ from above. By the largest reliable windows, all four fitted slopes are within a few percent of $\Sinst$. The pale dashed Bhattacharyya curve ($s=1/2$, labelled $S_{\rm Bhat}$) is shown as a diagnostic comparison only; the quoted Chernoff \emph{exponent} uses the true minimised value over $s\in[0,1]$ and agrees with the Bhattacharyya slope to $\Delta<0.0005$ (the $s^*$ shift moves the prefactor, not the exponent), even though the corresponding Chernoff and Bhattacharyya \emph{information values} differ by $\approx 4\%$ at our parameters (see Sec.~\ref{subsec:WKB}). \emph{(b)} Raw exponential decays of the four quantities (log scale). Over the reliable range, all four approach a common slope $\propto e^{-N\Sinst}$ (gray dashed) at large $N$. The total-variation curve flattens at $N\gtrsim 2500$ because of the double-precision subtractive-cancellation floor (about $10^{-13}$, well above true underflow at $10^{-308}$); these floor-limited points are automatically excluded from every fit by the answer-independent magnitude floor described in the methods note below, and contribute to none of the quoted slopes.
\label{fig:t1_convergence}}
\end{figure*}

We diagonalise the LMG Hamiltonian exactly up to $N=4500$ pseudospins at $g=0.95$ and extract four conceptually independent quantities --- the spectral doublet splitting $\Delta E$, and three distinguishability measures of the parity-conditioned magnetisation histograms (total variation, Jensen--Shannon divergence, Chernoff information) --- on a sliding sequence of logarithmically-spaced fitting windows. There is no fitting freedom in the WKB target: the instanton value predicted by the semiclassical mechanism, $\Sinst(0.95)=0.01079$, is computed analytically from the spin-coherent-state action of Eq.~\eqref{eq:Sinst_SCS} with no adjustable parameters. All four exponents converge onto that single number to within a few percent at the largest reliable system sizes, and the residual finite-$N$ deviations show an approximate collapse when rescaled by $N\Sinst(g)$ across six couplings $g\in\{0.70,0.80,0.85,0.90,0.95,0.98\}$ (App.~\ref{app:ed_supp}, Fig.~\ref{fig:multig}). Figure~\ref{fig:t1_convergence}(a) is the central ED result of the paper.

\paragraph*{Scope of the asymptotic claims.} A fully uniform asymptotic derivation of the nonlinear prefactors would require matching the WKB lobe, barrier, central-node, and turning-point regions while controlling the singular small-deviation expansion of each information functional. We do not attempt that uniform matched-asymptotic calculation here. The analytic contribution of the present work is the instanton-scale node/barrier mechanism (Sec.~\ref{subsec:WKB}) and the path-additivity exponent; the measure-dependent prefactors are treated numerically by ED with floor-controlled reliability masks.

By the largest reliable rolling-window centre ($N_c=2799$), the four fitted slopes agree with the analytical instanton action $\Sinst=0.01079$ to within a few percent:
\begin{equation}
\{S_{\rm gap},\,S_{\TV},\,S_{\JSD},\,S_{\rm Chernoff}\}=\{0.01041,\,0.01048,\,0.01106,\,0.01106\}.
\label{eq:four_exponent_headline}
\end{equation}
The spectral gap and total variation approach $\Sinst$ from below, the two nonlinear information functionals (JSD and Chernoff) from above. The full rolling-window history, the from-below/from-above approach pattern, the answer-independent floor mask and its sensitivity analysis, the per-bin JSD anatomy, the independent subleading-model ($M2$/$M3$) exponent extraction, and the cross-$g$ scaling collapse that disfavours a purely near-critical interpretation are documented in App.~\ref{app:ed_supp}. The central result here is the qualitative fact itself: all four exponents approach a single instanton scale, with the nonlinear functionals \emph{not} splitting off to the doubled exponent $2\Sinst$ that a naive lobe-based small-deviation expansion would predict.

\subsection{Backaction scaling and the mesoscopic benchmark}
\label{subsec:backaction_ed}
On the backaction side, ED confirms the semiclassical scaling $|J_{01}|\to N\mstar/2$ with a leading finite-$N$ offset,
\begin{equation}
|J_{01}| = \frac{N\mstar}{2}-\delta(N,g),\qquad \delta_\infty(g{=}0.95)\simeq 5.8,
\label{eq:J01_headline}
\end{equation}
the offset carrying a model spread of a few $\times 10^{-2}$ between two- and three-parameter fits (restricted-range fit and nested-model selection in App.~\ref{app:ed_supp}). The off-doublet leakage fraction $\ell_{\rm leak}=1-\eta_{\rm mismatch}$ decays algebraically with an effective exponent $\approx 1.0$--$1.2$, so the doublet weight $\eta_{\rm mismatch}=|J_{01}|^2/G_{01}\to 1$; this is a \emph{relative} matrix-element statement (the absolute off-doublet variance still grows with $N$ at fixed $\gamma_\phi$), not an absolute decoupling theorem, as detailed in Sec.~\ref{subsec:variance_budget} and App.~\ref{app:ed_supp}. Table~\ref{tab:benchmark} collects the complete set of benchmark quantities at the mesoscopic reference point $N=370$.

\subsection{Benchmark table at the mesoscopic reference point}

\begin{table}[ht]
\centering
\caption{Benchmark quantities at $N=370$, $g=\Gamma/J=0.95$, $\gamma_\phi=0.05\,\mathrm{s}^{-1}$, $J/\hbar=3.72\times 10^4\,\mathrm{rad\,s}^{-1}$. All values from exact diagonalisation in the symmetric Dicke sector. Within the projected doublet model, the rate scale $\Gamma_z^{(\rm par)}=2\gamma_\phi J_{01}^2$ exceeds $\Gamma_{01}$ by a factor $1.73$: the parity-population imbalance relaxes faster than the secular $E_0/E_1$ coherence decays. The percentages in the two leakage-rate rows are the variance-budget fractions $L^{(i)}_{\rm leak}/|J_{01}|^2$, equivalently the leakage rates quoted relative to $\gamma_\phi|J_{01}|^2$, not relative to $\Gamma_{01}$.}
\label{tab:benchmark}
\begin{tabular}{lcc}
\toprule
Quantity & Symbol & Value \\
\midrule
Order parameter & $\mstar$ & $0.3122$ \\
Tunnelling splitting & $\omega_{01}=\Delta E/\hbar$ & $1310\;\mathrm{rad\,s}^{-1}$ \\
Secular ratio & $\omega_{01}/\Gamma_{01}$ & $9.23$ \\
Off-diagonal $\Jz$ magnitude & $|J_{01}|$ & $49.51$ \\
Squared off-diagonal & $|J_{01}|^2$ & $2451$ \\
Doublet second moment & $G_{01}$ & $2839$ \\
Doublet weight & $\eta_{\rm mismatch}=|J_{01}|^2/G_{01}$ & $0.863$ \\
Doublet asymmetry & $\delta_{\rm var}$ & $0.187$ \\
Mixture reference & $G_{\rm mix}=(N\mstar)^2/4$ & $3337$ \\
Coherence rate & $\Gamma_{01}=\gamma_\phi G_{01}$ & $142\;\mathrm{s}^{-1}$ \\
Parity-population relaxation scale & $\Gamma_z^{(\rm par)}=2\gamma_\phi |J_{01}|^2$ & $245\;\mathrm{s}^{-1}$ \\
Rate ratio & $\Gamma_z^{(\rm par)}/\Gamma_{01}$ & $1.73$ \\
Coherence ratio (mixture reference) & $\Gamma_{01}/\Gamma_{\rm mix}$ & $0.851\;(\to 1)$ \\
Leakage rate $|E_0\rangle$ & $\gamma_\phi[(\Jz^2)_{00}-|J_{01}|^2]$ & $6.15\;\mathrm{s}^{-1}$ ($5\%$) \\
Leakage rate $|E_1\rangle$ & $\gamma_\phi[(\Jz^2)_{11}-|J_{01}|^2]$ & $32.6\;\mathrm{s}^{-1}$ ($27\%$) \\
Coherence time & $T_2=1/\Gamma_{01}$ & $7.04\;\mathrm{ms}$ \\
Static-channel Holevo (nats) & $\chi_\infty^{\rm static}=\JSD(p^{(0)},p^{(1)})$ & $0.0257$ \\
Total variation & $\TV$ & $0.137$ \\
\bottomrule
\end{tabular}
\end{table}

To summarise the numerical evidence: in the open LMG model in its broken phase, the same operator $\Jz$ that reads parity exponentially weakly --- with the spectral doublet splitting and the three histogram measures (TV, JSD, Chernoff) sharing the instanton exponent $\Sinst(g)$ --- acts on parity extensively strongly through the off-diagonal matrix element $|J_{01}|\to N\mstar/2$. This readout/backaction asymmetry is the core of the parity horizon: not a coincidence of unrelated scalings, but the two faces of a single parity-odd selection rule (Sec.~\ref{sec:two-roles}).


\FloatBarrier
\section{Operational consequences}
\label{sec:operational}
What does the readout--backaction separation imply for an operational measurement protocol --- specifically, continuous homodyne detection along $\Jz$?

In the Wiseman--Milburn unravelling~\cite{Wiseman2009}, the homodyne photocurrent is
\begin{equation}
dy_t = 2\sqrt{\gamma_\phi}\,\langle\Jz\rangle_{c}\,dt + dW_t,
\label{eq:photocurrent}
\end{equation}
where $\langle\Jz\rangle_c(t)\equiv\mathrm{Tr}[\Jz\rho_c(t)]$ is the conditional mean of $\Jz$ along the monitored trajectory with conditional state $\rho_c(t)$, and $dW_t$ is a standard Wiener increment.

The readout side of Proposition~\ref{prop:horizon} gives the asymptotic static-channel benchmark for what an outside observer can learn when the commuting $\Jz$ record is fully resolved ($\hat H=0$); a finite-time integrated homodyne current is a noisier, protocol-dependent version of this readout. Finite-time experiments with the full LMG dynamics are more subtle.

\paragraph*{Record-blindness of the strict doublet qubit.} In the strict doublet-qubit approximation, where the nontrivial part of $\Hdb$ is proportional to $\sigmaz$ and $\Jz^{\rm db}=J_{01}\sigmax$, both parity eigenstates have identical homodyne first moments: $\langle E_i|\Jz^{\rm db}|E_i\rangle = J_{01}\langle\sigmax\rangle_i = 0$, because $|E_{0,1}\rangle$ are eigenstates of $\sigmaz$, not $\sigmax$. The projected two-level mean signal is therefore record-blind to the initial parity pole. The dynamical excess found in Sec.~\ref{sec:dynamical} should not be read as following from the size of $J_{01}$ alone; it depends on the full monitored LMG dynamics, including the detailed wavefunction structure, causal conditioning of the record, and finite off-doublet seed terms absent from the strict projected model. This is consistent with Proposition~\ref{prop:horizon}, which concerns the static commuting-record benchmark rather than the full finite-time monitored dynamics.

\paragraph*{What $J_{01}$ supplies.} The Hamiltonian splitting $\omega_{01}=\Delta E/\hbar$ drives the doublet precession (intra-doublet Bohr oscillation), while the transverse response scale $J_{01}$ sets the strength of the measurement/backaction coupling. Both can enter finite-time temporal \emph{correlations} of the homodyne record. The Neyman--Pearson log-likelihood ratio classifier
\begin{equation}
\mathrm{LLR}(\{dy_t\}) = \ln\frac{P(\{dy_t\}|E_0)}{P(\{dy_t\}|E_1)}
\label{eq:LLR}
\end{equation}
can in principle exploit temporal and higher-order statistics of the record that are absent from the static commuting-record benchmark. The size of this \emph{dynamical excess} depends on the precession structure, the bath dephasing rate, the observation window, and the specific protocol. We quantify it directly in Sec.~\ref{sec:dynamical}: across the information window the continuous-monitoring excess over the static benchmark is $\plateauExcess\pm\plateauErr$, is organised by the single rate ratio $\xi=\omega_{01}/\Gamma_{01}$ more cleanly than by the tested observation-time variables, and closes in the Zeno tail at a location into which the same instanton action $\Sinst$ enters through $\omega_{01}$. The excess is thus not a theorem following from $J_{01}$ alone --- it is a property of the full monitored dynamics --- and Sec.~\ref{sec:dynamical} demonstrates it numerically with the validated simulation-and-classifier pipeline of App.~\ref{app:sme}.

What our ED analysis establishes independently of any dynamical calculation is that the extensive backaction strength $|J_{01}|^2$ provides the geometric \emph{raw material} for a dynamical excess. Whether a specific dynamical protocol remains close to the static-channel reference or exhibits an excess over it depends on how well that protocol's matched filter aligns with the time-correlation structure imprinted by the backaction.

\paragraph*{Bridge to the full dynamical problem.} For finite $\gamma_\phi$, the monitoring is not in the static channel because $[\Jz,\hat H_{\rm LMG}]\neq 0$ (i.e., the strict QND condition fails, as it must for the LMG). The static commuting-channel result should therefore be read as a benchmark for the ideal commuting-record protocol. In the full dynamical problem, the bath record can contain time correlations generated by $\omega_{01}$, measurement backaction through $J_{01}$, and leakage to the HP ladder. Inside the doublet, corrections are naturally controlled by ratios such as
\[
\frac{\omega_{01}}{\Gamma_z^{(\rm par)}}
= \frac{\omega_{01}}{2\gamma_\phi |J_{01}|^2}
= \frac{\xi}{2\,\eta_{\rm mismatch}},
\]
where $\xi=\omega_{01}/\Gamma_{01}$ is the secular ratio of Sec.~\ref{sec:scope_regimes} and $\eta_{\rm mismatch}=|J_{01}|^2/G_{01}$ is the doublet weight introduced in Eq.~\eqref{eq:eta_def}; this ratio compares coherent precession to measurement-induced parity-population mixing inside the doublet, while off-doublet corrections depend on the absolute leakage scale $\gamma_\phi L_{\rm leak}$. Section~\ref{sec:dynamical} studies this monitored problem numerically; the full Holevo limit for arbitrary measurements on the entire bath remains open.

\section{Scope of the static result and regime hierarchy}
\label{sec:scope_regimes}
Proposition~\ref{prop:horizon} is proved under the static $\Jz$-readout channel ($\hat H=0$); Sec.~\ref{sec:dynamical} studies the full monitored dynamics. This section places the two results on a common map, organised by the rate ratio $\xi=\omega_{01}/\Gamma_{01}$, and states cleanly what the static benchmark does and does not bound.

\subsection{The horizon as a suppression statement, not a bound on full dynamics}
The parity horizon is, formally, a statement about the static channel. The Holevo quantity $\chi_\infty^{\rm static}=\JSD(p^{(0)},p^{(1)})\sim e^{-N\Sinst}$ is the asymptotic classical information the bath records about parity \emph{when the system Hamiltonian is zero and the bath records over time become simultaneously diagonal in the $\Jz$-eigenbasis}. It is not, by itself, a bound on the Holevo quantity of the full LMG monitored channel ($\hat H_{\rm LMG}\neq 0$), which is a different mathematical object and remains open (Sec.~\ref{subsec:limitations}).

This distinction is consequential. When the full LMG dynamics are on, the doublet undergoes coherent Bohr oscillation at $\omega_{01}$, the photocurrent acquires temporal structure that the static $\Jz$-distribution by construction cannot encode, and finite-time monitored statistics can differ from $(1+\TV)/2$ in either direction. \emph{None of this constitutes a horizon violation.} A genuine horizon violation in the present framework would require the asymptotic full-LMG bath Holevo quantity to decay more slowly than $e^{-N\Sinst}$ as $N\to\infty$ --- a different statement entirely from any finite-$N$ deviation of an operational classifier from the static-TV reference. We do not identify such a mechanism in the present analysis: the lobe-overlap geometry that produces the $e^{-N\Sinst}$ exponent is generic to any operator-content statement about the parity doublet.

\subsection{Three regimes for the full LMG and the scope of the static result}
\label{subsec:three_regimes}
The ratio $\xi=\omega_{01}/\Gamma_{01}$ introduced in Sec.~\ref{sec:intro} organises the data over the tested parameter set, determining where the static channel is a natural reference for the dynamics, a reference under different averaging, or ceases to apply as a reference at all. The three regimes below are the three orderings of $\xi$ already described there; here we record only what each implies for the \emph{scope} of the static benchmark (Table~\ref{tab:regimes}).

\paragraph*{(S) Secular regime: $\xi\gg 1$, $N\lesssim N_\xi$.} First, an apparent paradox: setting $\hat H=0$ in the static channel might seem the \emph{opposite} of the fast-rotation regime $\xi\gg 1$. It is not, because the inputs of Proposition~\ref{prop:horizon} are energy eigenstates, whose single-time $\Jz$-histogram $p_m^{(i)}=|\langle m|E_i\rangle|^2$ is identical whether $\hat H=\hat H_{\rm LMG}$ or $\hat H=0$ at any $\xi$; the $\hat H=0$ channel is therefore exact for these inputs at the histogram level, not an approximation to the dynamics. The secular condition $\xi\gg 1$ is invoked separately, for the backaction-corrected problem: adiabatic (Bloch--Redfield secular) averaging is expected to suppress population--coherence cross-terms inside the doublet, making the static channel a natural benchmark for the slowly varying part of the record. We do not derive the analytic form of these corrections, but we test the secular-side expectation directly. We added six continuous-monitoring runs ($\Tobs=15$~ms, $\mathrm{d}t=12.5~\mu$s, $\secTrajPer$ trajectories each) below the original crossover scale, at $N=\secNlistdeep,\secNlistcross$ (i.e.\ $\xi\simeq\secXideep,\secXicross$). Because a finite-time QND homodyne record need not exactly saturate the idealised single-shot Helstrom reference $(1+\TV)/2$, we use the reference-free difference $P_{\rm full}-P_{\rm QND}$ between the monitored-record classifier and its own $\hat H=0$ control as the diagnostic. In the clearly secular regime ($\xi\gtrsim 6$; $N=\secNlistdeep$) this difference is statistically consistent with zero ($P_{\rm full}-P_{\rm QND}=\secFMQdeepA$ at $N=370$ and $\secFMQdeepB$ at $N=400$, both $\lesssim\secSigDeep\,\sigma$), confirming directly that the full monitored record carries no detectable dynamical advantage over the static $\Jz$-record control there. As $\xi$ approaches the crossover region the difference begins to trend positive, consistent with the finite-$N$ bypass picture of Sec.~\ref{sec:dynamical} (by $N=500$, $\xi\simeq 1.5$, the full record already shows a resolved positive excess over the static reference, $\secNfivehundredExc$ at $\secNfivehundredSig\,\sigma$); we do not use these data to assign a sharp onset threshold. The new runs thus sharpen, rather than contradict, the rate-ratio picture: the bypass is absent in the clearly secular regime and opens as $\xi$ approaches order unity. This is also where the distinct coherence-detectability window --- the ``Goldilocks coherence window'' testable by a Leggett--Garg inequality (LGI) --- lives; it is treated in a companion analysis~\cite{SweepPaper} and is separate from the information window of Sec.~\ref{sec:dynamical} (compared in Sec.~\ref{subsec:goldilocks_connection}).

\paragraph*{(C) Crossover and information window: intermediate $\xi$, $N\gtrsim N_\xi$.} Here neither fast-rotation nor strong-measurement averaging cleanly applies, so the static-channel histogram is not a controlled approximation to the monitored record. This is the regime in which a continuous record outperforms the single-shot histogram: Sec.~\ref{sec:dynamical} shows by direct simulation that the classifier exceeds the static benchmark $(1+\TV)/2$ by a statistically resolved finite-$N$ margin across the window.

\paragraph*{(Z) Doublet-projected Zeno side: $\xi\ll 1$.} At large $N$ measurement-induced dephasing dominates coherent doublet precession. In a doublet-projected description $\Jz$ acts as a transverse operator, $\Pdb\Jz\Pdb=J_{01}\sigmax$, so the monitored pointer basis is not the parity basis and the static-channel histogram should not be identified automatically with the full Zeno-limit record. What is recovered is the suppression of coherent Hamiltonian noncommutation; corrections from finite observation time and off-doublet leakage must still be controlled, in particular $T_{\rm obs}\Gamma_{\rm leak}\ll 1$ with $\Gamma_{\rm leak}=\gamma_\phi L_{\rm leak}$ (Sec.~\ref{subsec:leakage}).

\begin{table}[t]
\centering
\caption{Scope of the static-channel result across the three sub-regimes of the full LMG monitored dynamics. Crossover scales: $N_\xi$ where $\xi=\omega_{01}/\Gamma_{01}=1$ (secular boundary); $N_1$ where $\omega_{01}T_{\rm obs}=1$ (phase-unit crossing). The leakage-validity condition $T_{\rm obs}\Gamma_{\rm leak}\ll 1$ is a separate constraint on the doublet-projected description and is not labelled by $N_1$. For the benchmark parameters $g=0.95$, $\gamma_\phi=0.05\,\mathrm{s}^{-1}$, $T_{\rm obs}=15\,\mathrm{ms}$, one has $N_\xi\approx 530$ and $N_1\approx 690$.}
\label{tab:regimes}
\footnotesize
\setlength{\tabcolsep}{4pt}
\begin{tabular}{@{}p{0.19\textwidth}p{0.21\textwidth}p{0.30\textwidth}p{0.21\textwidth}@{}}
\toprule
Regime & Condition & Role of static channel & Status \\
\midrule
(S) Secular & $\xi\gg 1$, $N\lesssim N_\xi$ & natural reference after secular averaging & no detected dynamical excess \\
(C) Crossover/window & $\xi\sim 1$ to intermediate $\xi$ & not a controlled approximation to the record & finite-$N$ excess \\
(Z) Doublet-Zeno side & $\xi\ll 1$, $T_{\rm obs}\Gamma_{\rm leak}\ll 1$ & reference recovered for the tested classifier & leakage/time-scale caveats remain \\
\bottomrule
\end{tabular}
\end{table}

\subsection{The information window is a mesoscopic prefactor effect, not a horizon violation}
The continuous-monitoring advantage of Sec.~\ref{sec:dynamical} is exactly the mesoscopic-regime signature this map anticipates, and it must not be read as a horizon violation. The window sits around and beyond the crossover where coherent doublet dynamics remain dynamically relevant on the observation window, but before measurement-induced dephasing has fully frozen the monitored dynamics. Operationally, the organising variable is the rate ratio $\xi=\omega_{01}/\Gamma_{01}$, not either $\omega_{01}T_{\rm obs}$ or $\Gamma_{01}T_{\rm obs}$ alone (Sec.~\ref{subsec:xicontrol}). There the photocurrent has access to temporal structure --- coherent Bohr oscillation, second-order correlations of the record, intra-doublet rotations driven by $J_{01}$ --- that the static $\Jz$-histogram by construction does not encode. The static channel is a different channel, so the monitored photocurrent is not constrained by the information available in the static histogram.

The distinction that matters is asymptotic versus finite-$N$. A genuine horizon violation would require an appropriate parity-distinguishability or information measure for the full monitored channel to decay as $e^{-N\tilde S}$ with $\tilde S<\Sinst$ as $N\to\infty$ --- a statement about the exponent. The information window is consistent with a finite-$N$ prefactor effect: it opens at mesoscopic $N$ and closes again as $N$ grows (Sec.~\ref{subsec:closure}), without providing evidence for an asymptotic exponent below $\Sinst$: the lobe-overlap geometry that produces the $e^{-N\Sinst}$ rate persists in any operator-content statement about the parity doublet. The window adds a finite-$N$ dynamical layer to the horizon, rather than evidence for a change in its asymptotic exponent.

\subsection{Summary of scope}
Proposition~\ref{prop:horizon} is exact for the static channel at all $N$. For the full monitored dynamics, the static benchmark is the natural reference in the secular regime (where Sec.~\ref{sec:dynamical} finds no continuous-monitoring advantage), is beaten by a continuous record across the mesoscopic information window (Sec.~\ref{sec:dynamical}), and is recovered again deep on the Zeno side once the dynamics freeze. A genuine horizon violation would require an asymptotic exponent slower than $\Sinst$, which we do not find and do not expect in the broken phase; the information window is a finite-$N$ prefactor effect compatible with the asymptotic suppression. The static channel is the cleanest reference for the asymptotic content of the horizon; the mesoscopic window is where a time-resolved experiment can do better.

\section{Discussion}
\label{sec:discussion}
\subsection{Why basis mismatch is the right organising principle}
In the language of this paper, the parity horizon is a geometric mismatch between two bases in which the same operator $\Jz$ plays different roles: the $\Jz$ eigenbasis $\{|m\rangle\}$, which defines the readout distributions, and the parity/energy basis $\{|E_0\rangle,|E_1\rangle\}$, in which the same operator appears as a transverse backaction matrix element. The readout side uses the diagonal $\Jz$ statistics and sees only the exponentially small barrier/interference difference between the two parity-conditioned histograms. The backaction side uses the off-diagonal matrix element of $\Jz$ in the parity basis, which is large because the corresponding well combinations are macroscopically separated.

This mismatch is the reason for the exponential horizon: the parity-resolved difference between the two $\Jz$-histograms is controlled by forbidden-region interference between the two well tails, while the dominant lobe probabilities are almost parity-blind. By contrast, the off-diagonal matrix element of $\Jz$ in the parity basis is set by the macroscopic separation of the two well-localised combinations, $\sim N\mstar/2$. The horizon is not an accidental smallness of one matrix element; it is a generic consequence of the alignment between the bath operator's eigenbasis and the symmetry-breaking direction.

\subsection{Distinction from quantum error correction}
\label{subsec:not_qec}
The parity horizon should not be confused with parity protection in a quantum error-correction sense. Three standard frameworks are immediately excluded:

\emph{Not a decoherence-free subspace (DFS)}~\cite{Zanardi1997,Lidar1998}: A DFS requires $\hat L|\psi\rangle = c|\psi\rangle$ for all $|\psi\rangle$ in the subspace. For the doublet, $\Jz|E_0\rangle = J_{01}|E_1\rangle + \sum_{k\ge 2}\langle E_k|\Jz|E_0\rangle|E_k\rangle$, not proportional to $|E_0\rangle$.

\emph{Not a noiseless subsystem (NS)}~\cite{Knill2000}: Protection would require the projected jump operator to act as a multiple of the projector. From Eq.~\eqref{eq:Pdb_Jz_Pdb}, $\Pdb\Jz\Pdb=J_{01}\sigmax$, which is not a multiple of $\Pdb$: the noise acts as a logical $X$, and the Knill--Laflamme conditions fail at first order in $\sqrt{\gamma_\phi\,dt}$.

\emph{Not an AQEC code}~\cite{Faist2020}: The first-order KL deviation $J_{01}\sigmax$ is large (not small), so the doublet is not an approximate quantum error-correcting code for parity in any standard sense. The second-moment asymmetry $\delta_{\rm var}=0.187$ is a subleading correction that does not rescue the code structure.

What the parity horizon describes is something else entirely: the asymptotic Holevo information about the parity label recoverable from the static-channel reduced bath record is exponentially suppressed, while the bath's backaction on the reduced system can strongly scramble the parity bit through the transverse projected operator $\Pdb\Jz\Pdb=J_{01}\sigmax$. These are not contradictory --- they are two distinct consequences of the same basis mismatch.

\subsection{Relation to the Goldilocks coherence window}
\label{subsec:goldilocks_connection}
This paper's \emph{information window} (Sec.~\ref{sec:dynamical}) has a close cousin in a companion analysis of the same model: the \emph{Goldilocks coherence window}~\cite{SweepPaper}. The two are governed by the same underlying rate competition but are operationally and quantitatively distinct, and we state the distinction precisely so they are not conflated.

The shared root is the selection-rule structure of the doublet. The splitting $\omega_{01}=\Delta E/\hbar$ closes exponentially, $\omega_{01}\sim e^{-N\Sinst}$, while the dephasing and backaction scales grow polynomially, $\Gamma_{01}=\gamma_\phi G_{01}\sim\gamma_\phi N^2$ and $\Gamma_z^{(\rm par)}=2\gamma_\phi J_{01}^2\sim\gamma_\phi N^2\mstar^2$; their ratio is the control parameter $\xi=\omega_{01}/\Gamma_{01}$ that organises both windows. What differs is the \emph{observable} each window is defined by, and hence where each closes:
\begin{itemize}
\item The \emph{Goldilocks coherence window}~\cite{SweepPaper} is defined operationally by the Leggett--Garg correlator exceeding its classical bound, $K_3>1$: it is the region of $(\gamma_\phi/J,N)$ in which macroscopic quantum coherence remains \emph{dynamically detectable}. It concerns what coherence \emph{survives and is LGI-testable}, and at the benchmark dephasing it closes at $N_{\rm LGI}\sim 3\times 10^{3}$.
\item The \emph{information window} of the present work is defined by the discrimination excess $\Delta P>0$: the region in which a continuous monitoring record \emph{extracts} parity information that the static channel hides. It concerns what a record can \emph{read out}, and it closes near $N\sim 10^{3}$ (Sec.~\ref{subsec:closure}).
\end{itemize}
The two windows therefore answer different physical questions --- coherence \emph{survival} versus information \emph{extraction} --- with different operational criteria ($K_3>1$ versus $\Delta P>0$) and different closure scales, even though both are controlled by the same $\xi$ and the same instanton exponent $\Sinst$. The present paper supplies the static readout/backaction scalings that underlie both; Ref.~\cite{SweepPaper} builds the LGI-based coherence phase diagram, and Sec.~\ref{sec:dynamical} here demonstrates the monitoring-based information bypass. They are complementary faces of the same $\mathbb{Z}_2$ rate competition, not the same window.

The rate scale $\Gamma_z^{(\rm par)}=2\gamma_\phi J_{01}^2$ is the projected-doublet transverse-noise scale. Its operational interpretation depends on regime: on the secular side it can be viewed as parity-population relaxation in the energy/parity basis (at the benchmark $\Gamma_z^{(\rm par)}\simeq 1.73\,\Gamma_{01}$), while on the Zeno side the monitored $\sigmax$ pointer basis becomes the natural description and coherent doublet precession is suppressed. We do not use this rate as a standalone prediction of finite-time parity flips in the Zeno regime. This paper supplies the static readout/backaction scalings and, in Sec.~\ref{sec:dynamical}, the monitoring-based information window built on them; the companion analysis~\cite{SweepPaper} builds the distinct Leggett--Garg coherence phase diagram on the same scalings, identifying where finite-time experiments can witness macroscopic coherence.

\subsection{Platforms with the structural ingredients of the model}
\label{subsec:experimental}
The LMG Hamiltonian and its $\Jz$-dephasing realisation appear, with appropriate identification of parameters, in two collective-spin platforms: spinor BECs and trapped-ion or Rydberg arrays. We discuss them here as \emph{platforms that contain the structural ingredients of the model}, not as settings that automatically test the parity horizon. The horizon as stated in Proposition~\ref{prop:horizon} concerns the static $\Jz$-readout channel ($\hat H=0$) applied to parity-eigenstate inputs --- two restrictions that any real experiment must approximate, and that the natural broken-symmetry preparation on these platforms does not satisfy. We make the gap explicit so the reader can distinguish what is established (a structural mechanism in an idealised model) from what would be required to test it experimentally (substantive preparation and probe engineering that is not undertaken here).

\paragraph*{The parity-vs-well preparation distinction.} A real cooling protocol on a broken-symmetry platform produces a well state ($|P\rangle$ or $|R\rangle$, a magnetisation-localised state with definite sign of $\Jz$), not a parity-eigenstate mixture. By ``well state'' we mean the experimentally prepared symmetry-broken state, which approximates the exact orthogonal combination $|P\rangle=(|E_0\rangle+|E_1\rangle)/\sqrt{2}$ at large $N$; the Holstein--Primakoff saddle state $|\tilde P\rangle$ of Sec.~\ref{subsec:WKB} is a separate semiclassical approximation to $|P\rangle$, distinguished by the tilde. Well states are superpositions of $|E_0\rangle$ and $|E_1\rangle$ rather than energy eigenstates themselves. Under $\Jz$-readout on a well-state preparation, the discrimination problem becomes ``which well?''~--- a problem with $O(1)$ TV between the two distributions (the wells are macroscopically separated in $\Jz$) and no horizon-style suppression. This is operationally easy by design and is \emph{not} what Proposition~\ref{prop:horizon} addresses. The proposition's parity-discrimination problem requires inputs that the natural cooling protocol does not produce; bridging the two requires either (a) Gibbs cooling in the temperature window $\Delta E\ll k_BT\ll\hbar\omega_{\rm HP}$, where the doublet equilibrates to the symmetric prior used in Prop.~\ref{prop:horizon} --- but a Gibbs state realises only the symmetric mixed \emph{ensemble}; testing the binary discrimination statement would still require a way to label or post-select the parity sector across trials; or (b) a parent-Hamiltonian quench and adiabatic ramp that can prepare controlled parity-sector inputs, not merely one parity-symmetric ground-state branch; or (c) a non-demolition parity post-selection, which is itself constrained by the exponentially small $\Delta E$. None of these is straightforward, and we do not analyse a platform-specific implementation here.

\paragraph*{The static-channel approximation.} Even with the right preparation, the static $\Jz$-readout channel ($\hat H=0$) of Proposition~\ref{prop:horizon} is a mathematical idealisation. A real probe acts in the presence of $\hat H_{\rm LMG}\neq 0$; whether the proposition's prediction applies as a useful reference depends on the rate ratio $\xi=\omega_{01}/\Gamma_{01}$ being large (the secular regime of Sec.~\ref{sec:scope_regimes}), so that the static commuting-record channel benchmarks the slowly varying part of the monitored dynamics. We do not derive the analytic form of the corrections in this regime. Outside it --- across the mesoscopic window --- the monitored record does better than the static benchmark rather than worse, as Sec.~\ref{sec:dynamical} shows directly; the static result is then a baseline to beat, not an approximation to the dynamics.

\paragraph*{Spinor Bose--Einstein condensate.} A two-mode bosonic Josephson junction in the Fock regime with weak inter-mode tunnelling can realise an effective LMG-type Hamiltonian, with the on-site interaction playing the role of $J$ and the tunnel coupling that of $\Gamma$~\cite{StamperKurn1999,HardestyShaw2023}. Particle-number fluctuations between the two modes, for example through coupling to a thermal reservoir, can give an effective collective dephasing of the form $\hat L\propto\Jz$. The structural ingredients of the parity horizon are present: the doublet splitting and the on-doublet $J_{01}$ are both physical quantities in the Josephson description. Parameters in the range $N\sim 200$--$500$, $J/\hbar\sim 10^4~\mathrm{s}^{-1}$, $\gamma_\phi\sim 0.01$--$0.1~\mathrm{s}^{-1}$ would, within the model calibration used here, place $N_\xi$ in the mesoscopic range. Whether the natural cooling/initialisation protocol on such a system produces an approximate parity-eigenstate mixture (as opposed to a well state) is a substantive experimental question we do not address here; the proposition's relevance to a specific experimental run hinges on this preparation step.

\paragraph*{Trapped-ion or Rydberg arrays.} Collective spin systems of $N\sim 50$--$100$ ions with engineered Ising-like interactions and global laser dephasing~\cite{Monroe2021} (with the dissipative-LMG dynamics analysed in cavity-QED contexts in~\cite{MorrisonParkins2008}) fall in the small-$N$ regime where $\eta_{\rm mismatch}\approx 0.64$--$0.67$ (rising to $\approx 0.74$ by $N\sim 200$) and the asymptotic horizon scaling is not yet reached; the projected-doublet rate language ($\Gamma_z^{(\rm par)}$ and the $\sigmax$-pointer Zeno picture) should be read as interpretive at these sizes, since the doublet-projection validity parameter $\Tobs\Gamma_{\rm leak}$ is not small there --- the dynamical results of Sec.~\ref{sec:dynamical} do not rely on the projection, as the simulation evolves the full symmetric sector with leakage included. The structural ingredients are again present, and the Ramsey-type protocol that measures the doublet splitting $\omega_{01}=\Delta E/\hbar$ is a separate channel from one that samples the parity-conditioned $\Jz$-distribution; the two probe $\omega_{01}$ and $J_{01}=\langle E_0|\Jz|E_1\rangle$ respectively, and could in principle be combined into a joint measurement that exposes the readout--backaction asymmetry. A specific protocol that does this without contaminating $J_{01}$ measurement with the suppressed readout is not specified here.

\paragraph*{Summary of the experimental gap.} The paper establishes a structural mechanism in an idealised setting. The same model parameters that label the structural mechanism (the dimensionless coupling $g$, the doublet splitting $\Delta E$, the on-doublet $J_{01}$) also label the experimental platforms above, which is why these platforms are mentioned. But the proposition's preparation/channel idealisations are not automatically realised on those platforms; bridging the two requires preparation engineering and probe design that we do not undertake. The horizon is a theoretical mechanism whose direct experimental test requires this preparation-and-readout bridge; a fully specified experimental protocol is not provided here.

\subsection{Limitations and open questions}
\label{subsec:limitations}
The analysis combines an exact static characterisation (Secs.~\ref{sec:readout}--\ref{sec:separation}, from LMG eigenstates) with a direct numerical study of the monitored dynamics (Sec.~\ref{sec:dynamical}, via the SME/Girsanov pipeline of App.~\ref{app:sme}). Several questions remain open.

\textbf{(a) The full-bath Holevo limit\@.} Proposition~\ref{prop:horizon} gives the exact static-channel Holevo quantity under the $\hat H=0$ commuting-record restriction. The strict thermodynamic Holevo limit $\lim_{N\to\infty}\lim_{\Tobs\to\infty}\chi_{\rm E}^{\rm full}(\Tobs,N)$ for arbitrary measurements on the full bath is in principle a different mathematical object and, to our knowledge, has not been rigorously computed for the open LMG.

\textbf{(b) The closure exponent.} Section~\ref{sec:dynamical} demonstrates that the dynamical bypass closes in the Zeno tail, and that the closure is $\xi$-controlled and intrinsic. We do \emph{not} establish a closed-form closure law $\Delta P\sim\xi^p$: as discussed in App.~\ref{app:closurelaw}, the natural spectral heuristic is invalid for the causal conditional-filtering classifier, and the resolved knee is a finite-range descent rather than an asymptotic regime. Determining whether a universal closure exponent exists, and its value, requires a dedicated high-statistics study of the knee at two observation times; we state this as an open quantitative question.

\textbf{(c) First-principles $\delta_\infty(g)$.} The restricted-range two-parameter fit~\eqref{eq:J01_HP} gives $\delta_\infty\approx 5.8$ at $g=0.95$, with a model spread of a few $\times 10^{-2}$ between two- and three-parameter forms. A first-principles HP+Bogoliubov calculation of $\delta_\infty(g)$, including the orthogonalisation correction between $|\tilde P\rangle,|\tilde R\rangle$ and the exact $|P\rangle,|R\rangle$, is left open.

\textbf{(d) Operational $\Jz$-readout in experiments.} Proposition~\ref{prop:horizon} is proved under the static $\Jz$-readout channel ($\hat H=0$). Engineering a probe that approximates this channel in spinor-BEC or trapped-ion platforms --- without unwanted spin flips, added decoherence, or significant Hamiltonian-driven evolution during the readout window --- is a substantive experimental task we have not addressed; the spinor-BEC and trapped-ion settings should be read as candidate platforms that contain the ingredients, not as turnkey realisations. The static benchmark is most directly meaningful in the secular regime; the mesoscopic window of Sec.~\ref{sec:dynamical}, where a continuous record exceeds it, is the regime an experiment would most naturally target.

\section{Conclusion}
\label{sec:conclusion}
We have shown that the parity horizon in the open LMG model has more structure than the accessible-information (Holevo) framework alone reveals, and that this static horizon --- the commuting-record benchmark defined at $\hat H=0$ (Sec.~\ref{sec:readout}) --- is not a bound on time-resolved monitoring. The bath operator $\Jz$ plays two distinct geometric roles: as a static readout operator it sees the parity label only through $\Jz$-basis probabilities, whose parity-resolved differences are controlled by exponentially small forbidden-region overlap; as a backaction operator projected onto the parity doublet it acts as $J_{01}\sigmax$ with extensive strength $|J_{01}|^2 \sim (N\mstar/2)^2$. The exact static-channel identity, the WKB mechanism, and ED data on grids extending to $N=4500$ (with reliable readout-exponent fits restricted to the floor-limited subset summarised in Table~\ref{tab:reliability_windows}) together support this readout/backaction split.

On the readout side, the evidence supports the following asymptotic picture. The spectral doublet splitting and the three histogram distinguishability measures --- total variation, Jensen--Shannon divergence, and Chernoff information --- share the same leading exponent $\Sinst(g)$, governed by the LMG instanton action of Eq.~\eqref{eq:Sinst_SCS}. The reason is not simply that all quantities are small. Rather, path additivity in the WKB barrier, Eq.~\eqref{eq:path_additivity}, makes the actions accumulated by the two well-localised wavefunctions add to $\Sinst$ throughout the barrier interior. The linear total-variation distance inherits this exponent directly. For the nonlinear functionals the mechanism is more delicate: the node-bin contribution gives an instanton-scale lower-bound mechanism for JSD, while the barrier region and the observed interior Chernoff optimiser provide the corresponding deficit mechanism for Chernoff information. We therefore do not claim a uniform matched-asymptotic theorem for every bin and every functional (see Sec.~\ref{subsec:prop_horizon}). What we claim, and what the ED data support on grids extending to $N=4500$, is that the leading exponent extracted from the gap and from the three histogram measures is consistent, within finite-$N$ drift and reliability masks, with the same instanton action. The observed two-sided finite-$N$ approach --- from below for $S_{\rm gap}$ and $S_{\rm TV}$ and from above for $S_{\JSD}$ and $S_{\rm Chernoff}$ --- also shows an approximate collapse when plotted against $N\Sinst(g)$, which disfavors a purely near-critical explanation.

On the backaction side, exact diagonalisation confirms the semiclassical scaling $|J_{01}|\to N\mstar/2$. Over the asymptotic ED range the leading correction is well represented by $|J_{01}|=N\mstar/2-\delta(N,g)$, with the benchmark value $\delta_\infty(g=0.95)\simeq 5.8$ and only a small model spread between the two- and three-parameter fits. A first-principles derivation of this offset from the Holstein--Primakoff plus Bogoliubov expansion, including the orthogonalisation between saddle-point and exact well states, remains open. We have kept this limitation explicit because it is precisely the kind of subleading correction that matters when a finite-$N$ numerical claim is pushed close to the asymptotic regime.

The static horizon is nevertheless not the end of the story. When the full monitored LMG dynamics are restored, a time-resolved record contains temporal correlations that the frozen $\Jz$ histogram discards. Direct simulations of the continuously monitored model --- 1.48 million full-LMG trajectories, with matched QND controls across 77 independent settings and a Girsanov-optimal classifier for the specified homodyne record --- show a statistically resolved excess over the static benchmark across a finite mesoscopic window. The advantage is organised most cleanly by the rate ratio $\xi=\omega_{01}/\Gamma_{01}$, is statistically consistent with $\Tobs$-insensitivity over the tested paired comparison (Eq.~\ref{eq:tobs}) and the wider $N=700$ sweep (App.~\ref{app:audit}), and closes again deep in the strong-measurement regime. This is why the horizon is operationally useful: it gives the exponentially weak static benchmark which the monitored dynamics can beat in the appropriate finite-$N$ regime, while still leaving the asymptotic instanton suppression intact.

The main lesson is therefore not tied to a single information functional. In the static $\Jz$ channel, TV, JSD/Holevo, and Chernoff information all display instanton-controlled parity-readout suppression, whereas the projected backaction matrix element grows macroscopically. We expect analogous readout/backaction mismatches in other symmetry-broken open systems whose low-energy sector is a parity doublet and whose environmental coupling is parity-odd along the order-parameter direction. Establishing such analogues, however, is a model-specific problem: the LMG model gives a clean collective setting in which the mechanism can be stated sharply, tested numerically, and separated from the additional complications of preparation and platform-specific readout.


\begin{acknowledgments}
This work forms part of an ongoing study of coherence, decoherence, and macrorealism in the open Lipkin--Meshkov--Glick model, and builds on the author's companion analyses of the same system~\cite{SweepPaper,mouslopoulos2026basis}. The author acknowledges the use of AI assistants (Anthropic Claude and OpenAI ChatGPT) for editorial feedback and for assistance with code development; all calculations, numerical analyses, and conclusions are the author's own and remain the author's sole responsibility.
\end{acknowledgments}


\appendix

\section{Proof of Proposition~\ref{prop:horizon}}
\label{app:proof}
\noindent We give the proof of the static-channel parity-horizon identification $\chi_\infty^{\rm static}=\JSD(p^{(0)},p^{(1)})$ stated in Eq.~\eqref{eq:prop_horizon}. The proof has two ingredients: a physical identification of the asymptotic record-state structure, and a textbook reduction of the Holevo quantity for commuting density matrices.

\paragraph*{Setup.} Consider the open LMG~\eqref{eq:Lindblad_eq} restricted to the static $\Jz$-readout channel: the system Hamiltonian is set to zero ($\hat H=0$, condition (ii) of Sec.~\ref{subsec:qnd_scope}) and the bath is continuously monitored along $\Jz$ over an observation window $[0,\Tobs]$. In the static channel all $\Jz$-projectors commute with the measurement dynamics, so a long-time continuous record resolves the $\Jz$ eigenvalue without inducing transitions between $\Jz$ sectors~\cite{Wiseman2009}. Coarse-graining the asymptotic record into the resolved outcome $m$ therefore gives an orthogonal environment pointer alphabet $\{|m\rangle_E\}$, with weights given by the Born probabilities $p_m^{(i)}=\Tr_S\!\left[\hat\Pi_m^S\,|E_i\rangle_S\,{}_S\langle E_i|\right]$ where $\hat\Pi_m^S=|m\rangle_S\,{}_S\langle m|$ is the system-side $\Jz$-projector; the reduced record states are then the classical mixtures $\rho_E^{(i)}=\sum_m p_m^{(i)}|m\rangle_E\langle m|$ diagonal in the $\Jz$-pointer basis. Operationally, this is equivalent to an ideal projective $\Jz$-alphabet measurement on a single copy of the prepared parity-eigenstate input --- not to repeated independent $\Jz$ tomography (statement (B) of Sec.~\ref{subsec:prop_horizon}), and not to the full continuously monitored LMG record at $\hat H\neq 0$. (The conditional-state language used here and in continuous-monitoring frameworks generally is shorthand for the distributional content of the bath record over repeated trials; the paper's results depend only on this distributional content and are interpretation-neutral.)

\paragraph*{Effective record states (physical identification).} Conditioned on initial parity $i\in\{0,1\}$, the classical record asymptotically (\(\Tobs\to\infty\)) carries the parity-conditioned distribution $\{p_m^{(i)}\}_m$ defined in Eq.~\eqref{eq:p_m}. The effective bath record state, viewed as a classical mixture of orthogonal pointer states $\{|m\rangle_E\}$ for the record alphabet, is
\begin{equation}
\rho_E^{(i)} = \sum_m p_m^{(i)}\,|m\rangle_E\langle m|.
\label{eq:rho_E_def}
\end{equation}
\emph{This is the nontrivial physical identification underlying the proposition}: the static channel produces record states diagonal in a common pointer basis, with parity-conditioned classical distributions. The von Neumann entropy of $\rho_E^{(i)}$ equals the Shannon entropy of the classical distribution:
\begin{equation}
S(\rho_E^{(i)}) = -\sum_m p_m^{(i)}\ln p_m^{(i)} = H(p^{(i)}).
\end{equation}

\paragraph*{Parity ensemble at the bath.} For the prior parity ensemble $\rho=\tfrac12(|E_0\rangle\langle E_0|+|E_1\rangle\langle E_1|)$ on the system, the corresponding asymptotic bath ensemble is $\{1/2,\rho_E^{(0)};\, 1/2,\rho_E^{(1)}\}$. The average bath state is
\begin{equation}
\bar\rho_E = \tfrac12\rho_E^{(0)} + \tfrac12\rho_E^{(1)} = \sum_m \bar p_m\,|m\rangle_E\langle m|,
\qquad \bar p_m=\tfrac12(p_m^{(0)}+p_m^{(1)}).
\end{equation}

\paragraph*{Holevo collapse (textbook step).} The Holevo bound for this binary ensemble~\cite{NielsenChuang} is
\begin{equation}
\chi_\infty^{\rm static} = S(\bar\rho_E) - \tfrac12 S(\rho_E^{(0)}) - \tfrac12 S(\rho_E^{(1)}).
\label{eq:chi_def_app}
\end{equation}
Since all three states are diagonal in the same orthonormal basis (this is the physical identification step), the von Neumann entropies reduce to Shannon entropies:
\begin{equation}
\chi_\infty^{\rm static} = H(\bar p) - \tfrac12 H(p^{(0)}) - \tfrac12 H(p^{(1)}).
\label{eq:chi_classical_app}
\end{equation}
By definition, the Jensen--Shannon divergence with equal priors is exactly the right-hand side of \eqref{eq:chi_classical_app}, giving \eqref{eq:prop_horizon}.

\paragraph*{Asymptotic saturation.} For the static channel, the bath's classical record asymptotically resolves the $\Jz$-eigenvalue, so the accessible information --- the supremum of mutual information over measurements on the bath --- coincides with the Holevo quantity $\chi_\infty^{\rm static}$. The bound \eqref{eq:chi_def_app} is therefore saturated by the natural pointer measurement, and the asymptotic Holevo quantity equals the asymptotic accessible information, which equals $\JSD(p^{(0)},p^{(1)})$. This completes the proof of \eqref{eq:prop_horizon}. \hfill$\square$

\paragraph*{Remark.} The physical-identification step --- that the static $\Jz$-channel produces commuting record states diagonal in the pointer basis --- is the parity-horizon content of the proposition. The Holevo-collapse step is then standard~\cite{NielsenChuang,OhyaPetz}. The proof is short because both steps individually are simple; the substantive content lies in recognising that the readout horizon \emph{is} the basis-aligned, commuting-record-state structure of the static channel.

\section{LMG diagonalisation conventions}
\label{app:LMG}
The LMG Hamiltonian~\eqref{eq:H_LMG} in the symmetric Dicke sector $j=N/2$ takes the form, in the $|j,m\rangle$ basis with $m=-j,\ldots,+j$,
\begin{align}
H_{m,m}/J     &= -\frac{2}{N} m^2, \\
H_{m,m\pm 1}/J &= -2g\,\langle m\pm 1|\Jx|m\rangle = -g\sqrt{j(j+1)-m(m\pm 1)}.
\end{align}
We diagonalise the resulting $(N+1)\times(N+1)$ symmetric tridiagonal matrix using \texttt{scipy.linalg.eigh\_tridiagonal} with index selection (\texttt{select='i'}, \texttt{select\_range=(0,1)}) to extract the two lowest eigenpairs. At large $N$, the doublet splitting can fall below the numerical tolerance of the raw eigensolver, so the returned eigenvectors may span an arbitrary orthonormal basis of the nearly degenerate two-dimensional subspace. We therefore do not assign parity from the raw eigenvectors directly. Instead, after computing the two-dimensional low-energy subspace $V=(v_0,v_1)$, we form the restricted parity matrix $P_{\rm sub}=V^\dagger\hat{\mathcal{P}}V$ and diagonalise $P_{\rm sub}$. The resulting eigenvectors are used as the even and odd doublet states. We monitor the diagnostics $\|\hat{\mathcal{P}}|E_0\rangle-|E_0\rangle\|$, $\|\hat{\mathcal{P}}|E_1\rangle+|E_1\rangle\|$, $|\langle E_0|E_1\rangle|$, and the Hamiltonian residuals $\|\hat H|E_i\rangle-\varepsilon_i|E_i\rangle\|$ (where $\varepsilon_i$ denotes the numerical eigenvalue) throughout the sweep; all remain below $10^{-10}$ across the $N$-range used.

\section{WKB instanton action and path additivity}
\label{app:WKB}
\subsection{Derivation from the spin-coherent-state path integral}
The derivation that follows is standard semiclassical WKB applied to the LMG spin-coherent-state path integral; for general background see~\cite{ColemanInstantons,Landau3}.

Let $\mu\equiv 2m/N\in[-1,1]$ denote the dimensionless magnetisation per spin (distinct from the integer Dicke eigenvalue $m\in\{-j,\ldots,j\}$ used in the discrete sums of the main text). Although the Dicke label $m$ is discrete, the scaled coordinate $\mu$ has spacing $\Delta\mu=2/N\to 0$, so in the large-$N$ spin-WKB limit the Dicke-basis recurrence admits a continuum Hamilton--Jacobi description~\cite{DusuelVidalPRB05,GargKochetovParkStone03}; the instanton integral below is the continuum large-$N$ limit of the discrete under-barrier action. Twice the classical energy per spin in units of $J$, from the spin-coherent-state expectation value of Eq.~\eqref{eq:H_LMG} divided by $JN/2$, is
\begin{equation}
h(\mu,\phi)=-\mu^2-2g\sqrt{1-\mu^2}\cos\phi.
\label{eq:h_classical}
\end{equation}
(The prefactor of $2$ is a convention choice that cancels in the energy-conservation step below; only $\Sinst$ is physical.) The broken-symmetry minima are at $(\pm\mstar,0)$ with $h(\pm\mstar,0)=-(1+g^2)$. With the spin-coherent symplectic convention used here --- $(\mu,\phi)$ as canonical pair, with $\mu$ playing the role of position and $\phi$ of momentum --- the Euclidean equations of motion are $\dot\mu=\partial_\phi h$, $\dot\phi=-\partial_\mu h$; the opposite orientation convention (with $\phi$ as position and $\mu$ as momentum) reverses both signs and leaves the instanton action unchanged. Explicitly,
\begin{equation}
\dot\mu = 2g\sqrt{1-\mu^2}\sin\phi,\qquad \dot\phi = 2\mu-2g\mu\cos\phi/\sqrt{1-\mu^2}.
\end{equation}
The instanton connecting $(\mstar,0)$ to $(-\mstar,0)$ in imaginary time has $\phi=i\tilde\phi$ purely imaginary. Energy conservation $h(\mu,i\tilde\phi)=-(1+g^2)$ gives $\cosh\tilde\phi(\mu)=(1+g^2-\mu^2)/(2g\sqrt{1-\mu^2})$. The tunneling exponent equals the area of one quadrant of the closed bounce loop, $\int_{\rm quarter}\mu\,d\tilde\phi$ along the path from $(\mstar,\tilde\phi{=}0)$ to $(0,\tilde\phi_{\max})$ (so the one-way traversal is $2\Sinst$ and the full closed bounce $4\Sinst$); integration by parts (both boundary terms vanish since $\tilde\phi(\mstar)=0$ and $\mu(0)=0$) gives~\cite{DusuelVidalPRL04,Braun2007spin,Garg1998}
\begin{equation}
\boxed{\Sinst(g)=\int_0^{\mstar}\!\!d\mu\;\mathrm{arccosh}\!\left[\frac{1+g^2-\mu^2}{2g\sqrt{1-\mu^2}}\right].}
\label{eq:Sinst_app}
\end{equation}
With the spin prefactor $j=N/2$, the one-way under-barrier action is therefore $jW_1=(N/2)(2\Sinst)=N\Sinst$, giving $\Delta E\sim e^{-jW_1}=e^{-N\Sinst(g)}$. Numerical evaluation gives $\Sinst(0.95)=0.010787$.

\subsection{Path additivity: numerical consistency check}
\label{app:pathadd_numerics}
\begin{figure}[htbp]
\centering
\includegraphics[width=\textwidth]{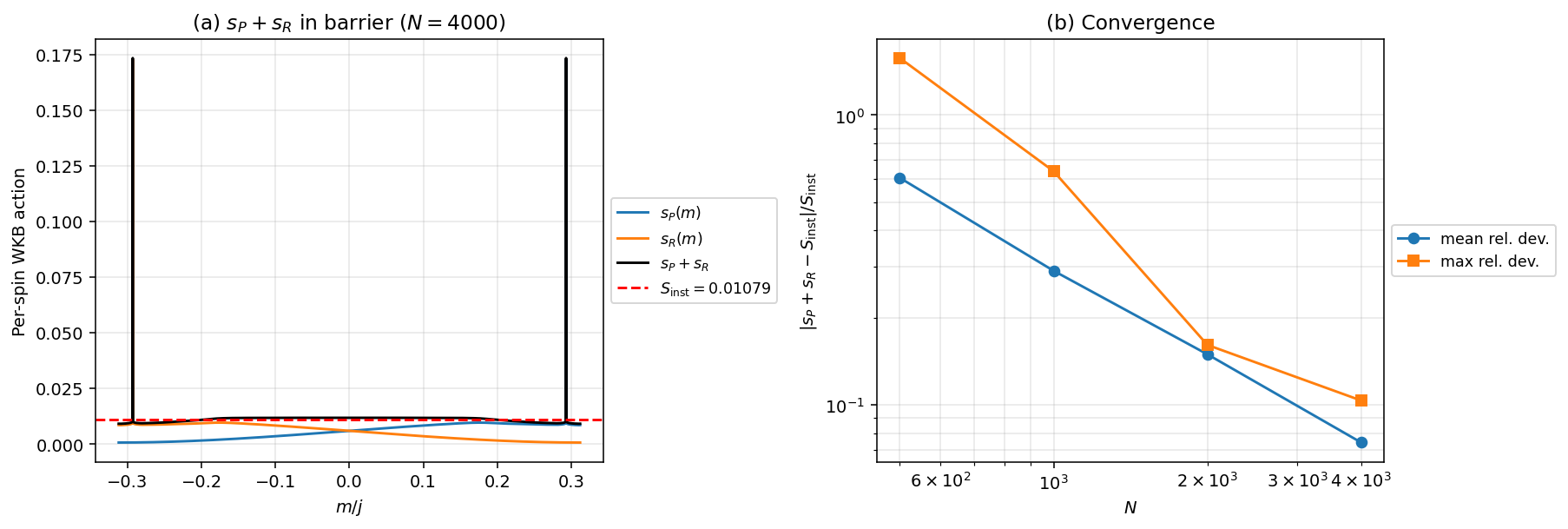}
\caption{\textbf{Path-additivity: numerical convergence evidence.} \emph{(a)} Per-spin WKB actions $s_P(\mu)$ (blue), $s_R(\mu)$ (orange), and their sum $s_P(\mu)+s_R(\mu)$ (black) versus normalised magnetisation $\mu=m/j$ in the barrier interior $|\mu|<\mstar$, computed from the exact LMG wavefunctions at the Dicke points at $N=4000$, $g=0.95$. The two actions cross at $\mu=0$ (the symmetry point) and their sum is approximately flat across the physical-regime plateau, approaching $\Sinst=0.01079$ (red dashed) up to a $\approx 9\%$ subleading correction at this $N$ in the physical-regime plateau (where $|\psi|$ is above the noise floor); the arithmetic mean over the full central-$80\%$ region, including noise-floor-dominated bins, is $\approx 5\%$ above $\Sinst$. The spikes at the boundaries reflect the WKB turning-point breakdown. \emph{(b)} Mean absolute relative deviation of $s_P+s_R$ from $\Sinst$ over the central $80\%$ of the barrier, as a function of $N$. The mean deviation decreases approximately as $\sim N^{-1.0}$ (empirical fit; the product $N\cdot(\text{rel.dev})\approx 300\pm 30$ is approximately constant across $N\in[500,4000]$), consistent with leading-order subleading-WKB amplitude corrections. Maximum deviation persists at $\sim 10$--$15\%$ and is dominated by the turning-point boundary layers.
\label{fig:pathadd}}
\end{figure}

The analytical statement of Sec.~\ref{subsec:WKB}, Eq.~\eqref{eq:path_additivity}, was that for the two well wavefunctions $\psi_{P,R}(m)$, the per-spin actions $s_{P,R}(\mu_m)=-\log|\psi_{P,R}(m)|/N$, evaluated at the Dicke points $\mu_m=2m/N$, satisfy
\begin{equation*}
s_P(\mu_m)+s_R(\mu_m) = \Sinst(g) \quad\text{(WKB, exact $N\to\infty$)}.
\end{equation*}
We provide direct evidence of convergence. Figure~\ref{fig:pathadd}(a) plots $s_P(\mu)$, $s_R(\mu)$, and $s_P(\mu)+s_R(\mu)$ versus the normalised magnetisation $\mu=2m/N=m/j$ for $N=4000$ at $g=0.95$. The sum is flat in the \emph{physical-regime plateau} of the barrier interior (where barrier wavefunctions are above the numerical noise floor): at $N=4000$, this plateau sits at $\approx 0.0117$, exceeding $\Sinst=0.01079$ by $\approx 9\%$. (The arithmetic mean over the full central $80\%$ region, which includes the $\approx 32\%$ of bins where $|\psi|$ falls below the double-precision noise floor and $s$ compresses toward $\sim 0.010$ under regularisation, is $\approx 0.0114$, a $+5\%$ signed excess. The two numbers describe different averaging domains; the visible plateau value of $\approx 0.0117$ is the physically meaningful figure.) Figure~\ref{fig:pathadd}(b) shows that the mean absolute relative deviation $|s_P+s_R-\Sinst|/\Sinst$ decays approximately as $\sim N^{-1.0\pm 0.1}$ from $\approx 61\%$ (maximum $\approx 160\%$) at $N=500$ to $\approx 7.5\%$ (max $\approx 12\%$) at $N=4000$, statistically compatible with an $O(1/N)$ correction. This is direct evidence of convergence \emph{toward} the path-additivity identity at the leading WKB order, not a precise quantitative verification within the available $N$-range: reaching $\le 1\%$ would require $N\gg 10^4$, beyond the double-precision reliability ceiling for the lowest doublet splitting. The maximum deviation persists at $\sim 10$--$15\%$ throughout the range and is dominated by the WKB turning-point boundary layers, not by path-additivity failure in the bulk.

\paragraph*{Methodology and precision limits.} The test uses the parity-symmetrised wavefunctions from \texttt{lmg\_doublet} with log-space regularisation ($\epsilon=10^{-300}$); no additional wavefunction floor filter is applied. At $N=4000$, approximately $32\%$ of central-$80\%$ barrier bins have $|\psi|\lesssim 10^{-14}$ and contribute through the regularised log, compressing $s$ toward $\sim 0.010$; this is why the arithmetic mean over all central-$80\%$ bins ($\approx 0.0114$, $+5\%$) is lower than the physical-regime plateau ($\approx 0.0117$, $+9\%$) visible in panel~(a). With the present double-precision ED implementation, the direct path-additivity test cannot be extended reliably to $g\lesssim 0.93$ over the tractable $N$ range: the barrier wavefunctions are entirely below the double-precision noise floor there ($|\psi|\sim e^{-N\Sinst(g)}\lesssim 10^{-15}$ even at $N=2000$ for $g=0.85$), so a direct consistency check of path additivity is only accessible in a narrow band of $g$ near criticality where $\Sinst(g)$ is small enough that $|\psi|$ in the barrier survives double-precision arithmetic. The four-exponent equality across a wider $g$-range (Sec.~\ref{subsec:multig}) is the indirect test of path additivity outside this band. Reaching $\le 1\%$ deviation at $g=0.95$ would require $N\gtrsim 3\times 10^4$, beyond the double-precision reliability ceiling. We therefore report the result as: convergence toward path additivity at the leading WKB order, with $\sim 1/N$ subleading corrections, directly checked at $g=0.95$ and consistent with leading-WKB path additivity within the accessible precision; the underlying leading-WKB relation is expected to hold across the full broken phase but is not directly testable elsewhere with double-precision arithmetic.

\section{Continuous-monitoring simulation and the Girsanov classifier}
\label{app:sme}
This appendix documents the stochastic-master-equation (SME) pipeline underlying Sec.~\ref{sec:dynamical}. A single validated integrator is reused for every reported number; the energy-basis observables it produces reproduce the exact-diagonalisation benchmarks of Sec.~\ref{sec:separation} to one part in $10^{10}$.

\paragraph*{Conditional dynamics.} An observer homodynes the collective channel $\hat L=\sqrt{\gamma_\phi}\hat J_z$. The conditional state obeys the diffusive SME of Eq.~\eqref{eq:sme}, equivalent to the density-matrix form
\begin{equation}
\mathrm{d}\rho_c=-\tfrac{i}{\hbar}[\hat H,\rho_c]\,\mathrm{d}t+\mathcal{D}[\hat L]\rho_c\,\mathrm{d}t+\mathcal{H}[\hat L]\rho_c\,\mathrm{d}W,
\end{equation}
with dissipator $\mathcal{D}[L]\rho=L\rho L^\dagger-\tfrac12\{L^\dagger L,\rho\}$ and innovation superoperator $\mathcal{H}[L]\rho=L\rho+\rho L^\dagger-\Tr[(L+L^\dagger)\rho]\rho$, and homodyne current $\mathrm{d}y=2\sqrt{\gamma_\phi}\langle\hat J_z\rangle_c\,\mathrm{d}t+\mathrm{d}W$. The full LMG Hamiltonian is retained and the conditional state is propagated in the complete symmetric Dicke sector ($j=N/2$, dimension $N+1$) under the exact $\hat J_z$ operator, with a free-evolution half-step and a measurement update; this is not a two-level truncation, and leakage out of the ground doublet into the rest of the sector is included by construction. Only the two discrimination hypotheses are the parity eigenstates $|E_0\rangle,|E_1\rangle$ within this full-sector evolution.

\paragraph*{Optimal classifier (Girsanov LLR).} The discrimination of the two parity hypotheses $H_0:|E_0\rangle$ and $H_1:|E_1\rangle$ from a record $\{\mathrm{d}y(t)\}_{0}^{\Tobs}$ is performed by the Neyman--Pearson optimal test, the log-likelihood ratio. For diffusive records the LLR has the Girsanov form
\begin{equation}
\Lambda=\int_0^{\Tobs}\!\big[\mu_0(t)-\mu_1(t)\big]\,\mathrm{d}y(t)
-\tfrac12\int_0^{\Tobs}\!\big[\mu_0(t)^2-\mu_1(t)^2\big]\,\mathrm{d}t,
\qquad \mu_i(t)=2\sqrt{\gamma_\phi}\,\langle\hat J_z\rangle_{c}^{(i)}(t),
\end{equation}
where $\langle\hat J_z\rangle_c^{(i)}(t)$ is the conditional mean produced by the hypothesis-$i$ quantum filter driven by the \emph{same} observed record. Two filters are propagated in parallel under each record; the decision is $\hat H=H_0$ if $\Lambda>0$. The success probability $P_{\rm full}=\tfrac12[\Pr(\Lambda>0\mid H_0)+\Pr(\Lambda<0\mid H_1)]$ is estimated by Monte Carlo over records generated from each true preparation. We stress that the classifier is causal and conditional: it does not reduce to a stationary power-spectrum statistic, a point that bears on the closure law (App.~\ref{app:closurelaw}).

\paragraph*{QND guardrail.} The same code with $\hat H=0$ implements a quantum-non-demolition classifier. With no coherent dynamics the record is a repeated noisy sampling of the static $\hat J_z$ distribution, whose optimal discrimination cannot exceed the single-shot Helstrom value $\tfrac12(1+\mathrm{TV})$. A QND excess consistent with zero therefore certifies that the pipeline does not manufacture spurious discrimination. Across all settings the mean QND excess is $\simeq 0$ and the number of cells exceeding the $2\sigma$ ceiling is consistent with chance (App.~\ref{app:audit}).

\paragraph*{Control parameter.} The window organiser is $\xi=\omega_{01}/\Gamma_{01}$ with $\omega_{01}=\Delta E/\hbar$ the doublet Bohr frequency and $\Gamma_{01}=\gamma_\phi G_{01}$, $G_{01}=\tfrac12(\langle E_0|\hat J_z^2|E_0\rangle+\langle E_1|\hat J_z^2|E_1\rangle)$, the measurement dephasing rate; both are computed from the ED doublet of Sec.~\ref{sec:separation}.

\section{Numerical audit of the dynamical runs}
\label{app:audit}
This appendix records the statistical discipline behind every dynamical number. All runs share the validated core of App.~\ref{app:sme}; they differ only in $(N,\Tobs,\mathrm{d}t)$ and random seed. Runs are independent by construction (disjoint seed blocks), so settings sharing $(N,\Tobs,\mathrm{d}t)$ across different runs are pooled directly.

\paragraph*{Conservative error bars.} For each setting the excess is estimated from $n_{\rm batch}$ independent batches. The reported standard error is the \emph{larger} of the binomial standard error $\sqrt{\bar p(1-\bar p)/n_{\rm tot}}$ and the batch-scatter standard error $\mathrm{std}(\{\hat p_b\})/\sqrt{n_{\rm batch}}$, and we report $\chi^2/\mathrm{dof}$ across batches. A setting whose batches disagree (large $\chi^2/\mathrm{dof}$) thereby inflates its own error bar rather than reporting a spuriously tight binomial interval. This rule is applied uniformly; in particular several apparently-significant deep-tail points seen at low batch count regressed to zero once the full batch set was included, and the rule prevents such partial points from being over-interpreted.

\paragraph*{Window statistic and QND control.} The plateau excess quoted in Sec.~\ref{subsec:window}, Eq.~\eqref{eq:plateau}, is the inverse-variance-weighted mean of the six plateau per-setting excesses ($N=625,650,675,700,725,750$), each setting treated as one independent unit pooled over its two observation times, with the quoted uncertainty taken from the setting-to-setting scatter rather than from naive trajectory pooling; the six are mutually consistent ($\chi^2/\mathrm{dof}=\plateauChiSq$), each individually significant at $4$--$10\sigma$, on $\plateauTraj$ trajectories in total. The wider $N=700$ observation-time sweep is a separate robustness check and is not what defines the six-setting plateau statistic. As an absolute control, the same pipeline run with $\hat H=0$ (the quantum-non-demolition classifier) cannot exceed the static Helstrom reference; across all $\totalCells$ pooled settings the mean QND excess is consistent with zero, and exactly one control cell (the $N=850$, $\Tobs=15$~ms setting) exceeds the $2\sigma$ ceiling --- consistent with chance for one excursion among $\totalCells$ controls. That cell sits inside the knee, not the plateau, and the window statistic excludes it from all six plateau settings, so it does not drive the reported excess.

\paragraph*{Timestep convergence.} Coarser $\mathrm{d}t=25~\mu$s tail runs were used diagnostically and indicate that coarse timesteps can overstate the deep-tail excess; the final quoted window and closure numbers therefore use $\mathrm{d}t=12.5~\mu$s. As a direct check that these quoted knee/tail values are not a timestep artefact, the four knee cells at $N=900,925$ (both $\Tobs$) were rerun at half the step, $\mathrm{d}t=6.25~\mu$s (Table~\ref{tab:convergence} and the open diamonds in Fig.~\ref{fig:dyn_knee}): the per-cell excess is statistically unchanged, with an inverse-variance-weighted half-step shift of $-0.0012\pm0.0027$, consistent with zero. This is a half-step convergence check for the quoted points, not a full $50\!\to\!25\!\to\!12.5\!\to\!6.25~\mu$s ladder; the coarse-step diagnostic remains qualitative.

\begin{table}[t]
\centering
\caption{\textbf{Half-step convergence check ($\mathrm{d}t=12.5\!\to\!6.25~\mu$s).} The four knee cells at $N=900,925$ (both $\Tobs$), rerun at half the integration step. The discrimination excess $P_{\rm full}-\tfrac12(1+\mathrm{TV})$ is unchanged within error; the inverse-variance-weighted shift across the four cells is $-0.0012\pm0.0027$, consistent with zero, confirming the quoted $\mathrm{d}t=12.5~\mu$s knee/tail values are not a timestep artefact. This is a fine-step check of the quoted points, not a full timestep ladder.}
\label{tab:convergence}
\begin{tabular}{r r c c c}
\toprule
$N$ & $\Tobs$ (ms) & excess ($12.5~\mu$s) & excess ($6.25~\mu$s) & shift ($6.25\!-\!12.5$) \\
\midrule
900 & 7.5 & $+0.0069\pm0.0031$ & $+0.0063\pm0.0040$ & $-0.0006\pm0.0050$ \\
900 & 15  & $+0.0073\pm0.0031$ & $+0.0070\pm0.0045$ & $-0.0003\pm0.0054$ \\
925 & 7.5 & $+0.0074\pm0.0028$ & $+0.0039\pm0.0040$ & $-0.0035\pm0.0048$ \\
925 & 15  & $+0.0065\pm0.0028$ & $+0.0072\pm0.0062$ & $+0.0007\pm0.0068$ \\
\bottomrule
\end{tabular}
\end{table}

\paragraph*{Audit table.} Table~\ref{tab:audit} lists, for each $(N,\Tobs)$ setting at the converged timestep, the pooled excess, conservative standard error, $\chi^2/\mathrm{dof}$, trajectory count, and QND-ceiling flag.
\begin{table*}[t]
\centering
\caption{Per-setting audit of the dynamical runs at the converged timestep ($\mathrm{d}t=12.5~\mu$s). Excess is $P_{\rm full}-\tfrac12(1+\mathrm{TV})$; s.e.\ is the conservative (max of binomial and batch-scatter) standard error; $\chi^2/$dof is computed across the independent batches of each setting; the QND flag marks settings whose quantum-non-demolition control exceeds the $2\sigma$ Helstrom ceiling. This table lists the converged-timestep ($\mathrm{d}t=12.5~\mu$s) main settings; the full $\totalCells$-setting corpus comprises $\totalTraj$ full-LMG trajectories, with matching QND controls and auxiliary timestep/validation runs. The displayed rows are arranged in two panels (left and right halves) read independently. Reading down in $N$ (equivalently down in $\xi$) traces the full arc: the plateau ($N\lesssim 750$), the knee ($775\lesssim N\lesssim 925$, including the $N=875,925$ shoulder), and the Zeno-side closure to within noise of zero ($N\gtrsim 1000$). Exactly one of the QND controls exceeds the $2\sigma$ ceiling, consistent with chance.}
\label{tab:audit}
\scriptsize
\setlength{\tabcolsep}{3pt}
\begin{tabular}{r r r r r r c @{\hspace{1em}} r r r r r r c}
\toprule
$N$ & $\xi$ & $\Tobs$ & excess & s.e. & $\chi^2/$dof & QND & $N$ & $\xi$ & $\Tobs$ & excess & s.e. & $\chi^2/$dof & QND \\
 & & (ms) & & & & & & & (ms) & & & & \\
\midrule
625 & 0.2902 & 15.0 & +0.0158 & 0.0028 & 0.87 &  & 775 & 0.0422 & 7.5 & +0.0079 & 0.0028 & 0.63 &  \\
625 & 0.2902 & 7.5 & +0.0139 & 0.0029 & 1.09 &  & 800 & 0.0308 & 15.0 & +0.0138 & 0.0028 & 0.31 &  \\
650 & 0.2095 & 15.0 & +0.0119 & 0.0028 & 0.79 &  & 800 & 0.0308 & 7.5 & +0.0129 & 0.0028 & 0.49 &  \\
650 & 0.2095 & 7.5 & +0.0128 & 0.0030 & 1.17 &  & 825 & 0.0225 & 15.0 & +0.0115 & 0.0031 & 1.21 &  \\
675 & 0.1516 & 15.0 & +0.0215 & 0.0033 & 1.36 &  & 825 & 0.0225 & 7.5 & +0.0107 & 0.0035 & 1.60 &  \\
675 & 0.1516 & 7.5 & +0.0169 & 0.0028 & 0.89 &  & 850 & 0.0164 & 15.0 & +0.0096 & 0.0028 & 0.76 & \checkmark \\
700 & 0.1098 & 50.0 & +0.0210 & 0.0039 & 0.44 &  & 850 & 0.0164 & 7.5 & +0.0115 & 0.0028 & 0.73 &  \\
700 & 0.1098 & 33.0 & +0.0102 & 0.0040 & 0.23 &  & 875 & 0.0120 & 15.0 & +0.0063 & 0.0028 & 0.72 &  \\
700 & 0.1098 & 22.0 & +0.0151 & 0.0045 & 1.28 &  & 875 & 0.0120 & 7.5 & +0.0074 & 0.0028 & 0.74 &  \\
700 & 0.1098 & 15.0 & +0.0120 & 0.0024 & 1.07 &  & 900 & 0.0088 & 15.0 & +0.0073 & 0.0031 & 1.22 &  \\
700 & 0.1098 & 10.0 & +0.0190 & 0.0039 & 0.86 &  & 900 & 0.0088 & 7.5 & +0.0069 & 0.0031 & 1.20 &  \\
700 & 0.1098 & 7.5 & +0.0173 & 0.0028 & 1.02 &  & 925 & 0.0065 & 15.0 & +0.0065 & 0.0028 & 0.50 &  \\
700 & 0.1098 & 6.0 & +0.0107 & 0.0045 & 1.29 &  & 925 & 0.0065 & 7.5 & +0.0074 & 0.0028 & 0.84 &  \\
700 & 0.1098 & 3.0 & +0.0101 & 0.0044 & 1.25 &  & 950 & 0.0048 & 15.0 & +0.0034 & 0.0025 & 0.54 &  \\
725 & 0.0797 & 15.0 & +0.0109 & 0.0028 & 0.74 &  & 1000 & 0.0026 & 15.0 & +0.0010 & 0.0056 & 0.54 &  \\
725 & 0.0797 & 7.5 & +0.0181 & 0.0028 & 0.85 &  & 1050 & 0.0014 & 15.0 & -0.0016 & 0.0056 & 0.96 &  \\
750 & 0.0580 & 15.0 & +0.0119 & 0.0028 & 1.04 &  & 1100 & 0.0008 & 15.0 & -0.0026 & 0.0056 & 0.92 &  \\
750 & 0.0580 & 7.5 & +0.0156 & 0.0032 & 1.31 &  & 1150 & 0.0004 & 15.0 & +0.0042 & 0.0056 & 0.39 &  \\
775 & 0.0422 & 15.0 & +0.0149 & 0.0031 & 1.24 &  & 1200 & 0.0002 & 15.0 & -0.0013 & 0.0056 & 0.23 &  \\
\bottomrule
\end{tabular}
\end{table*}

\paragraph*{Observation-time insensitivity: the $N=700$ wide sweep.} The paired $15$ vs $7.5$~ms comparison of Sec.~\ref{sec:dynamical} (Eq.~\ref{eq:tobs}) is corroborated by a wider single-$N$ sweep. At $N=700$ (converged timestep $\mathrm{d}t=12.5~\mu$s) we ran eight observation times spanning ${\sim}17\times$ in $\Tobs$ --- $3$, $6$, $7.5$, $10$, $15$, $22$, $33$, $50$~ms --- with per-setting excesses ranging from $+0.010$ to $+0.021$. A fit to a constant excess (inverse-variance weighted, each $\Tobs$ an independent unit) gives
\begin{equation}
\Delta P_{N=700}=+0.0144\pm 0.0012,\qquad \chi^2/\mathrm{dof}=1.29\ (\mathrm{dof}=7),
\end{equation}
statistically consistent with a $\Tobs$-independent excess across the full $3$--$50$~ms range. This is a single-$N$ check rather than a global fit, but together with the $13$-size paired comparison it supports reading $\xi$, not $\Tobs$, as the organising variable over the tested window.

\section{On the closure law and why no universal exponent is claimed}
\label{app:closurelaw}
The window excess descends to zero through the knee (Fig.~\ref{fig:dyn_knee}). It is natural to ask whether the approach to closure follows a power law $\Delta P\sim\xi^{p}$. We explain here why we report the closure qualitatively and treat the exponent as open.

\paragraph*{A heuristic $\xi^2$ argument, and why it fails for this classifier.} One may model the homodyne record as a zero-mean Gaussian process and estimate the parity information from the difference of the two hypotheses' power spectra. For a damped intra-doublet coherence the $\hat J_z$ power spectrum is a Lorentzian of width $\Gamma_{01}$ centred at $\pm\omega_{01}$; the parity-distinguishing feature is the displacement of this peak from zero frequency by $\omega_{01}$, which becomes unresolvable when $\Gamma_{01}\gg\omega_{01}$, i.e.\ as $\xi\to0$, with the leading (linear) term cancelled by the mirror symmetry $S(\omega)=S(-\omega)$ of a real record, suggesting $\Delta P\sim\xi^2$. This argument is \emph{not} valid for the classifier actually used. By parity the ensemble-averaged $\langle\hat J_z\rangle$ vanishes identically in both hypotheses, and the symmetric (classical) power spectrum is identical for the two hypotheses: the label resides in the antisymmetric, time-directed part of the two-time correlator, which a stationary power-spectrum statistic discards. The optimal classifier (App.~\ref{app:sme}) is causal and conditional --- it propagates hypothesis-conditioned filters along the record --- and is sensitive to precisely this time-directed information. The spectral heuristic therefore mis-identifies both the discriminating object and, even on its own terms, the exponent (a Gaussian-covariance Fisher information built from an $O(\xi^2)$ spectral difference scales as $\xi^4$, not $\xi^2$).

\paragraph*{What the data support.} The knee provides a resolved descent over which the excess falls from the plateau to the noise floor; the deep tail lies at the resolution floor, where no slope can be fit. A finite-range descent does not establish an asymptotic exponent. We therefore report the closure as a robust qualitative fact --- the bypass exists, is $\xi$-controlled (App.~\ref{app:audit}, Fig.~\ref{fig:dyn_tobs}), and returns to the static reference deep in the Zeno regime --- and we do not assign a closure exponent. Settling the exponent, if it is well defined, would require a dedicated high-statistics study of the knee region at two observation times (to maintain the $\xi$ versus $\omega_{01}\Tobs$ separation of Sec.~\ref{subsec:xicontrol} throughout the descent), targeting a per-point uncertainty well below the present plateau-to-floor contrast. We regard this as the natural next step and state it as an open quantitative question rather than a result of the present work.

\section{Reproducibility and computational pipeline}
\label{app:reproducibility}
\paragraph*{Parity convention used in the code.} In the symmetric Dicke basis $\{|j,m\rangle\}_{m=-j}^{+j}$ the spin-flip parity acts as pure reflection, $\hat{\mathcal{P}}|m\rangle=|-m\rangle$, with no Wigner phase factor. This matches the main-text definition $\hat{\mathcal{P}}=\prod_r\hat\sigma_x^{(r)}=\exp[i\pi(\hat J_x-N/2)]$ given in Eq.~\eqref{eq:parity_def}; it differs from the collective rotation $e^{i\pi\hat J_x}$ only by the global sign $(-1)^{j}=i^N$ (for even $N$), not an $m$-dependent factor (the $m$-dependent $(-1)^{j-m}$ is the Wigner-$d$ element for a $\pi$-rotation about $\hat y$, a different operation). The numerical projection used in the code is the natural one for this reflection convention.

\paragraph*{Methodological notes.} The numerical pipeline implements three robustness improvements relevant to a careful referee. (i) The parity doublet is resolved by diagonalising the parity operator in the two-dimensional low-energy subspace returned by the eigensolver (App.~\ref{app:LMG}); this is numerically robust at large $N$ where the bare doublet splitting falls below double-precision tolerance and the raw eigenvectors may be returned as arbitrary rotations. After the parity assignment, the eigenvectors are projected onto their exact parity sector via $(v\pm v[::-1])/2$ followed by renormalisation; with the spin-flip parity convention $\hat{\mathcal{P}}|m\rangle=|-m\rangle$ stated above, this is an idempotent operation that enforces the exact reflection symmetry the true eigenstates satisfy, cleaning floating-point noise asymmetries in the deep tails where $|\psi|\lesssim 10^{-15}$ without altering any physical content (bulk benchmark numbers in Table~\ref{tab:benchmark} reproduce to six significant figures with or without the projection step). (ii) The reliability mask is answer-independent: a data point is retained only if its value is finite and exceeds a conservative absolute floor (default $10^{-12}$), with no reference to the expected decay rate $\Sinst(g)$. This rejects only the cancellation-floor regime, where numerical noise replaces the physical exponential and the apparent decay slows (consecutive ratios approaching unity); it cannot select or reject points for their proximity to $\Sinst$, and the extracted exponents are stable whether or not any decay-rate criterion is additionally imposed (sensitivity table in App.~\ref{app:ed_supp}). (iii) All ED data use even $N$ exclusively: the integer $m=0$ Dicke eigenvalue and the exact parity-odd node $p_m^{(1)}|_{m=0}=0$ are sharp single-bin diagnostics only at even $N$. Multi-$g$ N-grids are constructed by snapping linspace points to the nearest even integer to avoid an integer-truncation parity pathology in which consecutive sampled $N$-values systematically alternate parity. The full reproduction package below records each of these choices.

The complete computational pipeline producing every numerical claim in Sec.~\ref{sec:separation} --- the scripts, cached \texttt{.npz} archives, in-text and appendix figures, and a SHA-256 manifest of all inputs and outputs --- is deposited in the public reproduction archive cited in the Data Availability statement. Wall time on a recent laptop (Apple M3 chip with 8 cores) is a few minutes for the full sweep, dominated by the multi-$g$ universality task.

\paragraph*{Reliability windows by observable.}
Each readout quantity is fitted only over the subset of the $N$-grid where its value lies above the double-precision cancellation floor ($10^{-12}$); the fit uses the answer-independent floor mask described in App.~\ref{app:reproducibility} (no decay-rate criterion). Table~\ref{tab:reliability_windows} summarises the retained data-point range for each observable. These are retained data-point ranges, not rolling-window-centre ranges: a rolling-window fit can have a centre $N_c$ above the upper retained data point when the window still spans enough retained points below it. The reliable range differs by observable: the exponentially small quantities ($\Delta E$, $\TV$, $\JSD$, Chernoff) are limited by eigensolver precision or subtractive cancellation, whereas the non-exponentially-suppressed backaction quantities $|J_{01}|^2$ and $\ell_{\rm leak}$ remain reliable to the largest ED sizes used here.

\begin{table*}[h]
\centering
\caption{Reliable fitting windows for each observable. The reliable-fit range is bounded above by the listed numerical limit; data points outside this range are excluded by the quality mask and never contribute to any quoted slope. Each upper $N$ is the largest retained grid point at which the observable remains above the conservative absolute floor of $10^{-12}$ applied uniformly throughout App.~\ref{app:reproducibility}. The exponentially small quantities ($\Delta E$, $\TV$, $\JSD$, Chernoff) are limited by eigensolver precision or subtractive cancellation, whereas the non-exponentially-suppressed backaction quantities $|J_{01}|^2$ and $\ell_{\rm leak}$ remain reliable to the largest ED sizes used here.}
\label{tab:reliability_windows}
\begin{tabular}{@{}p{0.31\textwidth}cp{0.46\textwidth}@{}}
\toprule
Observable & Reliable-fit $N$ range & Limit set by\\
\midrule
$\Delta E$ (doublet splitting)    & $[100, 2900]$ & eigensolver/cancellation floor near the $10^{-12}$ mask\\
TV (total variation)              & $[100, 2500]$ & subtractive-cancellation floor in $|p^{(0)}-p^{(1)}|$\\
JSD (Jensen--Shannon)             & $[100, 2500]$ & per-bin log-space cancellation near the $10^{-12}$ mask\\
Chernoff information $C$          & $[100, 2500]$ & per-bin $(p^{(0)})^s(p^{(1)})^{1-s}$ cancellation near the $10^{-12}$ mask\\
$|J_{01}|^2$ (backaction)         & $[100, 6000]$ & no exponential suppression; lobe geometry\\
$\ell_{\rm leak}$ (leakage frac.) & $[50, 6000]$  & no exponential suppression; algebraic decay\\
\bottomrule
\end{tabular}
\end{table*}

\noindent The rolling-window exponent fits in Sec.~\ref{sec:separation} automatically respect these limits via the quality mask. Each computation is documented with its purpose, parameter grid, output schema, and producing script in the bundle's \texttt{README.md}. The bundle uses only \texttt{numpy}, \texttt{scipy}, \texttt{matplotlib}, and \texttt{joblib}; library versions are recorded in the manifest.

\section{Supplementary exact-diagonalisation analyses}
\label{app:ed_supp}
This appendix collects the detailed exact-diagonalisation analyses summarised in Sec.~\ref{sec:separation}: the full rolling-window convergence of the four readout exponents and its floor-mask sensitivity, the cross-$g$ scaling collapse, the per-bin JSD anatomy, the independent subleading-model exponent extraction, the $\delta(N,g)$ backaction-offset fit, the off-doublet leakage scaling, and the dynamical regime markers. All numbers use the canonical pipeline of App.~\ref{app:reproducibility}.

\subsection{Rolling-window convergence: detailed observations and mask sensitivity}
\label{app:ed_rollingwindow}
Three observations:

\paragraph*{(i) From-below convergence of $S_{\rm gap}$ and $S_{\rm TV}$.} At $N_c=450$, $S_{\rm gap}=0.00898$ and $S_{\rm TV}=0.00761$, lying $17\%$ and $29\%$ below $\Sinst$. As $N_c$ grows the slopes rise monotonically: $S_{\rm gap}(N_c{=}949)=0.01019$ ($-5.6\%$); $S_{\rm gap}(N_c{=}1999)=0.01045$ ($-3.2\%$). The same pattern holds for $S_{\rm TV}$ with a slightly slower convergence.

\paragraph*{(ii) Convergence of $S_{\JSD}$ and $S_{\rm Chernoff}$ via a low-$N$ undershoot then from above.} At $N_c=450$, $S_{\JSD}=0.01044$ and $S_{\rm Chernoff}=0.01054$, $-3.2\%$ and $-2.3\%$ below $\Sinst$. After this small low-$N$ undershoot, both rise above $\Sinst$: at $N_c=649$ they reach $0.0117$ (an excess of $\approx 8\%$) and then descend monotonically: $S_{\JSD}(N_c{=}1999)=0.01113$ ($+3.1\%$), $S_{\JSD}(N_c{=}2799)=0.01106$ ($+2.5\%$). The Chernoff slope is computed with proper minimisation over $s\in[0,1]$, returning $s^*\simeq 0.60$. Importantly, the leading \emph{exponent} is insensitive to this optimisation: the Chernoff and Bhattacharyya ($s=1/2$) slopes agree to $\Delta<0.0005$ at every $N_c$ (Fig.~\ref{fig:t1_convergence}), even though the corresponding \emph{information values} differ by $(C-\mathcal{B})/C\approx 3$--$4\%$. The drift of $s^*$ away from $1/2$ therefore shifts the prefactor, not the exponent --- which is exactly why $S_{\rm Chernoff}$ tracks $S_{\JSD}$: both are controlled by the same parity-odd barrier structure and the common per-bin exponent $e^{-N\Sinst}$, robustly to the choice of $s$.

\paragraph*{(iii) Common asymptote.} By $N_c=2799$ (the largest window remaining after the cancellation-floor filter), all four slopes are within a few percent of $\Sinst=0.01079$:
\begin{equation*}
\{S_{\rm gap},\,S_{\TV},\,S_{\JSD},\,S_{\rm Chernoff}\}=\{0.01041,\,0.01048,\,0.01106,\,0.01106\}.
\end{equation*}
(The JSD and Chernoff slopes coincide to the displayed precision at this window centre because, as noted in Sec.~\ref{subsec:WKB}, the $s^*$ optimisation shifts the Chernoff prefactor but not its leading exponent, which is the same parity-odd-node exponent that controls JSD; at other window centres they differ in the fourth--fifth figure, e.g.\ $0.01101$ vs $0.01102$ at $N_c\simeq 2799$ under the M3 fit.)
The next log-spaced window centre at $N_c\simeq 3999$ contains only $1$ reliable point (and any centre beyond contains $0$ reliable points) (the cancellation-floor filter rejects the rest); they are excluded from the figure. The larger-$N$, floor-controlled analysis therefore does not support an asymptotic JSD excess over $\Sinst$ as a quadratic-functional effect: any apparent excess at moderate $N_c$ relaxes monotonically toward $\Sinst$ within the reliable fitting range.

\paragraph*{Methods note (data quality).} For each $(N,g)$, the readout quantities are computed in double precision with all sums in their natural representation. At large $N$, $\JSD$ and $\TV$ approach the subtractive-cancellation floor of the eigensolver and the per-bin probability arithmetic (around $10^{-13}$, well above true double-precision underflow at $10^{-308}$). To avoid biasing the extracted slopes, we use an \emph{answer-independent} floor mask: a point is retained only if its value is finite and exceeds a conservative absolute floor (default $10^{-12}$), with no criterion that references the expected decay rate $\Sinst(g)$. This rejects only cancellation-floor noise --- points whose value has saturated near the arithmetic floor, where the apparent decay has stopped (consecutive ratios approaching unity) and the true exponential is replaced by noise --- and cannot, by construction, select or reject points for their proximity to $\Sinst$. The mask is applied identically to every fit reported in this paper. We verify in the sensitivity check below that varying the floor across three orders of magnitude leaves the extracted exponents stable, and that the four-exponent result is unchanged whether or not any decay-rate criterion is imposed; floor-contaminated tail points are excluded automatically.

\paragraph*{Mask sensitivity.} To confirm that the floor mask does not shape the result, we re-extracted all four exponents under pure absolute-magnitude floors at $10^{-12}$, $10^{-11}$, and $10^{-10}$ (no decay-rate criterion of any kind), and compared with the floor-plus-decay-rate variant. The extracted exponents are stable across all variants:
\begin{center}
\small
\setlength{\tabcolsep}{5pt}
\begin{tabular}{@{}l c c c c@{}}
\toprule
& floor $+$ slope & floor $10^{-12}$ & floor $10^{-11}$ & floor $10^{-10}$ \\
\midrule
$S_{\rm gap}$ & $0.0100$ & $0.0100$ & $0.0099$ & $0.0099$ \\
$S_{\rm TV}$ & $0.0097$ & $0.0094$ & $0.0094$ & $0.0092$ \\
$S_{\JSD}$ & $0.0110$ & $0.0109$ & $0.0109$ & $0.0108$ \\
$S_{\rm Chernoff}$ & $0.0110$ & $0.0109$ & $0.0109$ & $0.0108$ \\
\bottomrule
\end{tabular}
\end{center}
Across the pure absolute-floor variants the shifts are at the few-percent level; the TV exponent is the most sensitive, while JSD and Chernoff are essentially unchanged. Removing the decay-rate criterion entirely changes the values by less than $1\%$ for $\JSD$ and Chernoff and leaves $S_{\rm gap}$ unchanged ($\Sinst=0.01079$ for comparison). The qualitative four-exponent pattern --- gap and TV approaching $\Sinst$ from below, JSD and Chernoff from above, with no mask-induced change in ordering or trend --- is therefore independent of the masking choice and is not an artifact of any decay-rate-dependent filter.

\subsection{Approximate scaling collapse across $g$: disfavouring a purely near-critical interpretation}
\label{subsec:multig}
The most direct test of whether the finite-$N$ deviations seen above are a near-critical effect (a possibility we test here) is to repeat the analysis at multiple values of $g$ and check whether the deviations vanish as $g$ moves away from criticality. We run the rolling-window analysis at $g\in\{0.70,0.80,0.85,0.90,0.95,0.98\}$, with the $N$-grid for each $g$ adapted to keep $N\Sinst(g)$ in the same range so that the asymptotic regime is reachable.

\begin{figure*}[t]
\centering
\includegraphics[width=0.96\textwidth]{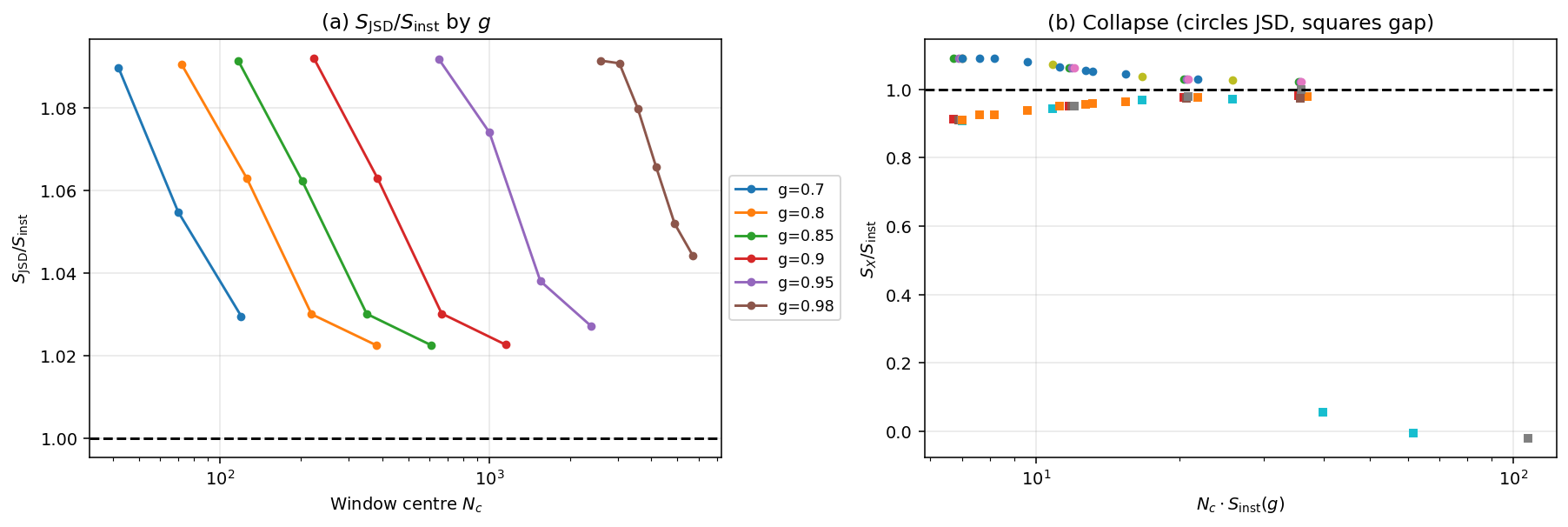}
\caption{\textbf{Approximate scaling collapse across $g$.} \emph{(a)} $S_{\JSD}/\Sinst(g)$ versus window centre $N_c$ for six values of $g\in[0.70,0.98]$. Each curve descends toward unity (black dashed) as $N_c$ grows, but within the accessible window --- bounded above by $N\Sinst(g)\lesssim 25$--$30$, where the histogram functionals remain above the double-precision floor --- a residual offset of a few percent remains; the offset is the WKB-prefactor correction and is \emph{not} claimed to vanish within this range. The centre at which a given relative deviation is reached scales like $\Sinst(g)^{-1}$. \emph{(b)} The same data plotted against the scaling variable $x=N_c\cdot\Sinst(g)$. Circles: $S_{\JSD}/\Sinst$ (approach from above). Squares: $S_{\rm gap}/\Sinst$ (approach from below). The collapse onto a common curve, with weak residual $g$-dependence at the percent level, indicates that the dominant finite-$N$ deviation is controlled primarily by $N\Sinst(g)$, set by the WKB amplitude prefactor structure, rather than by proximity to the critical point $g_c=1$. The downward trend of the squares ($S_{\rm gap}/\Sinst$ falling below unity) at large $x$ is the slope bias of the local-exponent estimator in the presence of an unremoved polynomial prefactor $N^\alpha$ (which contributes $-\alpha/N$ to the fitted slope, with $\alpha>0$ for the spectral gap); it is not a breakdown of the collapse, and is removed when the prefactor is fitted explicitly (model M3, Fig.~\ref{fig:subleading}).
\label{fig:multig}}
\end{figure*}

Figure~\ref{fig:multig}(b) shows the resulting collapse. The two-sided structure (circles for $S_{\JSD}$ converging from above, squares for $S_{\rm gap}$ converging from below) overlays well across $g\in\{0.70,\ldots,0.95\}$ when plotted against the scaling variable $N_c\Sinst(g)$. The collapse is approximate (residual spread $\sim 1\%$, comparable to the residual finite-$N$ deviation itself), not perfect; the strongest deviation is at $g=0.98$, very close to criticality, where the available $N$ range is the most stretched. With that caveat, the data disfavour the near-critical attribution: at fixed $N\Sinst(g)$ the slope deviation is approximately $g$-independent, whereas a purely near-critical correction would be expected to show a systematic residual dependence on $1-g$ and to single out the near-critical cases. A plausible interpretation is that the residual finite-$N$ offset is dominated by WKB-prefactor corrections, such as those associated with $A_P(\mu)A_R(\mu)$, rather than by a separate near-critical mechanism.

\subsection{Per-bin JSD anatomy: parity-node dominance}
\label{subsec:jsd_anatomy}
\begin{figure}[htbp]
\centering
\includegraphics[width=\textwidth]{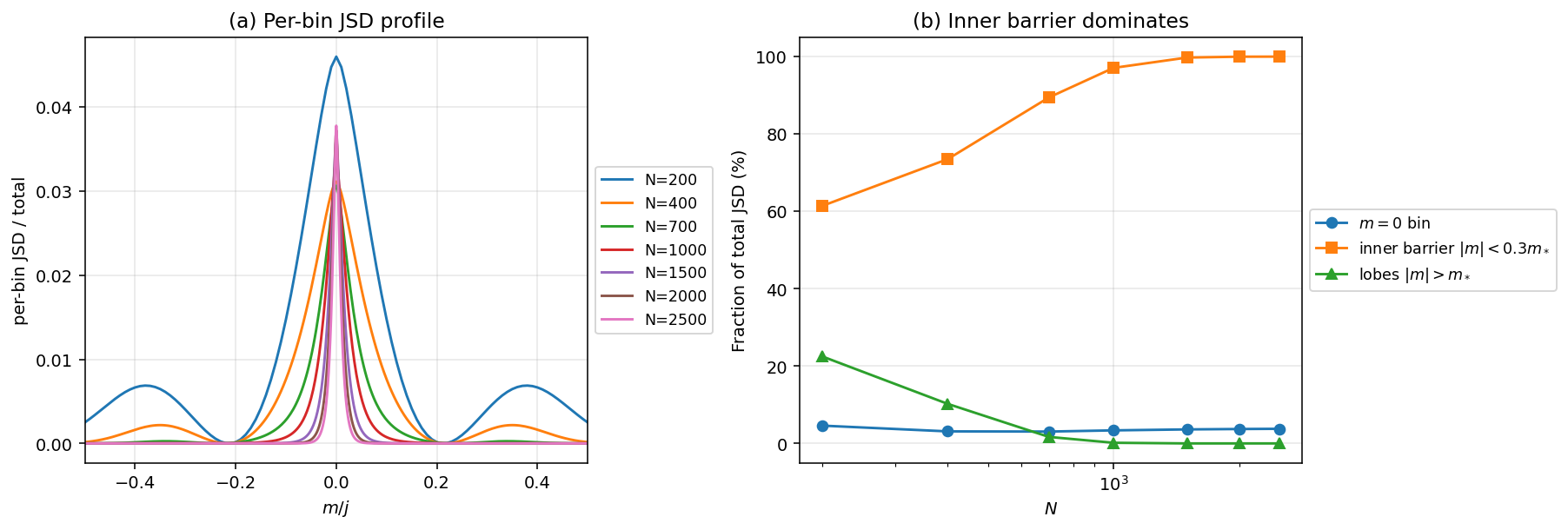}
\caption{\textbf{Per-bin JSD anatomy versus $N$.} \emph{(a)} Per-bin JSD profile (normalised to the total) at six values of $N\in[200,2500]$. The profile narrows as $N$ grows; the contribution concentrates within a barrier-region band centred on $m=0$ but \emph{not} into the single $m=0$ bin alone. \emph{(b)} Fractional contributions: $m=0$ single bin (blue) stays at $\approx 3$--$5\%$ across the entire range --- a finite-bin diagnostic rather than the fundamental object --- while the inner-barrier region $|m/j|<0.3\mstar$ (orange) rises from $61\%$ at $N=200$ to $\approx 100\%$ ($99.99\%$) at $N=2500$, and the lobes $|m/j|>\mstar$ (green) drop from $22\%$ to below $0.01\%$. The robust feature is the concentration of total JSD in the central barrier band, not the single $m=0$ bin.
\label{fig:jsd_anatomy}}
\end{figure}

Path additivity gives $p_m^{(0)}-p_m^{(1)}=2\psi_P(m)\psi_R(m)\sim e^{-N\Sinst}$ throughout the barrier interior, away from WKB boundary layers. This is precisely the quantity that controls the total variation, which therefore inherits the leading exponential directly. For the JSD and Chernoff, the per-bin contribution is a nonlinear function of $p^{(0)}$ and $p^{(1)}$, and the leading-exponent inheritance is more subtle: the dominant contribution comes from the central region around the parity node, where the two well tails $\psi_P$ and $\psi_R$ are comparable and both exponentially small. This does not mean $p_m^{(0)}\approx p_m^{(1)}$ at the node; indeed $p_0^{(1)}=0$ by parity while $p_0^{(0)}>0$. Away from this central equal-tail region, one well tail dominates the local probability while the other enters only as a small correction. The JSD then enters its small-deviation Taylor regime, so the per-bin contribution is suppressed by the square of the smaller relative tail. Schematically, in the local WKB notation this gives an off-node per-bin contribution scaling as $e^{-2N\max(s_P(\mu),s_R(\mu))}$, and using $s_P+s_R=\Sinst$ with $\max(s_P,s_R)\ge\Sinst/2$, this is no slower than $e^{-N\Sinst}$ and typically faster away from the central core. Summing across the spectrum, the dominant contribution to the total JSD comes from a narrow central core of Dicke bins around $m=0$; the $m=0$ bin alone carries $\approx 3$--$5\%$ of the total at all $N$ in our range, and the surrounding central core (a few neighbouring bins) makes up the bulk. The Chernoff information follows the same picture, with its deficit from unity governed by the same central-core mechanism. Figure~\ref{fig:jsd_anatomy} below verifies this directly.

The inner-barrier fraction rises from $61.4\%$ at $N=200$ to $99.99\%$ at $N=2500$ (effectively $\approx 100\%$), while the lobe fraction drops from $22.5\%$ to below $0.01\%$. The single $m=0$ bin contributes a roughly constant $\approx 3$--$5\%$ throughout; this is consistent with a central-node core that is localised on a narrow band of Dicke bins, with the surrounding inner-barrier region picking up further $e^{-N\Sinst}$ contributions from the few bins still inside the central region where the two well-wavefunction tails $\psi_P(m)$, $\psi_R(m)$ are comparably small (\emph{not} where $p_m^{(0)}$ and $p_m^{(1)}$ are equal, since by parity $p_0^{(1)}=0$ while $p_0^{(0)}>0$). The lobe contribution is pushed down by the Taylor-regime suppression and is exponentially smaller than the central-core contribution, becoming negligible for $N\gtrsim 1000$ in the ED data.

\subsection{Independent leading-exponent extraction}
\label{subsec:subleading}
Beyond rolling-window slope fits, we can extract the leading exponent independently from each readout quantity by fitting a parametrised subleading-correction model. We use the WKB-motivated three-parameter model
\begin{equation}
\log Q(N) = -N\cdot S + \alpha\log N + \log A,
\label{eq:subleading_model}
\end{equation}
with $S$, $\alpha$, $\log A$ all free. Here $\alpha$ is an effective polynomial-prefactor exponent; spin-WKB amplitudes generically produce polynomial prefactors, but we do not derive the expected $\alpha$ values analytically here. We perform a sliding lower-cutoff analysis: for each $N_{\min}\in\{100,200,\dots,2000\}$, we fit the model on the subset $N\ge N_{\min}$ and report the best-fit $S$.

\begin{figure}[htbp]
\centering
\includegraphics[width=\textwidth]{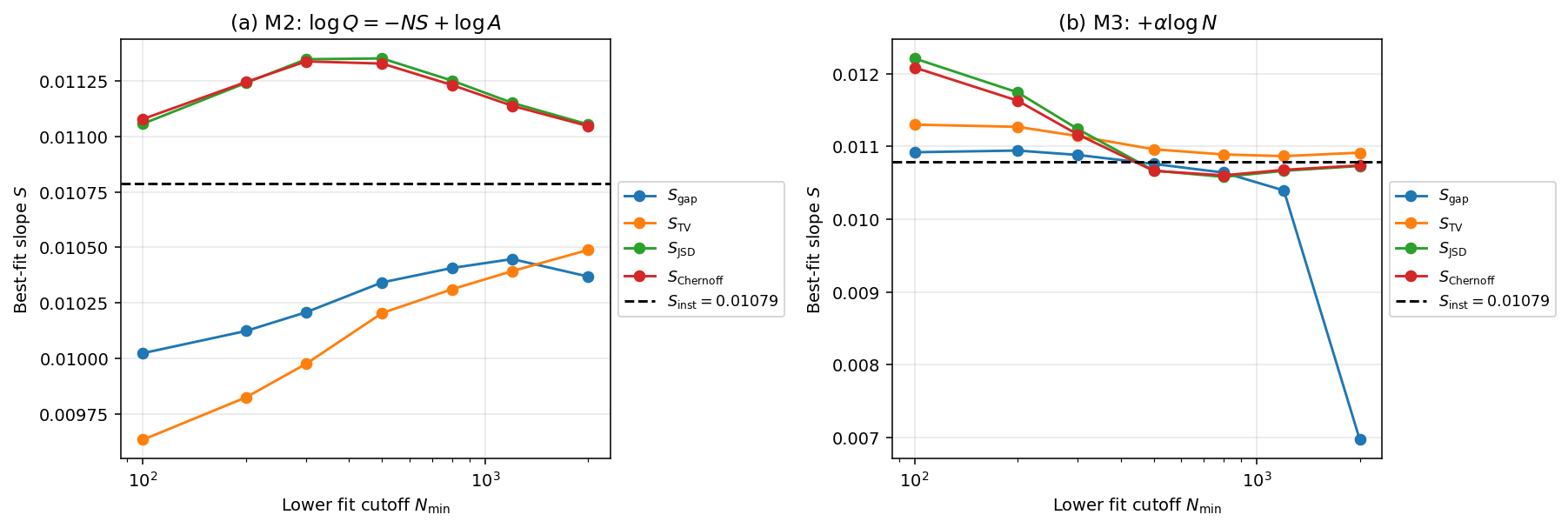}
\caption{\textbf{Sliding lower-cutoff leading-exponent extraction.} For each readout quantity $Q\in\{\Delta E,\TV,\JSD,\text{Chernoff}\}$, the leading exponent $S$ is fitted independently using model M2 ($\log Q=-NS+\log A$, panel a) and M3 ($+\alpha\log N$, panel b) on the subset $N\ge N_{\min}$. \emph{(a)} M2 converges slowly because the omitted polynomial prefactor distorts the slope. \emph{(b)} For intermediate lower cutoffs, the M3 fits remain stable and their error bands are consistent with the instanton value $\Sinst=0.01079$ (black dashed) for every readout quantity. Polynomial-prefactor exponents $\alpha$ depend on the amplitude structure of each quantity: the spectral splitting, TV, and the nonlinear information functionals (JSD, Chernoff) carry different effective prefactors reflecting their distinct functional dependence on the well wavefunctions. At $N_{\min}=2000$ the fit becomes poorly constrained because too few reliable high-$N$ points remain after the magnitude-floor mask; the large error bar is compatible with $\Sinst$, but the central value at this cutoff should not be interpreted as an independent asymptotic estimate.
\label{fig:subleading}}
\end{figure}

The result is summarised in Fig.~\ref{fig:subleading}. The two-parameter (M2, $\alpha\equiv 0$) and three-parameter (M3, $\alpha$ free) fits both yield $S\to\Sinst$ as $N_{\min}\to\infty$, but the M3 fit converges much faster: at $N_{\min}=800$, M3 gives
\begin{equation*}
\{S_{\rm gap}, S_{\rm TV}, S_{\JSD}, S_{\rm Chernoff}\}_{\rm M3,N_{\min}=800} = \{0.01064, 0.01089, 0.01058, 0.01060\},
\end{equation*}
each within about $2\%$ of $\Sinst=0.01079$; the extracted polynomial-prefactor exponents are
\begin{equation*}
\alpha\in\{+0.37,\;+0.97,\;-0.98,\;-0.95\}\quad\text{for}\quad\{\Delta E,\TV,\JSD,\text{Chernoff}\}\text{ respectively.}
\end{equation*}
The nonlinear information functionals (JSD and Chernoff) cluster at $\alpha\approx -1$ in the M3 fits; the doublet splitting and TV have positive $\alpha$, consistent with a polynomial amplitude growth from the pre-asymptotic WKB prefactor. We interpret these fitted $\alpha$ values as empirical prefactor diagnostics, not as first-principles prefactor exponents. The total variation $\TV$ is linear in $|p^{(0)}-p^{(1)}|$, and the WKB cross term $p_m^{(0)}-p_m^{(1)}=2\psi_P(m)\psi_R(m)\sim e^{-N\Sinst}$ (path additivity, Eq.~\eqref{eq:path_additivity}) gives the $\TV$ exponent \emph{directly}, not via any Pinsker-type inequality (which would be too loose by a square root). The Pinsker--Lin bound $\TV^2\le 2\JSD$~\cite{Lin1991} is satisfied by the data (ratio $\TV^2/\JSD\approx 0.73$ at the benchmark, well below the bound of $2$) but is consequence not cause. These polynomial signatures are consistent with the expected spin-WKB prefactor structure~\cite{Braun2007spin}, although a rigorous prefactor calculation is beyond our present analytic scope.

\subsection{$\delta$ effective parameter and its drift}
\label{subsec:delta_fit}
We characterise the backaction-side correction $\delta(g,N)$ defined in \eqref{eq:J01_HP}. We compute $|J_{01}|$ over $N\in[100,6000]$ at $g=0.95$ and fit the residual $N\mstar/2-|J_{01}|$ to three nested models:
\begin{align}
\text{M1:} \quad &\delta(N)=\delta_\infty,\\
\text{M2:} \quad &\delta(N)=\delta_\infty+c/N,\\
\text{M3:} \quad &\delta(N)=\delta_\infty+c/N+d/N^2.
\end{align}

\paragraph*{Restricted-range fit.} The function $\delta(N)=N\mstar/2-|J_{01}|$ is non-monotonic at small $N$: it rises from $\delta\approx 2.6$ at $N=100$ (where the well states are not yet effectively orthogonal and the HP+orthogonalisation expansion does not apply) to a peak $\delta\approx 8.3$ near $N\sim 350$, then decreases monotonically. The asymptotic intercept is therefore extracted by restricting the fit to $N\ge 1000$, where the HP saddle-state overlap is already exponentially small in $N$ and the HP+orthogonalisation picture is well within its apparent asymptotic regime:
\begin{itemize}
\item M1 (constant): $\delta_\infty=6.12$, AIC $\approx -91$.
\item M2 ($\delta_\infty+c/N$): $\delta_\infty=5.810\pm 0.004$, $c=+656\pm 7$, AIC $\approx -236$; $\Delta\mathrm{AIC}(M1\to M2)\approx +144$.
\item M3 ($+d/N^2$): $\delta_\infty=5.846\pm 0.001$, $c=+490$, $d=+1.5\times 10^5$, AIC $\approx -362$.
\end{itemize}
The AIC strongly favours the $1/N$ correction over a constant $\delta$; both M2 and M3 give consistent asymptotes $\delta_\infty\approx 5.81$--$5.85$. The dominant uncertainty on the asymptotic intercept is the M2/M3 model-spread (a few $\times 10^{-2}$), which exceeds the formal $\pm$ values; the quoted formal uncertainties are least-squares model-condition estimates on deterministic ED residuals, not statistical sampling errors. An independent stability check at $N_{\min}=2000$ returns $\delta_\infty=5.83\pm 0.001$ (M2), consistent with the $N_{\min}=1000$ result within the model-spread. We therefore adopt $\delta_\infty(g{=}0.95)=5.8\pm 0.1$ as a conservative rounded value, with the M2/M3 model spread dominating over formal least-squares errors.

Extension to $N\gtrsim 10^4$ is numerically limited because extracting $\delta=N\mstar/2-|J_{01}|$ requires subtracting two $O(N)$ quantities whose difference is $O(1)$. This is a finite-precision limitation, not a physical one.

A first-principles HP+Bogoliubov derivation of $\delta_\infty(g)$, including the orthogonalisation correction between saddle-point and exact well states, would close this loose end. We flag it as the most natural open analytical question on the backaction side.

\subsection{Doublet leakage scaling}
\label{subsec:leakage}
We characterise the off-doublet leakage $\ell_{\rm leak}=L_{\rm leak}/G_{01}$ over $N\in[50,6000]$ at $g=0.95$. Throughout, ``leakage'' here means \emph{intra-sector} leakage: $\hat J_z$-matrix-element weight escaping the ground doublet into the higher Holstein--Primakoff phonon rungs \emph{within} the symmetric Dicke sector $j=N/2$. This is distinct from out-of-sector leakage into lower-$j$ multiplets, which vanishes identically in the ideal collective model (where $[\hat{\mathbf J}^2,\hat H_{\rm LMG}]=0$) and arises only from the non-collective perturbations discussed in Sec.~\ref{sec:intro}; we do not quantify the latter here. A power-law fit on the asymptotic range $N\ge 200$ gives
\begin{equation}
\ell_{\rm leak}(N) \;\sim\; N^{-p_{\rm leak}},\qquad p_{\rm leak}\approx 1.0\text{--}1.2,\qquad N\in[200,6000].
\label{eq:leak_fit}
\end{equation}
The exponent $p_{\rm leak}$ is an empirical effective exponent, not a claimed exact rational value, and it is estimator- and window-dependent: a two-parameter $A/N^{p}$ fit over the full grid returns $p\approx 1.25$, whereas local log-slopes over the asymptotic range give $p\approx 1.0$--$1.1$, drifting toward $\approx 1$ at the largest $N$. We therefore quote the conservative range $p_{\rm leak}\approx 1.0$--$1.2$ rather than a sharp value. The leading $1/N$ end of this range is the natural scale of an ordinary Holstein--Primakoff matrix-element correction (rather than a tunnelling-instanton scale), consistent with the algebraic, non-instanton character of the decay; a first-principles HP derivation of the exponent is left open. The fractional doublet weight $\eta_{\rm mismatch}=1-\ell_{\rm leak}\to 1$ as $N\to\infty$ with an approximately $1/N$-to-slightly-faster algebraic rate. Sample values: $\eta_{\rm mismatch}(N=370)=0.863$, $\eta(N=1000)=0.966$, $\eta(N=5000)=0.994$. This supports a controlled strict-doublet approximation in the relative matrix-element sense, increasingly accurate as $N$ grows for $N\gtrsim 500$, with residual leakage tracked explicitly by $\ell_{\rm leak}$. The \emph{absolute} off-doublet leakage budget is a separate question addressed below.

\paragraph*{Spectral isolation vs.\ absolute leakage.} We use ``isolated'' in this paper in the spectral sense: the ground doublet is separated from the phonon ladder by the HP gap $E_2-E_1\to\hbar\omega_{\rm HP}=O(J)$, while the doublet splitting itself closes exponentially. This does not imply that the full Lindblad dynamics is closed within the doublet at fixed $\gamma_\phi$. Since $G_{01}\sim N^2$ and $\ell_{\rm leak}\sim N^{-p_{\rm leak}}$ with $p_{\rm leak}\approx 1.0$--$1.2$, the \emph{absolute} off-doublet variance $L_{\rm leak}=G_{01}\,\ell_{\rm leak}$ scales as $N^{2-p_{\rm leak}}\approx N^{0.8}$--$N^{1.0}$ (sublinear-to-linear growth), and the absolute off-doublet pumping rate $\gamma_\phi L_{\rm leak}$ grows with $N$ at fixed $\gamma_\phi$; the important point is the growth, not the precise effective exponent. Different thermodynamic scaling conventions give different conclusions: fixed $\gamma_\phi$ gives a growing $\Gamma_{\rm leak}\sim\gamma_\phi N^{2-p_{\rm leak}}$; if $\gamma_\phi=\gamma_\phi^{(1)}/N$ with fixed $\gamma_\phi^{(1)}$, then $\Gamma_{\rm leak}\sim\gamma_\phi^{(1)}N^{1-p_{\rm leak}}$ (vanishes for $p_{\rm leak}>1$, marginal near $p_{\rm leak}\approx 1$), while the dominant intra-doublet rate $\Gamma_z^{(\rm par)}=2\gamma_\phi|J_{01}|^2$ still scales as $N$; if $\gamma_\phi=\gamma_\phi^{(2)}/N^2$ with fixed $\gamma_\phi^{(2)}$, then the dominant intra-doublet rate is $O(1)$ and the absolute leakage vanishes. We deliberately do not commit to a single thermodynamic scaling. The large-$N$ leakage result of this paper should therefore be read as a relative matrix-element statement ($\eta_{\rm mismatch}\to 1$), not as an absolute dynamical decoupling theorem. At the $N=370$ benchmark with $\gamma_\phi=0.05~\mathrm{s}^{-1}$, the absolute off-doublet rate is $\gamma_\phi L_{\rm leak}\approx 19~\mathrm{s}^{-1}$, which is $\approx 13\%$ of the intra-doublet coherence-decay rate $\Gamma_{01}=142~\mathrm{s}^{-1}$: significant but not dominant. In the relative matrix-element sense, the strict-doublet approximation used in Sec.~\ref{sec:operational} \emph{improves} with $N$ because $\gamma_\phi L_{\rm leak}/\Gamma_{01} = L_{\rm leak}/G_{01} = \ell_{\rm leak} \to 0$. At fixed $\gamma_\phi$, however, the absolute leakage rate $\gamma_\phi L_{\rm leak} \sim \gamma_\phi N^{2-p_{\rm leak}}$ grows with $N$, so for protocols with a fixed laboratory observation time the absolute leakage timescale must still be checked separately.

\begin{figure}[!ht]
\centering
\includegraphics[width=0.96\columnwidth]{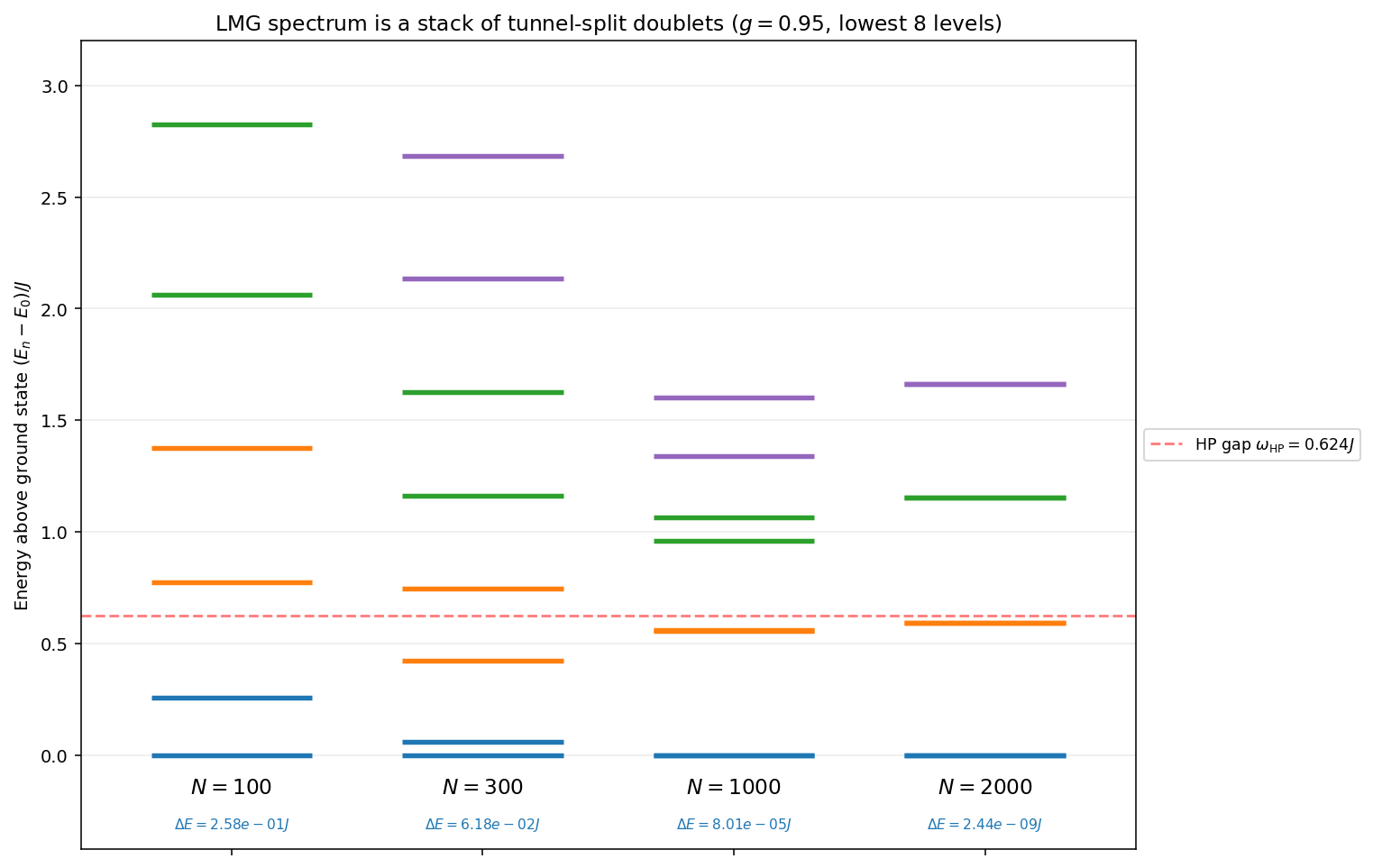}
\caption{\textbf{LMG spectrum is a stack of tunnel-split doublets ($g=0.95$, lowest 8 levels).} Energy levels above the ground state, at four representative $N$. Each color is a doublet pair: ground doublet (blue), 1st phonon doublet (orange), 2nd (green), 3rd (purple). The ground-doublet splitting $\Delta E_{\rm doublet}$ shrinks from $0.26\,J$ at $N=100$ to $2.4\times 10^{-9}\,J$ at $N=2000$ --- eight orders of magnitude --- while the phonon-ladder spacing approaches the asymptotic $\hbar\omega_{\rm HP}=0.624\,J$. The off-doublet leakage $\ell_{\rm leak}$ characterised in this section is the aggregate $\hat J_z$-matrix-element weight from the ground doublet outside the doublet subspace; over the fitted ED range it decays algebraically, with an effective exponent $\approx 1.0$--$1.2$ (estimator-dependent), consistent with a leading $1/N$ Holstein--Primakoff matrix-element correction.\label{fig:spectrum_levels}}
\end{figure}

Figure~\ref{fig:spectrum_levels} makes the relevant off-doublet spectral structure concrete. At $N=100$ the parity doublet is two visibly split levels at the bottom of the spectrum; by $N=2000$ the doublet has collapsed onto a single line at the resolution of the figure ($\Delta E_{\rm doublet}=2.4\times 10^{-9}\,J$). The next pair of levels (orange) is the first \emph{phonon doublet}: each rung of the Holstein--Primakoff ladder is itself a tunnel-split pair, with intra-rung splitting also exponentially small, while the rung-to-rung spacing approaches the asymptotic value $\hbar\omega_{\rm HP}=0.624\,J$ with finite-$N$ corrections. Off-doublet $\hat J_z$ leakage is naturally interpreted as matrix-element weight from the ground doublet into the HP phonon ladder. A rung-by-rung decomposition is not attempted here; the empirical algebraic decay $\ell_{\rm leak}\sim N^{-p_{\rm leak}}$ ($p_{\rm leak}\approx 1.0$--$1.2$) characterises the aggregate weight outside the doublet rather than its bin-by-bin destination.

\subsection{Dynamical regime markers}
\begin{figure*}[t]
\centering
\includegraphics[width=0.96\textwidth]{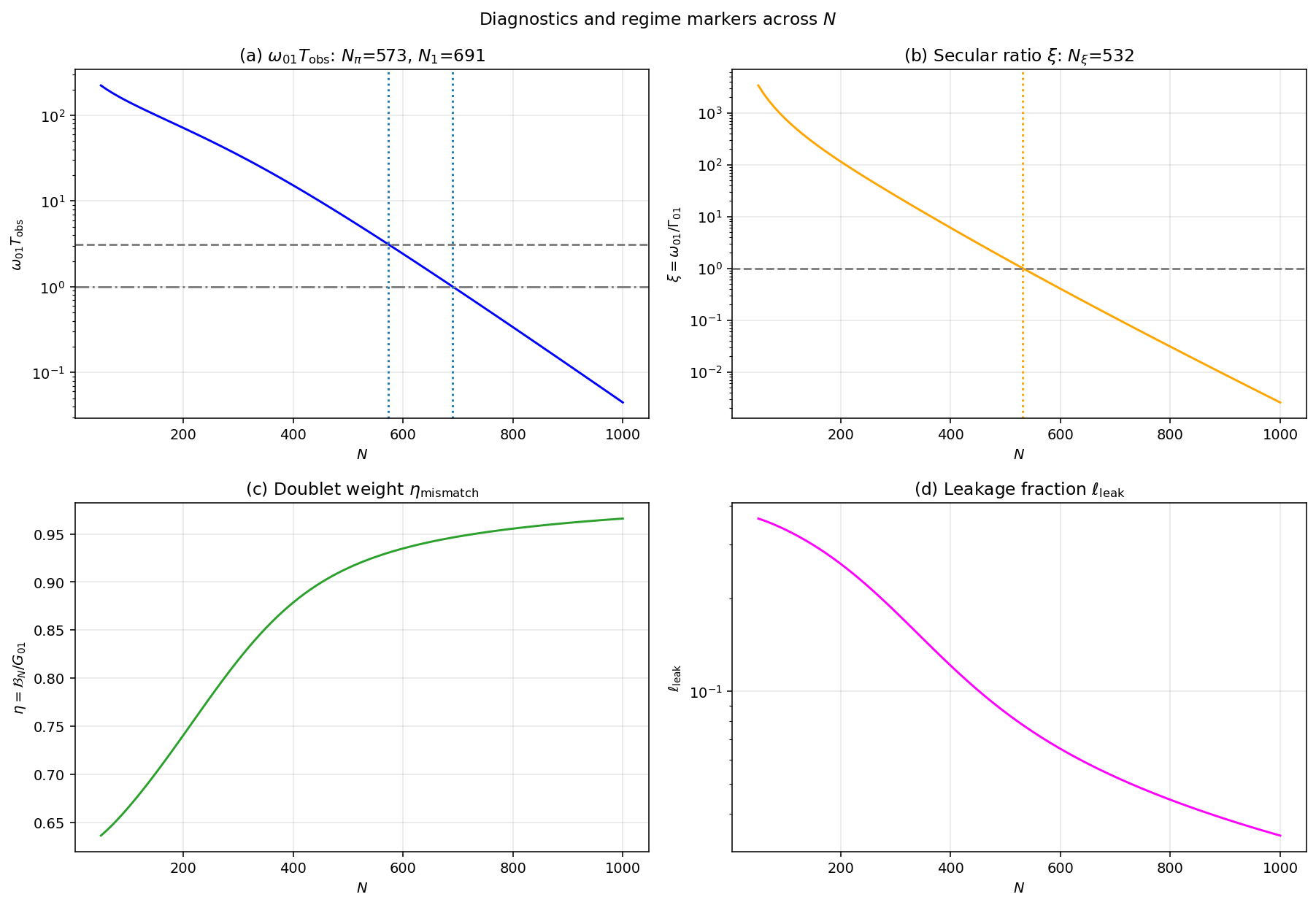}
\caption{\textbf{Diagnostic quantities and regime markers across $N$.}
\emph{(a)} Phase per observation window $\omega_{01}\Tobs$ for $\Tobs=15~\mathrm{ms}$: crosses $\pi$ at $N_\pi\simeq 575$ and $1$ at $N_1\simeq 690$. \emph{(b)} Secular ratio $\xi=\omega_{01}/\Gamma_{01}$: crosses unity at $N_\xi\simeq 530$, the boundary between the secular side and the bath-dominated/post-secular side. \emph{(c)} Doublet weight $\eta_{\rm mismatch}=\BN/G_{01}$ rises from $0.63$ to $0.97$. \emph{(d)} Leakage fraction $\ell_{\rm leak}$ on log scale, falling from $0.37$ to $0.03$ as the off-doublet $\Jz$-variance becomes a negligible correction.
\label{fig:diagnostics}}
\end{figure*}

Three additional scales characterise the dynamics of the doublet at the benchmark parameters introduced above. Two of these are window-dependent (set by the choice of finite observation window $T_{\rm obs}$), the third is intrinsic. We use $T_{\rm obs}=15~\mathrm{ms}\approx 2T_2$ at the benchmark as the reference value. The phase $\omega_{01}T_{\rm obs}$ accumulated at fixed $N$ scales linearly with $T_{\rm obs}$, but the crossing values $N_\pi$ and $N_1$ shift only approximately logarithmically with $T_{\rm obs}$ because $\omega_{01}(N)$ is exponentially small in $N$; the intrinsic scale $N_\xi$ does not depend on $T_{\rm obs}$.
\begin{itemize}
\item Secular ratio crosses unity at $N_\xi$, where $\xi=\omega_{01}/\Gamma_{01}=1$. For $g=0.95$ we find $N_\xi\simeq 530$. For $N<N_\xi$ the doublet precession is fast compared with the bath rate (the secular side); for $N>N_\xi$ the bath-dominated/post-secular side begins, with Zeno-side behaviour becoming more pronounced as $\omega_{01}T_{\rm obs}$ falls below unity.
\item Phase-half-period crossing at $N_\pi$, where $\omega_{01}\Tobs=\pi$. For our $\Tobs=15~\mathrm{ms}$ we find $N_\pi\simeq 575$.
\item Phase-unit crossing at $N_1$, where $\omega_{01}\Tobs=1$. For $\Tobs=15~\mathrm{ms}$, $N_1\simeq 690$.
\end{itemize}
These three scales (Table~\ref{tab:regime_markers}) define the regime in which the readout--backaction separation has its most pronounced dynamical consequences. The diagnostics figure (Fig.~\ref{fig:diagnostics}) plots the underlying quantities.

\begin{table}[h]
\centering
\caption{Regime markers and key crossover scales at $g=0.95$, $\gamma_\phi=0.05~\mathrm{s}^{-1}$, $J/\hbar=3.72\times 10^4~\mathrm{rad\,s}^{-1}$, $\Tobs=15~\mathrm{ms}$. $N_\xi\simeq 530$ marks the loss of secularity ($\xi=1$); $N_1\simeq 690$ marks the phase-unit crossing $\omega_{01}\Tobs=1$ and is a diagnostic marker, \emph{not} the Zeno closure. The dynamical information window (Sec.~\ref{sec:dynamical}) is broad: the full-record advantage opens in the crossover region as $\xi$ approaches order unity (a resolved positive excess is already present by $N\approx 500$, $\xi\simeq 1.5$), strengthens through $N\approx 950$, and closes (excess consistent with zero) only near $N_{\rm close}\sim 1000$ ($\xi\lesssim$ a few $\times 10^{-3}$), well beyond $N_1$. We do not assign a sharp lower onset threshold.}
\label{tab:regime_markers}
\begin{tabular}{l c l}
\toprule
Scale & Value & Condition \\
\midrule
$N_\xi$        & 530 & $\xi=\omega_{01}/\Gamma_{01}=1$ \\
$N_\pi$        & 575 & $\omega_{01}\Tobs=\pi$ \\
$N_1$          & 690 & $\omega_{01}\Tobs=1$ \\
\bottomrule
\end{tabular}
\end{table}

\clearpage
\section*{Data availability}
The complete computational pipeline and the cached numerical results that
reproduce every figure and the benchmark table of this paper --- the
exact-diagonalisation code for the static readout and backaction quantities and
the stochastic-master-equation/Girsanov-classifier code for the dynamical
bypass, together with the cached \texttt{.npz} archives, the figure scripts, a
SHA-256 manifest, and the pinned software environment --- are openly available
in a Zenodo archive~\cite{ParityHorizonData}. The in-text numerical values are
single-sourced from these data via the included macro file, so the values
reported here are generated from the archived results rather than transcribed.

\end{document}